\documentclass[final,3p]{elsarticle}
\usepackage{amsmath}
\usepackage{amssymb}
\usepackage{amsfonts}
\usepackage{amsthm}
\usepackage{color}
\usepackage[T1]{fontenc}
\usepackage{bm}
\usepackage{tikz}
\usepackage{adjustbox}
\usetikzlibrary{matrix,calc,arrows.meta}
\usepackage{tabulary}
\usepackage{tabularx}
\usepackage{booktabs}   
\usepackage{multirow}   
\usepackage{array}
\usepackage{graphicx}
\usepackage{subcaption}
\usepackage{pgfplots}
\pgfplotsset{compat=1.18}
\pgfplotsset{colormap={coolwarmC}{rgb=(0.2298,0.2987,0.7537) rgb=(0.3042,0.4069,0.8453) rgb=(0.3837,0.5102,0.9178) rgb=(0.4677,0.6056,0.9685) rgb=(0.5543,0.6901,0.9955) rgb=(0.6408,0.7608,0.9978) rgb=(0.7240,0.8149,0.9757) rgb=(0.8006,0.8504,0.9300) rgb=(0.8674,0.8644,0.8626) rgb=(0.9256,0.8255,0.7711) rgb=(0.9595,0.7670,0.6741) rgb=(0.9697,0.6905,0.5751) rgb=(0.9567,0.5980,0.4773) rgb=(0.9214,0.4914,0.3834) rgb=(0.8654,0.3711,0.2958) rgb=(0.7906,0.2314,0.2162) rgb=(0.7057,0.0156,0.1502)}}
\pgfplotsset{colormap={greensRevC}{rgb=(0.0000,0.2667,0.1059) rgb=(0.0000,0.3423,0.1373) rgb=(0.0000,0.4230,0.1707) rgb=(0.0651,0.4833,0.2191) rgb=(0.1340,0.5423,0.2683) rgb=(0.1935,0.6051,0.3156) rgb=(0.2526,0.6681,0.3629) rgb=(0.3514,0.7179,0.4120) rgb=(0.4518,0.7671,0.4612) rgb=(0.5407,0.8087,0.5333) rgb=(0.6293,0.8500,0.6061) rgb=(0.7044,0.8817,0.6790) rgb=(0.7792,0.9132,0.7518) rgb=(0.8385,0.9370,0.8149) rgb=(0.8976,0.9606,0.8779) rgb=(0.9332,0.9745,0.9194) rgb=(0.9686,0.9882,0.9608)}}
\pgfplotsset{colormap={plasmaC}{rgb=(0.0504,0.0298,0.5280) rgb=(0.1934,0.0184,0.5903) rgb=(0.2999,0.0096,0.6316) rgb=(0.3994,0.0009,0.6561) rgb=(0.4949,0.0120,0.6579) rgb=(0.5844,0.0686,0.6328) rgb=(0.6651,0.1386,0.5856) rgb=(0.7360,0.2094,0.5279) rgb=(0.7982,0.2802,0.4695) rgb=(0.8533,0.3516,0.4137) rgb=(0.9018,0.4251,0.3597) rgb=(0.9426,0.5026,0.3058) rgb=(0.9734,0.5858,0.2515) rgb=(0.9914,0.6754,0.1985) rgb=(0.9930,0.7717,0.1548) rgb=(0.9744,0.8746,0.1441) rgb=(0.9400,0.9752,0.1313)}}
\usepackage{placeins}
\usepackage{xfrac}
\usepgfplotslibrary{groupplots}
\usepackage{tcolorbox}
\usepackage{algorithm}
\usepackage[noend]{algpseudocode}
\usepackage{standalone}
\usepackage{hyperref}
\usepackage[capitalise]{cleveref}
\usepackage{parskip}
\makeatletter
\renewcommand{\paragraph}{%
  \@startsection{paragraph}{4}{\z@}%
  {3.25ex \@plus1ex \@minus.2ex}%
  {-0.6em}%
  {\normalfont\normalsize\bfseries\paragraphdot}%
}
\newcommand{\paragraphdot}[1]{#1\ }
\makeatother
\definecolor{clr1}{RGB}{0,84,159}       
\definecolor{clr2}{RGB}{161,16,53}      
\definecolor{clr3}{RGB}{0,97,101}       
\definecolor{clr4}{RGB}{246,168,0}      
\definecolor{clr5}{RGB}{87,171,39}      
\definecolor{clr6}{RGB}{156,158,159}    
\definecolor{clr7}{RGB}{100,101,103}    
\definecolor{clr8}{RGB}{122,111,172}    
\definecolor{clr9}{RGB}{0,152,161}      
\definecolor{clr10}{RGB}{227,0,102}     
\definecolor{rwth1}{RGB}{0,84,159}      
\definecolor{rwth2}{RGB}{142,186,229}   
\definecolor{rwth3}{RGB}{0,97,101}      
\definecolor{rwth4}{RGB}{0,152,161}     
\definecolor{rwth5}{RGB}{87,171,39}     
\definecolor{rwth6}{RGB}{189,205,0}     
\definecolor{rwth7}{RGB}{255,237,0}     
\definecolor{rwth8}{RGB}{246,168,0}     
\definecolor{rwth9}{RGB}{227,0,102}     
\definecolor{rwth10}{RGB}{204,7,30}     
\definecolor{rwth11}{RGB}{161,16,53}    
\definecolor{rwth12}{RGB}{97,33,88}     
\definecolor{rwth13}{RGB}{122,111,172}  
\definecolor{loss1}{RGB}{0,114,178}   
\definecolor{loss2}{RGB}{213,94,0}    
\definecolor{loss3}{RGB}{0,158,115}   
\definecolor{loss4}{RGB}{204,121,167} 
\definecolor{loss5}{RGB}{230,159,0}   
\definecolor{ekA}{RGB}{5,48,97}      
\definecolor{ekB}{RGB}{67,147,195}   
\definecolor{ekC}{RGB}{161,230,242}
\definecolor{ekD}{RGB}{244,165,130}  
\definecolor{ekE}{RGB}{178,24,43}    
\newcommand{\RN}[1]{\uppercase\expandafter{\romannumeral#1}}
\usepackage{lineno}
\modulolinenumbers[100]
\journal{}

\definecolor{matrixfill}{gray}{0.90}
\definecolor{phasefill}{HTML}{8FD19E}
\definecolor{phaseline}{HTML}{8FD19E}

\newcommand{\hdim}[4]{%
  \draw[thin] (#1,-0.3) -- (#1,#3-0.3);
  \draw[thin] (#2,-0.3) -- (#2,#3-0.3);
  \draw[<->,>={Stealth[length=2mm]},thin] (#1,#3) -- (#2,#3)
     node[midway,fill=white,inner sep=1pt,font=\small] {$#4$};
}
\newcommand{\vdim}[4]{%
  \draw[thin] (-0.3,#1) -- (#3-0.3,#1);
  \draw[thin] (-0.3,#2) -- (#3-0.3,#2);
  \draw[<->,>={Stealth[length=2mm]},thin] (#3,#1) -- (#3,#2)
     node[midway,fill=white,inner sep=1pt,font=\small,rotate=90] {$#4$};
}
\newcommand{\cmark}[2]{%
  \draw[thin,dash dot] (#1-0.9,#2) -- (#1+0.9,#2);
  \draw[thin,dash dot] (#1,#2-0.9) -- (#1,#2+0.9);
}
\newcommand{\beq}{\begin{equation}}
\newcommand{\eeq}{\end{equation}}

\begin{document} 
\begin{frontmatter}
\title{Hyperreduction accelerates finite element model updating for the discovery of neural materials}
\cortext[cor1]{Corresponding author}
\author[add1]{Hagen Holthusen\corref{cor1}} 
\ead{holthusen@ame.rwth-aachen.de}
\author[add1]{Matthias Knipper}
\ead{knipper@ame.rwth-aachen.de}
\author[add1]{Kevin Linka}
\ead{linka@ame.rwth-aachen.de}
\address[add1]{Computational Mechanics in Medicine, Institute of Applied Medical Engineering, \\RWTH Aachen University, Pauwelsstra{\ss}e 20, 52074 Aachen, Germany}
\begin{abstract}
Finite element model updating is among the most general and robust approaches to discover constitutive models from structural experiments: since the displacement field is computed rather than measured, it accommodates arbitrary geometries, boundary conditions, and sparse or noisy data.
However, every evaluation of the loss requires a nonlinear finite element solve per load step, and the training of neural materials with a large number of weights and biases requires many such evaluations.
Intrusive model order reduction is well suited to lower this cost, since it reduces the finite element model rather than replacing it and thus retains its boundary conditions and variational structure.
This raises the question: How can we construct a reduced model that remains accurate while the weights of the network change at every step of the training?
Here, we combine proper orthogonal decomposition and energy-conserving sampling and weighting with adjoint sensitivities of the reduced model, and adapt both the trial space and the sampled elements along the optimization path.

The constitutive law is represented by a polyconvex neural network.
To this end, we introduce a logarithmic normalization term which, together with a zero-anchored input convex architecture and a growth term, renders the energy polyconvex, coercive, and non-negative by construction, independent of the trained weights.

We demonstrate the approach on two experimental datasets and one synthetic three-dimensional example.
For a two-phase specimen, the training time decreases from nine hours to $41$ minutes compared with a full-order identification, without loss of accuracy.
For the three-dimensional example, six modes replace $48\,672$ degrees of freedom, $116$ of $15\,625$ elements suffice to evaluate the internal forces, and the entire training takes $13$ minutes.
In all cases, the identified material reproduces the measured reaction force and predicts the displacement field from the force alone.
Our results show that hyperreduction makes the discovery of neural materials from structural data affordable, while retaining the advantages of finite element model updating: it accommodates arbitrary geometries and boundary conditions as well as sparse, noisy, or force-only data.
\end{abstract}
\end{frontmatter}
\section{Introduction}
\label{sec:intro}

A finite element simulation is only as predictive as its constitutive model.
Neural materials, that is, constitutive models in which neural networks parameterize the energy and the dissipation potential \citep{fuhg2024review}, have changed how we obtain these models:
constitutive artificial neural networks \citep{linka2021,linka2023}, physics-augmented and input-convex architectures \citep{klein2022,flaschel2025convex}, and their extensions to inelasticity \citep{holthusen2024,rosenkranz2024,holthusen2026generalized,holthusen2026complement,kalina2026} and multiphysics \cite{klein2024,fuhg2024,ortigosa2025,holthusen2026thermo} discover energies and dissipation potentials that satisfy thermodynamic principles by construction.
The data to train them, however, are rarely pairs of stress and strain.
What we measure in a structural experiment are displacement fields and reaction forces.
This raises the question: How do we discover a neural material from measurements of a structure, and can we do so at a cost that is practical?

\paragraph{Discovery from structural data.}
Full-field measurements probe a rich set of deformation states within a single specimen, but stresses remain inaccessible, such that the material law has to be inferred from indirect observables \citep{fuhg2024review,avril2008}.
The available methods differ in whether they solve the forward problem \citep{avril2008,roemer2025,riemer2026,friedrichs2026}.
Methods that do not, such as the equilibrium gap \citep{avril2008,claire2004} and the virtual fields method \citep{gsediac1989,pierron2012}, evaluate balance of linear momentum directly on the measured displacement field.
This idea underlies the unsupervised EUCLID framework \citep{flaschel2021euclid}, which has been extended to neural network models \citep{thakolkaran2022}, constitutive artificial neural networks \citep{alheit2026}, experimental \citep{abbasi2026} and three-dimensional data \citep{bourdyot2026}, heterogeneous specimens \citep{chaurasiya2026,tac2026}, sparse and noisy data \citep{knauf2026}, and optimized specimen design \citep{ghouli2025}.
The equilibrium gap method has recently been used to calibrate physics-augmented neural networks for viscoelasticity \citep{riemer2026} and finite-strain hyperelasticity \citep{friedrichs2026}, and the modified constitutive relation error offers a related route for thermodynamically consistent networks \citep{benady2024}.
Material Fingerprinting \citep{flaschel2026fingerprint,flaschel2026adaptive} and neural operators \citep{flaschel2026neuralop} go one step further and replace the optimization by a database lookup or by a single forward pass of a network trained on many specimens.
These methods are fast, yet they rely on a dense displacement field, from which strains are computed, and on assumptions such as plane stress.
It was shown that an under-resolved surface measurement and a non-negligible specimen thickness can affect the calibration \cite{friedrichs2026}, and that the coverage of deformation states governs how well the identified model generalizes.
Simulation-based approaches instead compare measured and simulated responses in the spirit of finite element model updating \citep{mahnken1996}.
Since the displacement field is the solution of a boundary value problem rather than an input, these approaches accommodate sparse and noisy data and, in principle, any specimen geometry, load path, and boundary condition that the finite element model resolves.
They have been used to calibrate spline-based models \citep{wiesheier2024,wiesheier2026}, physics-augmented neural networks \citep{wu2025,tan2026,knipper2026}, constitutive artificial neural networks \citep{knipper2026}, neural elastoplasticity \citep{gavris2025,ferreira2026}, and cohesive zone laws \citep{gavris2026}.
The same route extends to coupled problems, in which displacements, temperature, and reaction forces enter a single loss and the adjoint of the coupled residual provides the gradient \citep{spencer2026}.
Differentiable finite element solvers make this route practical, with sensitivities obtained by automatic differentiation or adjoints \citep{xue2023jaxfem,knipper2026,ferreira2026,regazzoni2026}, and beyond finite elements for image-based observations with particle methods \citep{xu2026}.
In the terminology of \citep{roemer2025}, these are reduced approaches, in which the state is eliminated by solving the boundary value problem in every iteration.

\paragraph{Cost of simulation-based identification.}
The robustness of this route comes at a computational price.
Each evaluation of the loss requires a nonlinear finite element solution for every load step, and each gradient additionally requires an adjoint solve with the tangent stiffness.
For neural materials, whose parameters are weights and biases, this procedure is repeated over many optimization steps.
Since the generalization of an identified model improves with the combination of specimens and load cases \citep{friedrichs2026}, the number of such solves per loss evaluation grows with the richness of the data.
GPU acceleration and automatic differentiation reduce the cost of a single solve \citep{xue2023jaxfem,knipper2026}, but not the number of degrees of freedom that enter it.
This motivates us to ask: Can we reduce the cost of the forward model itself, while retaining the finite element model, its boundary conditions, and the physical structure of the constitutive network?

\paragraph{Model order reduction and hyperreduction.}
Projection-based model order reduction addresses this question in two steps \citep{benner2015}.
First, a reduced basis, typically obtained by proper orthogonal decomposition (POD) of solution snapshots, restricts the displacement field to a low-dimensional subspace.
Second, because the nonlinear internal forces must still be assembled over all elements, hyperreduction approximates their evaluation on a small subset of the mesh.
Hyperreduction follows two lines.
Interpolation-based methods, such as the discrete empirical interpolation method \citep{chaturantabut2010} and the gappy-POD-based GNAT \citep{carlberg2013}, approximate the nonlinear term by a few of its entries.
Cubature-based methods select and weight a subset of elements.
They start from the pioneering works of hyperreduction and cubature \citep{ryckelynck2005,an2008}, and lead to energy-conserving sampling and weighting (ECSW) \citep{farhat2015} and the empirical cubature method (ECM) \citep{hernandez2017}.
A recent variant does not select a subset of the integration points but places new ones in strain space, by clustering and subsequent empirical correction (E3C) \citep{wulfinghoff2025e3c}.
In solid mechanics, POD with empirical interpolation was applied to hyperelasticity \citep{radermacher2016}, and recent studies extend reduced models to finite-strain homogenization \citep{hernandez2014,kunc2019,lange2024}, nearly incompressible materials \citep{shamim2024}, viscoelasticity \citep{trainotti2024}, and plasticity and damage \citep{agouzal2024,kehls2026,ashkpour2026}.
Comparisons of hyperreduction schemes on nonlinear elasticity \citep{larsson2026,faust2026strain} and component-wise reduced models coupled by mortar tying \citep{ritzert2026} address accuracy and reusability.
In parallel, nonlinear manifolds replace the linear subspace \citep{lee2020,scheunemann2025,faust2025manifold}, and non-intrusive surrogates learn the reduced model from data, with autoencoders \citep{kehls2025ae} or physics-augmented neural networks \citep{schuetz2026}.
The latter report a rapid divergence outside the training range, which shows that learned reduced models are reliable only within the range they were trained on.

\paragraph{Reduced models in inverse problems.}
Reduced models have a long record in inverse problems and PDE-constrained optimization \citep{ghattas2021}.
Hyperreduced models have been embedded in gradient-based design optimization \citep{amsallem2015}, and the basis can be progressively constructed along the optimization path to keep it valid where the optimizer goes \citep{zahr2015}.
In solid mechanics, POD- or PGD-based models identify the parameters of hydrogels \citep{agarwal2025}, cardiac tissue \citep{pfaller2020}, elastoplastic materials \citep{buljak2018}, and history-dependent laws \citep{bhattacharyya2026}. 
While these works identify a few parameters of an explicit constitutive law, it was noted that the reduced model is valid only within the range of parameters it was trained on \citep{agarwal2025}.
A different route replaces the boundary value problem by a learned surrogate, for example by parametric physics-informed neural networks for calibration from full-field data \citep{anton2025}.
With this at hand, a neural constitutive model poses a harder question.
Its parameters are a large number of weights and biases that change at every optimization step, so that a basis and a set of elements built at a finite number of parameter vectors must remain reliable along the entire optimization path.
This raises the question: Can we combine POD and ECSW with the adjoint-based discovery of a neural constitutive model, and does the reduced model remain accurate while the parameters evolve?

\paragraph{Research gap.}
Reduced-order models accelerate the identification of parametric constitutive laws \citep{agarwal2025,pfaller2020,buljak2018,bhattacharyya2026}, and hyperreduction is established for forward simulations of hyperelastic and inelastic solids \citep{radermacher2016,kehls2026,ashkpour2026}.
In contrast, the discovery of neural constitutive models from structural data relies on full-order solvers \citep{knipper2026,tan2026,wu2025,gavris2025,ferreira2026,regazzoni2026}, on local equilibrium evaluations of the measured field \citep{thakolkaran2022,alheit2026,riemer2026,friedrichs2026}, or on surrogates that replace the boundary value problem \citep{anton2025,flaschel2026neuralop}. 
To the best of our knowledge, the combination of projection-based reduction with hyperreduction and gradient-based constitutive discovery has not been studied. 

\paragraph{Aim of the study.}
We aim to make inverse material discovery from structural data affordable by reducing the finite element model rather than replacing it.
For this purpose, we combine three ingredients.
First, we construct the reduced model by projecting the equilibrium onto a POD subspace, which for hyperelastic materials corresponds to minimizing the potential on this subspace.
Second, we apply ECSW, which approximates the internal energy by a weighted sum over a small subset of elements.
Third, we derive adjoint sensitivities of the reduced model with respect to the network parameters.
The reduced basis is built from snapshots at different load steps and sampled parameter vectors, and the loss compares measured and predicted reaction forces and, if desired, displacements.

\paragraph{Outline.}
The remainder of this manuscript is organized as follows. 
In \cref{sec:methods}, we formulate the inverse material discovery problem with its adjoint, the reduction techniques employed, and the neural elasticity model, which we aim to discover.
In \cref{sec:results}, we demonstrate the approach on two experimental and one synthetic examples, while \cref{sec:discussion} discusses our findings.
Finally, we conclude our work in \cref{sec:conclusion}.
\section{Methods}
\label{sec:methods}

In this section, we develop a framework to identify the constitutive behavior of a hyperelastic material from structural measurements, i.e., from reaction forces and, optionally, displacement fields. 
We first formulate the discovery as a bilevel optimization problem and derive its gradient by the adjoint method in \cref{sec:inverse_discovery}. 
Since every evaluation of the loss requires the solution of a nonlinear finite element problem, we then employ a subspace projection and a hyperreduction in \cref{sec:reduction}, which lower the computational cost. 
With this at hand, we specify the constitutive model in \cref{sec:neural_elasticity}, which is represented by a physics-augmented-like neural network, i.e. polyconvex neural material, and whose weights and biases are the unknowns of the inverse problem. 
We conclude with the workflow in \cref{sec:workflow}, which adapts the reduction to the evolving weights during training and decides when the reduced model can be trusted.


\subsection{Inverse material discovery}
\label{sec:inverse_discovery}

On the structural level, material parameters cannot be identified from stress--strain pairs directly. Instead, we observe the response of a specimen or component, typically in terms of reaction forces and, if available, displacement fields. For hyperelastic materials, this inverse problem can be elegantly written as a bilevel optimization problem, i.e., as two nested minimization problems,
\begin{equation}
    \bm{\theta}^\ast = \arg \underset{\bm{\theta}}{\min}\, \mathcal{L}\big(\bm{u}^\ast(\bm{\theta}),\bm{\theta}\big)
    \qquad \text{with} \qquad
    \bm{u}^\ast(\bm{\theta}) = \arg \underset{\bm{u}\in\mathcal{U}}{\min}\, \Pi(\bm{u},\bm{\theta}),
\label{eq:bilevel}
\end{equation}
where $\bm{\theta}$ collects the material parameters, $\bm{u}$ denotes the displacement field, $\mathcal{U}$ the set of kinematically admissible displacements, $\Pi$ the total potential energy of the structure, and $\mathcal{L}$ a loss function that quantifies the discrepancy between the predicted and the observed structural response. The inner problem determines the equilibrium state for a given set of material parameters, while the outer problem adjusts the parameters such that the predicted response matches the observations. Note that the loss depends on the parameters in two ways: implicitly, through the equilibrium displacement $\bm{u}^\ast(\bm{\theta})$, and explicitly, for instance through the reaction forces. In the context of neural networks, $\bm{\theta}$ comprises the weights and biases, and solving the outer problem corresponds to training.

In the following, we first state the inner problem and its discretization. We then define the loss function and conclude with the gradient of the loss, which is required to solve the outer problem by means of gradient-based optimization.

\paragraph{Inner problem.}
For hyperelastic materials, the constitutive behavior is fully characterized by the strain energy density $\psi(\bm{F};\bm{\theta})$, where the semicolon indicates that the material parameters act solely as parameters of the potential and $\bm{F} = \bm{I} + \mathrm{Grad}(\bm{u})$ denotes the deformation gradient. The total potential energy of the structure reads
\begin{equation}
    \Pi(\bm{u},\bm{\theta}) = \underbrace{\int_{\mathcal{B}} \psi\big(\bm{F}(\bm{u});\bm{\theta}\big)\,\mathrm{d}V}_{=:\, \Pi_{\mathrm{int}}(\bm{u},\bm{\theta})} - \;\Pi_{\mathrm{ext}}(\bm{u}),
\label{eq:global_potential}
\end{equation}
where $\Pi_{\mathrm{int}}$ denotes the internal and $\Pi_{\mathrm{ext}}$ the external potential. We assume dead loads, such that the external potential is linear in the displacement and independent of the material parameters, e.g., $\Pi_{\mathrm{ext}} = \int_{\mathcal{B}} \bm{f}\cdot\bm{u}\,\mathrm{d}V + \int_{\partial_{\bm{t}}\mathcal{B}} \bar{\bm{t}}\cdot\bm{u}\,\mathrm{d}A$ with body force $\bm{f}$ and prescribed traction $\bar{\bm{t}}$. The boundary conditions read
\begin{equation}
    \bm{u} = \bar{\bm{u}}\ \ \text{on}\ \partial_{\bm{u}}\mathcal{B},
    \qquad
    \bm{P}\cdot\bm{n} = \bar{\bm{t}}\ \ \text{on}\ \partial_{\bm{t}}\mathcal{B},
\end{equation}
where $\bm{P} = \partial\psi/\partial\bm{F}$ is the Piola stress and $\bm{n}$ the outward unit normal in the reference configuration. The essential boundary conditions define the admissible set $\mathcal{U} := \{\bm{u} \,|\, \bm{u} = \bar{\bm{u}} \text{ on } \partial_{\bm{u}}\mathcal{B}\}$, while the natural boundary conditions follow from the stationarity of $\Pi$.\footnote{The existence of a minimizer in \cref{eq:bilevel} requires suitable convexity and growth conditions of $\psi$, e.g., polyconvexity \cite{ball1977}. Numerically, we solve the stationarity condition $\delta_{\bm{u}}\Pi = 0$.}

In the present work, we prescribe only Dirichlet boundary conditions and neglect body forces. Consequently, the external potential vanishes, and the structure is loaded exclusively through the prescribed displacements $\bar{\bm{u}}$. The forces acting on the structure then emerge as reaction forces on $\partial_{\bm{u}}\mathcal{B}$, i.e., as the quantities work-conjugate to the prescribed displacements. These reaction forces are precisely what we measure in the experiment.

\paragraph{Spatial and temporal discretization.}
In time, we discretize the loading path into $N_{\mathrm{t}}$ load steps $t_n$ with prescribed displacements $\bar{\bm{u}}_n$. Since the potential depends neither on rates nor on internal variables, the equilibrium state at $t_n$ is fully determined by $\bar{\bm{u}}_n$. Hence, for quasi-static hyperelasticity, the inner problem decouples into $N_{\mathrm{t}}$ independent problems, and time merely parameterizes the loading path.

In space, we adopt an isoparametric finite element discretization. Within each element $e$, the displacement field and the reference geometry are approximated as
\begin{equation}
    \bm{u}^h = \mathbf{N}\,\mathbf{d}^e,
    \qquad
    \bm{X}^h = \mathbf{N}\,\mathbf{X}^e,
\end{equation}
where $\mathbf{N}$ denotes the shape function matrix, and $\mathbf{d}^e$ and $\mathbf{X}^e$ refer to the nodal displacements and reference coordinates of element $e$, which are extracted from their global counterparts $\mathbf{d}$ and $\mathbf{X}$. Inserting this approximation into \cref{eq:global_potential} yields the discrete potential $\Pi^h(\mathbf{d},\bm{\theta})$, whose stationarity condition defines the residual
\begin{equation}
    \mathbf{r}(\mathbf{d},\bm{\theta}) := \frac{\partial\Pi^h}{\partial\mathbf{d}}
    = \overset{n_{\mathrm{el}}}{\underset{e=1}{\mathsf{A}}} \left[ \mathbf{r}^e_{\mathrm{int}} - \mathbf{r}^e_{\mathrm{ext}} \right],
    \qquad
    \mathbf{r}^e_{\mathrm{int}} := \int_{\mathcal{B}^e} \frac{\partial\psi\big(\bm{F}(\bm{u}^h);\bm{\theta}\big)}{\partial\mathbf{d}^e}\,\mathrm{d}V,
\label{eq:residual}
\end{equation}
where $\mathsf{A}$ denotes the assembly operator over all $n_{\mathrm{el}}$ elements. By the chain rule, the integrand equals $\bm{P} : \partial\bm{F}/\partial\mathbf{d}^e$, such that $\mathbf{r}^e_{\mathrm{int}}$ represents the discrete virtual work of the stresses. In practice, it is evaluated directly from $\psi$ by algorithmic differentiation.

To account for the boundary conditions, we partition the global vector of nodal displacements into free and prescribed components, $\mathbf{d} = [\mathbf{d}_{\mathrm{f}}, \mathbf{d}_{\mathrm{D}}]$, and the residual accordingly. Equilibrium requires only the free components to vanish,
\begin{equation}
    \mathbf{r}_{\mathrm{f}}(\mathbf{d}_{\mathrm{f}}, \mathbf{d}_{\mathrm{D}}, \bm{\theta}) = \mathbf{0},
\label{eq:equilibrium}
\end{equation}
which we solve for $\mathbf{d}_{\mathrm{f}}^\ast$ by a Newton--Raphson scheme with the tangent stiffness $\mathbf{K}_{\mathrm{ff}} := \partial\mathbf{r}_{\mathrm{f}}/\partial\mathbf{d}_{\mathrm{f}} = \partial^2\Pi^h/\partial\mathbf{d}_{\mathrm{f}}\partial\mathbf{d}_{\mathrm{f}}$. The remaining components do not vanish; instead, they are exactly the nodal reaction forces required to maintain the prescribed displacements,
\begin{equation}
    \mathbf{f}_{\mathrm{D}} := \mathbf{r}_{\mathrm{D}}(\mathbf{d}_{\mathrm{f}}^\ast, \mathbf{d}_{\mathrm{D}}, \bm{\theta}).
\label{eq:reaction_forces}
\end{equation}
Hence, the discrete potential provides both the equilibrium equations and the observable forces, which will play a central role in the definition of the loss.

\paragraph{Outer problem.}
The outer problem defines the discrepancy between the observed and the predicted structural response. Experimentally, we typically record the total reaction force on a set of boundary nodes $\mathcal{F}$, e.g., the loaded surface, in the loading direction $\bm{e}$. Its model counterpart at load step $t_n$ follows from \cref{eq:reaction_forces} as
\begin{equation}
    \mathcal{H}(\mathbf{d}_n,\bm{\theta}) := \bm{e}\cdot\sum_{i\in\mathcal{F}} \mathbf{f}^{\,i}_{\mathrm{D},n},
\end{equation}
where $\mathbf{f}^{\,i}_{\mathrm{D},n}$ denotes the reaction force vector at node $i$. If, in addition, displacements are measured at $N_{\mathrm{obs}}$ observation points $\bm{X}_k$, e.g., by digital image correlation, we compare them to the interpolated finite element displacements $\bm{u}^h_n(\bm{X}_k)$. 
To give an idea at this point, the loss then may reads
\begin{equation}
    \mathcal{L} = \frac{1}{N_{\mathrm{t}}} \sum_{n=1}^{N_{\mathrm{t}}} \left[ \big[ \mathcal{H}(\mathbf{d}_n,\bm{\theta}) - f^{\mathrm{exp}}_n \big]^2 + \frac{\alpha}{N_{\mathrm{obs}}} \sum_{k=1}^{N_{\mathrm{obs}}} \big\lVert \bm{u}^h_n(\bm{X}_k) - \bm{u}^{\mathrm{exp}}_n(\bm{X}_k) \big\rVert^2 \right],
\label{eq:loss}
\end{equation}
where $f^{\mathrm{exp}}_n$ and $\bm{u}^{\mathrm{exp}}_n$ denote the measured force and displacements at $t_n$. The weighting factor $\alpha \geq 0$ balances the different units and magnitudes of both contributions; for $\alpha = 0$, we recover a purely force-based identification.

\paragraph{Gradient of the loss.}
Training is typically carried out by gradient-based optimization, which requires the derivative of the loss with respect to the parameters. Since the equilibrium state $\mathbf{d}_{\mathrm{f}}^\ast(\bm{\theta})$ is only implicitly defined through \cref{eq:equilibrium}, we differentiate the equilibrium condition with respect to $\bm{\theta}$. By the implicit function theorem, which holds as long as $\mathbf{K}_{\mathrm{ff}}$ is regular, this yields
\begin{equation}
    \mathbf{K}_{\mathrm{ff}}\,\frac{\mathrm{d}\mathbf{d}_{\mathrm{f}}^\ast}{\mathrm{d}\bm{\theta}} + \frac{\partial\mathbf{r}_{\mathrm{f}}}{\partial\bm{\theta}} = \mathbf{0}
    \qquad \Rightarrow \qquad
    \frac{\mathrm{d}\mathbf{d}_{\mathrm{f}}^\ast}{\mathrm{d}\bm{\theta}} = -\mathbf{K}_{\mathrm{ff}}^{-1}\,\frac{\partial\mathbf{r}_{\mathrm{f}}}{\partial\bm{\theta}},
\label{eq:sensitivity}
\end{equation}
with the mixed derivative $\partial\mathbf{r}_{\mathrm{f}}/\partial\bm{\theta} = \partial^2\Pi^h/\partial\mathbf{d}_{\mathrm{f}}\partial\bm{\theta}$. At a strict local minimum of $\Pi^h$, $\mathbf{K}_{\mathrm{ff}}$ is positive definite, and regularity is guaranteed. Applying the chain rule, the directional derivative of the loss in direction $\mathbf{v}$ reads
\begin{equation}
    \left.\frac{\mathrm{d}}{\mathrm{d}\varepsilon}\,\mathcal{L}\big(\mathbf{d}_{\mathrm{f}}^\ast(\bm{\theta}+\varepsilon\mathbf{v}),\bm{\theta}+\varepsilon\mathbf{v}\big)\right|_{\varepsilon=0}
    = \underbrace{\left[ \frac{\partial\mathcal{L}}{\partial\bm{\theta}} - \frac{\partial\mathcal{L}}{\partial\mathbf{d}_{\mathrm{f}}}\,\mathbf{K}_{\mathrm{ff}}^{-1}\,\frac{\partial\mathbf{r}_{\mathrm{f}}}{\partial\bm{\theta}} \right]}_{=:\, \mathrm{d}\mathcal{L}/\mathrm{d}\bm{\theta}} \cdot\, \mathbf{v},
\label{eq:loss_gradient}
\end{equation}
where, for the sake of readability, we omit the sum over the load steps. Here, the explicit derivative $\partial\mathcal{L}/\partial\bm{\theta}$ stems from the reaction forces in \cref{eq:reaction_forces}, which depend on the material parameters directly.

The inverse of the tangent stiffness is, of course, never formed explicitly. Evaluating \cref{eq:sensitivity} directly would require one linear solve per parameter, which becomes prohibitive for neural networks with many parameters. 
Instead, we solve a single adjoint problem per load step,
\begin{equation}
    \mathbf{K}_{\mathrm{ff}}^{T}\,\bm{\lambda} = \left[\frac{\partial\mathcal{L}}{\partial\mathbf{d}_{\mathrm{f}}}\right]^{T},
    \qquad
    \frac{\mathrm{d}\mathcal{L}}{\mathrm{d}\bm{\theta}} = \frac{\partial\mathcal{L}}{\partial\bm{\theta}} - \bm{\lambda}^{T}\,\frac{\partial\mathbf{r}_{\mathrm{f}}}{\partial\bm{\theta}},
\label{eq:adjoint}
\end{equation}
where $\bm{\lambda}$ denotes the adjoint variable. Inserting $\bm{\lambda} = \mathbf{K}_{\mathrm{ff}}^{-T}\,[\partial\mathcal{L}/\partial\mathbf{d}_{\mathrm{f}}]^{T}$ confirms that \cref{eq:adjoint} is equivalent to \cref{eq:loss_gradient}, while its cost is independent of the number of parameters.\footnote{For hyperelastic materials, the tangent stiffness is the Hessian of $\Pi^h$ and thus symmetric, i.e., $\mathbf{K}_{\mathrm{ff}}^{T} = \mathbf{K}_{\mathrm{ff}}$. The transposed formulation in \cref{eq:adjoint}, however, holds for arbitrary regular tangents.} This procedure is known as the adjoint method in PDE-constrained optimization. 

The present formulation belongs to the class of finite element model updating approaches \cite{avril2008}. In contrast to residual-based approaches such as the Equilibrium Gap Method \cite{claire2004} or unsupervised discovery \cite{flaschel2021euclid}, in which the measured displacements are inserted into the discretized weak form and its residual serves as the loss, we solve the discrete equilibrium problem in every forward evaluation. Residual-based approaches thus require full-field displacements at all nodes, complemented by global reaction forces to fix the scale of the material response. In the present formulation, by contrast, global reaction forces are in principle sufficient, while displacement data, ranging from a few observation points to full fields, enter only optionally. This reduced data requirement comes at the cost of solving the inner problem, which motivates the reduction techniques discussed in the following.


\subsection{Reduction techniques}
\label{sec:reduction}

The inverse material discovery outlined above is general in its application, yet computationally demanding: in every epoch, the inner problem has to be solved for all $N_{\mathrm{t}}$ load steps, each requiring several Newton iterations on the full finite element mesh. Model order reduction techniques aim to lower exactly this cost. However, a reduced model is usually constructed for a specific problem, i.e., for a given geometry, set of boundary conditions, and loading path, and it remains accurate only in the vicinity of the material parameters used to construct it. In general, it does not generalize to arbitrary material parameters.

During training, precisely this limitation becomes critical, since the outer problem continuously updates the material parameters. In the early epochs, the parameters may change substantially, and any reduced model built beforehand is likely to become inaccurate. As training proceeds, however, the parameter updates decrease, and the inner problem is repeatedly solved for similar parameters at identical load steps.

In the following, we introduce two complementary strategies to reduce the cost of solving the inner problem. First, a subspace projection reduces the size of the linear system to be solved in each Newton iteration. While this lowers the cost of the linear solve, the residual and the tangent stiffness still have to be assembled over all elements. Second, we therefore employ a hyperreduction method that reduces the assembly cost itself.

\paragraph{Subspace projection.}
Within one epoch, we solve \cref{eq:equilibrium} for the free nodal displacements $\mathbf{d}_{\mathrm{f}} \in \mathbb{R}^{N_{\mathrm{f}}}$ at all $N_{\mathrm{t}}$ load steps, whose solutions differ due to the different prescribed displacements. Across epochs, we solve the same problem at identical load steps, whose solutions differ due to the updated material parameters. We collect a finite number of these equilibrium states in the snapshot matrix
\begin{equation}
    \mathbf{S} = \left[\mathbf{d}_{\mathrm{f},1} - \mathbf{d}_{\mathrm{f},0},\ \dots,\ \mathbf{d}_{\mathrm{f},N_{\mathrm{snap}}} - \mathbf{d}_{\mathrm{f},0}\right] \in \mathbb{R}^{N_{\mathrm{f}} \times N_{\mathrm{snap}}},
\label{eq:snapshot_matrix}
\end{equation}
where $N_{\mathrm{f}}$ denotes the number of free degrees of freedom, $N_{\mathrm{snap}}$ the number of snapshots, and $\mathbf{d}_{\mathrm{f},0}$ a reference state, e.g., the undeformed configuration $\mathbf{d}_{\mathrm{f},0} = \mathbf{0}$. Each snapshot $s$ is associated with the material parameters $\bm{\theta}_s$ and the load step at which it was computed.

A common approach to extract a reduced basis from these snapshots is the proper orthogonal decomposition \cite{sirovich1987}, which relies on the thin singular value decomposition
\begin{equation}
    \mathbf{S} = \mathbf{U}\,\bm{\Sigma}\,\mathbf{Z}^T,
\label{eq:snapshot_matrix_decomposition}
\end{equation}
where $\mathbf{U} \in \mathbb{R}^{N_{\mathrm{f}} \times p}$ and $\mathbf{Z} \in \mathbb{R}^{N_{\mathrm{snap}} \times p}$ have orthonormal columns, and $\bm{\Sigma} = \mathrm{diag}(\sigma_1, \dots, \sigma_p)$ contains the singular values $\sigma_1 \geq \sigma_2 \geq \dots \geq \sigma_p \geq 0$ with $p = \min\{N_{\mathrm{f}}, N_{\mathrm{snap}}\}$. The left singular vectors are ordered by their contribution to the snapshot data, i.e., the first columns of $\mathbf{U}$ capture the dominant deformation modes. We therefore retain only the first $N_{\mathrm{r}} \ll N_{\mathrm{f}}$ columns as reduced basis,
\begin{equation}
    \mathbf{V} = \left[\mathbf{U}_{:,1},\ \dots,\ \mathbf{U}_{:,N_{\mathrm{r}}}\right] \in \mathbb{R}^{N_{\mathrm{f}} \times N_{\mathrm{r}}}.
\label{eq:pod_basis}
\end{equation}
With this at hand, we approximate the free nodal displacements within the affine subspace
\begin{equation}
    \mathbf{d}_{\mathrm{f}} \approx \mathbf{d}_{\mathrm{f},0} + \mathbf{V}\,\mathbf{q},
\label{eq:subspace}
\end{equation}
where $\mathbf{q} \in \mathbb{R}^{N_{\mathrm{r}}}$ denotes the vector of reduced coordinates. Instead of solving \cref{eq:equilibrium} in the full space, we project the residual onto the subspace in a Galerkin sense. In each Newton iteration $i$, this yields the reduced linear system
\begin{equation}
    \underbrace{\mathbf{V}^T\,\mathbf{K}_{\mathrm{ff}}\,\mathbf{V}}_{=:\, \mathbf{K}_{\mathrm{r}}}\,\Delta\mathbf{q}^{i} = -\underbrace{\mathbf{V}^T\,\mathbf{r}_{\mathrm{f}}}_{=:\, \mathbf{r}_{\mathrm{r}}},
    \qquad
    \mathbf{q}^{i+1} = \mathbf{q}^{i} + \Delta\mathbf{q}^{i},
\label{eq:reduced_newton}
\end{equation}
where $\mathbf{K}_{\mathrm{ff}}$ and $\mathbf{r}_{\mathrm{f}}$ are evaluated at $\mathbf{d}_{\mathrm{f},0} + \mathbf{V}\mathbf{q}^{i}$ and $\bm{\theta}$. For hyperelastic materials, this Galerkin projection has a clear variational interpretation: it corresponds to minimizing the discrete potential $\Pi^h$ over the affine subspace in \cref{eq:subspace}. Consequently, the reduced tangent $\mathbf{K}_{\mathrm{r}}$ inherits the symmetry.

While the size of the linear system reduces from $N_{\mathrm{f}}$ to $N_{\mathrm{r}}$, both $\mathbf{r}_{\mathrm{r}}$ and $\mathbf{K}_{\mathrm{r}}$ are still assembled over all elements. For nonlinear problems, this assembly quickly dominates the overall cost, which motivates the hyperreduction discussed next.

\paragraph{Hyperreduction.}
We employ the Energy-Conserving Sampling and Weighting (ECSW) method \cite{farhat2014,farhat2015}, which operates directly on the projected quantities. To this end, we note that the reduced residual and tangent are exact sums of element contributions,
\begin{equation}
    \mathbf{r}_{\mathrm{r}} = \sum_{e\in\mathcal{E}} \left[\mathbf{V}^e\right]^T \mathbf{r}^e,
    \qquad
    \mathbf{K}_{\mathrm{r}} = \sum_{e\in\mathcal{E}} \left[\mathbf{V}^e\right]^T \mathbf{K}^e\,\mathbf{V}^e,
\label{eq:reduced_sums}
\end{equation}
where $\mathcal{E}$ denotes the set of all $n_{\mathrm{el}}$ elements, $\mathbf{r}^e$ and $\mathbf{K}^e$ the element residual and tangent, and $\mathbf{V}^e$ the submatrix of $\mathbf{V}$ that contains the rows associated with the free degrees of freedom of element $e$. ECSW replaces these sums by weighted sums over a small subset $\tilde{\mathcal{E}} \subset \mathcal{E}$,
\begin{equation}
    \mathbf{r}_{\mathrm{r}} \approx \tilde{\mathbf{r}}_{\mathrm{r}} = \sum_{e\in\tilde{\mathcal{E}}} w^e \left[\mathbf{V}^e\right]^T \mathbf{r}^e,
    \qquad
    \mathbf{K}_{\mathrm{r}} \approx \tilde{\mathbf{K}}_{\mathrm{r}} = \sum_{e\in\tilde{\mathcal{E}}} w^e \left[\mathbf{V}^e\right]^T \mathbf{K}^e\,\mathbf{V}^e,
\label{eq:ecsw}
\end{equation}
with non-negative weights $w^e \geq 0$. Hence, the cost of assembly scales with the number of sampled elements $|\tilde{\mathcal{E}}| \ll n_{\mathrm{el}}$ rather than with $n_{\mathrm{el}}$. 
Noteworthy, the hyperreduced problem remains the minimization of an approximate potential $\tilde{\Pi}^h = \sum_{e\in\tilde{\mathcal{E}}} w^e\,\Pi^{h,e}$, a weighted sum of element energies.

In both cases, with or without hyperreduction, the gradient computation remains in the reduced space. Since the
implicit differentiation acts on the residual that is actually solved, the adjoint
problem in \cref{eq:adjoint} becomes
\begin{equation}
    \tilde{\mathbf{K}}_{\mathrm{r}}^{T}\,\bm{\lambda}_{\mathrm{r}}
    = \mathbf{V}^{T}\left[\frac{\partial\mathcal{L}}{\partial\mathbf{d}_{\mathrm{f}}}\right]^{T},
    \qquad
    \frac{\mathrm{d}\mathcal{L}}{\mathrm{d}\bm{\theta}}
    = \frac{\partial\mathcal{L}}{\partial\bm{\theta}}
    - \bm{\lambda}_{\mathrm{r}}^{T}\,\frac{\partial\tilde{\mathbf{r}}_{\mathrm{r}}}{\partial\bm{\theta}},
\label{eq:reduced_adjoint}
\end{equation}
with $\bm{\lambda}_{\mathrm{r}} \in \mathbb{R}^{N_{\mathrm{r}}}$, and with
$\mathbf{K}_{\mathrm{r}}$ and $\mathbf{r}_{\mathrm{r}}$ in place of
$\tilde{\mathbf{K}}_{\mathrm{r}}$ and $\tilde{\mathbf{r}}_{\mathrm{r}}$ without
hyperreduction. Hence, the adjoint inherits the symmetry and definiteness of the
reduced tangent, and $\mathbf{K}_{\mathrm{ff}}$ is never factorized during training.
Note that \cref{eq:reduced_adjoint} yields the exact gradient of the reduced model,
not an approximation of the full order gradient.

\paragraph{Computation of the ECSW weights.}
For each snapshot $s$ within the snapshot matrix in \cref{eq:snapshot_matrix}, we evaluate every element at the projected state
\begin{equation}
    \tilde{\mathbf{d}}_{\mathrm{f},s} = \mathbf{d}_{\mathrm{f},0} + \mathbf{V}\mathbf{V}^T\left[\mathbf{d}_{\mathrm{f},s} - \mathbf{d}_{\mathrm{f},0}\right]
\end{equation}
and the corresponding parameters $\bm{\theta}_s$, i.e., at states the reduced model can actually reach. The projected element residual and tangent are stacked into
\begin{equation}
    \mathbf{g}^e_s =
    \begin{bmatrix}
        \left[\mathbf{V}^e\right]^T \mathbf{r}^e(\tilde{\mathbf{d}}_{\mathrm{f},s}, \bm{\theta}_s) \\[2pt]
        \operatorname{vec}\!\left(\left[\mathbf{V}^e\right]^T \mathbf{K}^e(\tilde{\mathbf{d}}_{\mathrm{f},s}, \bm{\theta}_s)\,\mathbf{V}^e\right)
    \end{bmatrix}
    \in \mathbb{R}^{N_{\mathrm{r}} + N_{\mathrm{r}}^2},
\end{equation}
where $\operatorname{vec}(\bullet)$ denotes column-wise vectorization. Arranging these vectors with one column per element and one block of rows per snapshot yields
\begin{equation}
    \mathbf{G} =
    \begin{bmatrix}
        \mathbf{g}^1_1 & \cdots & \mathbf{g}^{n_{\mathrm{el}}}_1 \\
        \vdots & \ddots & \vdots \\
        \mathbf{g}^1_{N_{\mathrm{snap}}} & \cdots & \mathbf{g}^{n_{\mathrm{el}}}_{N_{\mathrm{snap}}}
    \end{bmatrix}
    \in \mathbb{R}^{N_{\mathrm{snap}}[N_{\mathrm{r}} + N_{\mathrm{r}}^2] \times n_{\mathrm{el}}},
    \qquad
    \mathbf{b} = \mathbf{G}\,\mathbf{1} = \sum_{e\in\mathcal{E}} \mathbf{G}_{:,e},
\label{eq:ecsw_system}
\end{equation}
where $\mathbf{1}$ denotes a vector of ones. Thus, $\mathbf{b}$ contains the exact reduced residual and tangent of every snapshot, i.e., the full mesh with all weights equal to one. The weights follow from the non-negative least-squares (NNLS) problem
\begin{equation}
    \mathbf{y}^\ast = \arg \underset{\mathbf{y}}{\min}\, \lVert \mathbf{G}\mathbf{y} - \mathbf{b} \rVert_2^2
    \qquad \text{subject to} \qquad
    \mathbf{y} \geq \mathbf{0},
\label{eq:ecsw_nnls}
\end{equation}
where $\mathbf{y} = [w^1, \dots, w^{n_{\mathrm{el}}}]^T$ collects one weight per element. Note that the trivial solution $\mathbf{y} = \mathbf{1}$ satisfies \cref{eq:ecsw_nnls} exactly, but yields no reduction. The aim is therefore a sparse solution, in which most weights vanish. The non-negativity constraint naturally promotes such sparsity, since many weights are driven to the boundary $w^e = 0$ of the admissible set. All elements with $w^e > 0$ form the reduced set $\tilde{\mathcal{E}}$.


\subsection{Neural Elasticity}
\label{sec:neural_elasticity}

The inverse material discovery and the reduction techniques developed so far are independent of the explicit constitutive model.
We now specify the energy $\psi(\bm{F};\bm{\theta})$ whose parameters $\bm{\theta}$ are identified in the remainder of this work.
In recent years, several physics-constrained neural network material models were introduced, which all perform well in finite strain hyperelasticity \cite{holthusen2026benchmarking}.
Here, we follow the Physics-Augmented Neural Network approach of \cite{klein2022}, which is polyconvex \cite{ball1977} by construction.
In the present work, we restrict ourselves to isotropic, hyperelastic material behavior.
In the following, we first introduce the neural network, then the growth term, and conclude with the logarithmic normalization.

To this end, we decompose the energy into three contributions
\begin{equation}
    \psi(\bm{F};\bm{\theta}) := \mathcal{N}(\bm{x}(\bm{F});\bm{\theta}) + \psi_{\mathrm{gr}}(\bm{F}) + \psi_{\mathrm{nrm}}^{\ln}(J;\bm{\theta}),
    \label{eq:energy}
\end{equation}
where $\mathcal{N}$ denotes the physics-constrained neural network with input vector $\bm{x}(\bm{F})$, $\psi_{\mathrm{gr}}$ adds a growth contribution to ensure coercivity of the overall energy, and $\psi_{\mathrm{nrm}}^{\ln}$ depends on the determinant $J$ of the deformation gradient and normalizes the energy to ensure a stress-free undeformed reference state.
Recall that $\psi$ is polyconvex if there exists a convex function $\mathcal{P}$ such that $\psi(\bm{F}) = \mathcal{P}(\bm{F},\operatorname{cof}\bm{F},J)$ \cite{ball1977}.
Each of the three terms in \cref{eq:energy} is constructed to satisfy this property.

\paragraph{Neural network.}
To obtain a network that is objective, isotropic, and polyconvex, we feed it with the invariants
\begin{equation}
    I_1 := \tfrac13 \operatorname{tr}\bm{C}, \qquad
    I_2 := \tfrac13 \operatorname{tr}\left(\operatorname{cof}\bm{C}\right), \qquad
    J := \det\bm{F},
    \label{eq:invariants}
\end{equation}
where $\bm{C}=\bm{F}^T\bm{F}$ denotes the right Cauchy-Green tensor and $\operatorname{cof}\bm{C}$ its cofactor.
Each of these invariants satisfies objectivity and material symmetry for isotropic materials, i.e., for a symmetry group equal to the full group of proper orthogonal matrices.
Moreover, $I_1$ and $I_2$ are convex functions of $\bm{F}$ and $\operatorname{cof}\bm{F}$, respectively, while $J$ enters the energy directly.
From these invariants, we define the input vector as
\begin{equation}
    \bm{x}(\bm{F}) := \big[I_1-1,\ I_2-1,\ J-1,\ -[J-1]\big], \qquad \bm{x}(\bm{I}) = \bm{0}.
    \label{eq:input}
\end{equation}
Since all weights of the network are constrained to be non-negative (see below), $\mathcal{N}$ is non-decreasing in each entry of $\bm{x}$.
The mirrored entry $-[J-1]$ therefore allows the network to also decrease with $J$, such that both volumetric expansion and compression can be represented.

The network follows the Input Convex Neural Network ($\mathrm{ICNN}$) architecture of \cite{amos_input_2017}, which is convex with respect to its input.
In line with \cite{holthusen2026complement}, we slightly adapt the original architecture to be zero anchored, i.e., $\mathcal{N}(\bm{0};\bm{\theta})=0$.
The first hidden layer is obtained via
\begin{equation}
    \bm{h}_1 := \varphi_1\big(\bm{S}_1\,\bm{x} + \bm{b}_1\big) - \varphi_1\big(\bm{b}_1\big),
\end{equation}
while all subsequent layers, including the output layer $L$, are obtained via
\begin{equation}
    \bm{h}_\ell := \varphi_\ell\big(\bm{W}_\ell\,\bm{h}_{\ell-1}+\bm{S}_\ell\,\bm{x} + \bm{b}_\ell\big) - \varphi_\ell\big(\bm{b}_\ell\big), \qquad \ell=2,\hdots,L,
\end{equation}
where $\bm{h}_\ell$ denotes the output of layer $\ell$, $\varphi_\ell$ the activation function applied component-wise, $\bm{W}_\ell$ and $\bm{S}_\ell$ the weight matrices acting on the previous layer and on the input, respectively, and $\bm{b}_\ell$ the bias vector.
The network output is the scalar $\mathcal{N}(\bm{x};\bm{\theta}) := \bm{h}_L$, and $\bm{\theta}$ collects all weights and biases.
The subtraction of $\varphi_\ell(\bm{b}_\ell)$ shifts each layer such that it vanishes at $\bm{x}=\bm{0}$, which yields the zero anchoring.
In contrast to the original ICNN architecture, we constrain all weights $\bm{W}$ and $\bm{S}$ to be non-negative, since the invariants $I_1$ and $I_2$ are already convex functions of $\bm{F}$ and $\operatorname{cof}\bm{F}$, cf. \cite{linden_neural_2023}.
Together with convex and non-decreasing activation functions $\varphi$, this renders $\mathcal{N}$ convex and non-decreasing in $\bm{x}$, and hence $\mathcal{N}(\bm{x}(\bm{F});\bm{\theta})$ polyconvex in $\bm{F}$.

\paragraph{Growth term.}
The network alone does not guarantee coercivity of the energy, see \cref{app:polyconvexity_coercivity}.
To this end, we adopt the growth term of \cite{holthusen2026thermo} and adapt it to the invariants in \cref{eq:invariants},
\begin{equation}
    \psi_{\mathrm{gr}}(\bm{F}) := \epsilon\Big[\big[I_1 - 1 - \tfrac23\ln J\big] + \big[I_2 - 1 - \tfrac43\ln J\big] + \big[J - 1 - \ln J\big]\Big],
    \label{eq:growth}
\end{equation}
where $0<\epsilon\ll 1$ denotes a scaling parameter.
Note that each summand is greater than or equal to zero\footnote{This follows from the inequality of arithmetic and geometric means, $I_1 \geq J^{2/3}$ and $I_2 \geq J^{4/3}$, together with $\ln x \leq x-1$.}, and polyconvex, since $-\ln J$ is convex in $J$.
Moreover, the growth term does not contribute to the stress in the reference configuration,
\begin{equation}
    \frac{\partial\psi_{\mathrm{gr}}}{\partial \bm{F}}\bigg|_{\bm{F} = \bm{I}} = \bm{0},
\end{equation}
which is discussed in greater detail in \cref{app:polyconvexity_coercivity}.

\paragraph{Logarithmic normalization term.}
In general, the network $\mathcal{N}$ yields a non-zero stress at $\bm{F}=\bm{I}$, which has to be compensated by the normalization.
To this end, we define the derivatives of the network with respect to the components of $\bm{x}$ at the undeformed configuration,
\begin{equation}
    a_1 := \frac{\partial\mathcal{N}}{\partial x_1}\bigg|_{\bm{F} = \bm{I}},
    \qquad
    a_2 := \frac{\partial\mathcal{N}}{\partial x_2}\bigg|_{\bm{F} = \bm{I}},
    \qquad
    a_J := \frac{\partial\mathcal{N}}{\partial x_3}\bigg|_{\bm{F} = \bm{I}} - \frac{\partial\mathcal{N}}{\partial x_4}\bigg|_{\bm{F} = \bm{I}},
\end{equation}
where $a_1,a_2\geq 0$ due to the monotonicity of $\mathcal{N}$ with respect to its inputs, while $a_J\in\mathbb{R}$.
With these derivatives at hand, we introduce the normalization term
\begin{equation}
    \psi_{\mathrm{nrm}}^{\ln}(J;\bm{\theta}) := -\big[\tfrac{2}{3}a_1 + \tfrac{4}{3}a_2\big]\ln J - a_J\,[J-1],
    \qquad
    \frac{\partial\psi_{\mathrm{nrm}}^{\ln}}{\partial \bm{F}}\bigg|_{\bm{F} = \bm{I}} = -\big[\tfrac{2}{3}a_1 + \tfrac{4}{3}a_2+a_J\big]\bm{I},
    \label{eq:normalization}
\end{equation}
which is convex in $J$, and thus polyconvex, and cancels the stress contribution of $\mathcal{N}$ at the reference state.
Since all three terms in \cref{eq:energy} vanish at $\bm{F}=\bm{I}$, the energy is zero and the stress vanishes in the undeformed configuration.
In contrast to the affine normalization introduced in \cite{linden_neural_2023}, the logarithmic normalization, together with the zero-anchored ICNN architecture and the growth energy, ensures a non-negative energy, which is discussed in greater detail in \cref{app:polyconvexity_coercivity}.
Nevertheless, both normalizations satisfy polyconvexity and coercivity.


\subsection{Workflow during hyperreduced inverse discovery}
\label{sec:workflow}

The reduction techniques introduced above lower the cost of a single solve, yet a fixed projection $\mathbf{V}$ and a fixed set of sampled elements $\tilde{\mathcal{E}}$ cannot be expected to capture the structural response for all states of the weights $\bm{\theta}$ visited during training.
To this end, we adapt the reduction to the evolving weights.
In the following, we first introduce the bootstrap epochs, then the canary epochs, and conclude with the error measures that govern the switch between the reduced and the full order model.

\paragraph{Bootstrap epochs.}
The first $N_{\mathrm{boot}}$ epochs of the training are solved in the full space for all load steps $t_n$.
Their purpose is to collect representative snapshots in the snapshot matrix $\mathbf{S}$.
Subsequently, the projection $\mathbf{V}$ and the set of sampled elements $\tilde{\mathcal{E}}$ for the ECSW are computed from the snapshots collected so far.
Provided the resulting reduced solution does not deviate too strongly from the full order solution, the following epochs are solved in the reduced space.

\paragraph{Canary epochs.}
Since $\mathbf{V}$ and $\tilde{\mathcal{E}}$ are based on past snapshots, they are in general not suitable throughout the entire training.
Hence, we interleave canary epochs, in which the full order problem is solved at the current weights $\bm{\theta}$.
After the bootstrap epochs, every $N_{\mathrm{canary}}$-th epoch is a canary epoch.\footnote{In principle, $N_{\mathrm{canary}}$ could be chosen adaptively. Since the weights typically change less in later stages of the training, fewer canary epochs can be expected to be necessary there. This is not investigated in the present work.}
Each canary epoch consists of three steps.
\textit{(i) Snapshot update.}
Based on the full order solution, we decide whether the current state is representable by the present projection $\mathbf{V}$.
If it is not, the corresponding snapshots are added to $\mathbf{S}$.
Between two canary epochs, $\mathbf{S}$ remains unchanged, since no full order solution is available for comparison.
\textit{(ii) Update of the reduction.}
If new snapshots were added to $\mathbf{S}$ or if the reduced solution was found to deviate too strongly at the previous canary epoch, $\mathbf{V}$ and $\tilde{\mathcal{E}}$ are recomputed from the updated $\mathbf{S}$.
Otherwise, both are retained, as they remain constant between two canary epochs.
\textit{(iii) Selection of the model.}
Independent of whether the reduction was updated, we compare the full order solution with the reduced solution at the same weights $\bm{\theta}$.
If both deviate too strongly, the epochs up to the next canary epoch are solved with the full order model; otherwise, they are solved with the reduced order model.

The error measures underlying steps (i)-(iii) are discussed next.

\paragraph{Dimension of the reduced space.}
The projection $\mathbf{V}$ is defined by the first $N_{\mathrm{r}}$ columns of $\mathbf{U}$, a procedure known as proper orthogonal decomposition (POD) \citep{sirovich1987,berkooz1993proper,radermacher2016}.
To choose $N_{\mathrm{r}}$, we employ the relative projection error of the snapshots,
\begin{equation}
    \mathrm{error}_{\mathrm{r}}(k) = \sqrt{\frac{\sum_{i=k+1}^{p} \hat{\sigma}_i^2}{\sum_{i=1}^{p} \hat{\sigma}_i^2}}, \qquad \hat{\sigma}_i := \frac{\sigma_i}{\sigma_1},
    \label{eq:pod_error}
\end{equation}
where $\sigma_i$ denotes the $i$-th singular value of $\mathbf{S}$ and $\hat{\sigma}_i$ its normalization by the largest one.
Note that the normalization cancels in the quotient and merely renders the measure independent of the scaling of $\mathbf{S}$.
We choose the smallest dimension for which this error stays below a prescribed tolerance $\varepsilon_{\mathrm{POD}}$, i.e.,
\begin{equation}
    N_{\mathrm{r}} = \min\big\{k : \mathrm{error}_{\mathrm{r}}(k) \leq \varepsilon_{\mathrm{POD}}\big\}.
    \label{eq:Nr}
\end{equation}

\paragraph{Snapshot update.}
In each canary epoch, we add a new full order state $\mathbf{d}_{\mathrm{f},\mathrm{new}}$ to $\mathbf{S}$ only if it cannot be sufficiently well represented by the current projection $\mathbf{V}$.
This is quantified by the relative projection error
\begin{equation}
    \mathrm{error}_{\mathrm{proj}}(\mathbf{d}_{\mathrm{f},\mathrm{new}}) = \frac{\lVert [\mathbf{d}_{\mathrm{f},\mathrm{new}}-\mathbf{d}_{\mathrm{f},0}] - \mathbf{V}\mathbf{V}^T [\mathbf{d}_{\mathrm{f},\mathrm{new}}-\mathbf{d}_{\mathrm{f},0}] \rVert}{\max\big\{\lVert \mathbf{d}_{\mathrm{f},\mathrm{new}}-\mathbf{d}_{\mathrm{f},0} \rVert,\beta\big\}},
    \label{eq:proj_error}
\end{equation}
where $\beta > 0$ is a small number that avoids a division by zero.
If $\mathrm{error}_{\mathrm{proj}}(\mathbf{d}_{\mathrm{f},\mathrm{new}}) \geq \varepsilon_{\mathrm{proj}}$, the state is added to $\mathbf{S}$.
In the present work, we limit the number of snapshots to $N_{\mathrm{snap}}^{\max}$ and replace the oldest snapshot whenever the limit is reached, such that $N_{\mathrm{r}} \leq N_{\mathrm{snap}}^{\max}$.

\paragraph{Number of ECSW elements.}
To obtain a sparse solution, we solve \cref{eq:ecsw_nnls} with a greedy algorithm\footnote{Each restricted non-negative least-squares subproblem is solved exactly using the Lawson--Hanson active-set algorithm. Here, we use the python package \texttt{scipy.optimize.nnls}}, which successively adds the column of $\mathbf{G}$ most correlated with the current residual to the support.
We terminate once the relative training error
\begin{equation}
    \mathrm{error}_{\mathrm{ECSW}} = \frac{\lVert \mathbf{G}\mathbf{y}-\mathbf{b}\rVert}{\lVert \mathbf{b}\rVert}
    \label{eq:ecsw_error}
\end{equation}
falls below a prescribed tolerance $\varepsilon_{\mathrm{ECSW}}$ or the support reaches the maximum size $n_{\mathrm{el}}$.
The elements with $w^e>0$ form the reduced set $\tilde{\mathcal{E}}$.

\paragraph{Model selection.}
Finally, we decide in each canary epoch whether the reduced model remains sufficiently accurate.
To this end, we compare the reduced and the full order loss, $\mathcal{L}^{\mathrm{red}}$ and $\mathcal{L}^{\mathrm{full}}$, evaluated at identical weights $\bm{\theta}$\footnote{Note that we only evaluate the discrepancy of the data here. Any additional contribution by regularization to the loss is neglected.}.
With the error measures $e^{\mathrm{full}} = \sqrt{\mathcal{L}^{\mathrm{full}}}$ and $e^{\mathrm{red}} = \sqrt{\mathcal{L}^{\mathrm{red}}}$, we define the absolute and the relative deviation
\begin{equation}
    \mathrm{error}_{\mathcal{L}} = \sqrt{\big\lvert \mathcal{L}^{\mathrm{red}} - \mathcal{L}^{\mathrm{full}} \big\rvert}, \qquad
    \mathrm{error}_{\mathcal{L},\mathrm{rel}} = \frac{\big\lvert e^{\mathrm{red}} - e^{\mathrm{full}} \big\rvert}{\max\big\{e^{\mathrm{full}}, e^{\mathrm{red}}, \beta\big\}}.
    \label{eq:gate_errors}
\end{equation}
The reduced model is accepted if
\begin{equation}
    \mathrm{error}_{\mathcal{L}} \leq \varepsilon_{\mathcal{L}} \quad \text{or} \quad \mathrm{error}_{\mathcal{L},\mathrm{rel}} \leq \varepsilon_{\mathcal{L},\mathrm{rel}}.
    \label{eq:gate_condition}
\end{equation}
Both criteria are required: a purely relative tolerance tightens uncontrollably as $\mathcal{L}^{\mathrm{full}} \to 0$ during convergence, whereas a purely absolute tolerance admits large relative deviations while the loss is still large.
If neither condition holds, the following epochs are solved with the full order model until a subsequent canary epoch accepts the reduction.
\section{Results}
\label{sec:results}

We now examine whether the hyperreduced inverse discovery of \cref{sec:methods} lowers the cost of the training without degrading the identified material.
To this end, we consider three examples that probe different aspects of the approach.
In \cref{sec:case1}, we identify the two phases of a heterogeneous plate of the Mechanical MNIST Challenge from the reaction force alone, for which a full-order reference training allows a direct comparison of accuracy and cost.
In \cref{sec:case2}, we turn to experimental data of a perforated membrane and compare a force-only loss with a loss that additionally contains the measured displacement field.
In \cref{sec:case3}, we consider a synthetic, three-dimensional block with two stiff inclusions, for which the ground truth is known and the full-order solve dominates the cost.
All settings are collected in \cref{tab:hyperparameters}.
Throughout this section, the reported errors are evaluated by a full-order solve at the identified parameters.

The approach is implemented in our in-house code ASIMOV, which builds on JAX \cite{bradbury2018}.
We run all computations in double precision on the CPU of a MacBook Pro (Apple M4, ten cores, $24$~GB memory, macOS 26.6), to which all reported wall-clock times refer.

\begin{table}[tbp]
    \centering
    \caption{\textbf{Settings of the inverse identification for all three cases.} In each case, the full-order solves (bootstrap, canary and final evaluation) use the same solver settings as the reduced runs. The reduction parameters of Case~1 define the base configuration, around which the study in \cref{fig:mnist_sweeps} varies one parameter at a time.}
    \label{tab:hyperparameters}
    \footnotesize
    \begin{tabularx}{\textwidth}{@{}l X l l l@{}}
        \toprule
        Category & Parameter & Case~1 & Case~2 & Case~3 \\
        \midrule
        Data
            & specimen & MNIST sample 22 & TT1 \cite{abbasi2026} & synthetic block \\
            & load steps $N_{\mathrm{t}}$ & $21$ & $51$ (DIC at $11$) & $10$ \\
            & loss & force & force; force + displ. & force \\
            & normalization of the force & target, $\bar{f}$ & target and model, $\bar{f}$ & target, $\bar{f}$ \\
            & $\bar{f}$ & $15.94$~N & $236.9$~N & $85.61$~N \\
        \midrule
        Discretization
            & element & hexahedron & hexahedron, Q1/P0 & hexahedron \\
            & elements $n_{\mathrm{el}}$ & $2\,500$ & $1\,518$ & $15\,625$ \\
            & degrees of freedom, free $N_{\mathrm{f}}$ & $15\,606$, $14\,994$ & $10\,206$, $9\,234$ & $52\,728$, $48\,672$ \\
        \midrule
        Network
            & networks & one per phase & one & one per phase \\
            & hidden layers, activation & \multicolumn{3}{l}{$[8, 8]$, softplus; output activation softplus} \\
            & weight constraint & \multicolumn{3}{l}{$\bm{W}, \bm{S} \geq \bm{0}$, enforced by projection after each update} \\
            & growth parameter $\epsilon$, \cref{eq:growth} & \multicolumn{3}{l}{$10^{-4}$} \\
        \midrule
        Optimization
            & optimizer, weight decay, precision & \multicolumn{3}{l}{AdamW \cite{loshchilov2019}, $10^{-4}$, float64} \\
            & epochs & $1\,000$ & $700$ & $400$ \\
            & learning rate & $2\cdot 10^{-3}$ & $2\cdot 10^{-2}$ & $2\cdot 10^{-3}$ \\
            & global norm clipping \cite{pascanu2013} & $1$ & $1$ & $1$ \\
        \midrule
        Solver
            & Newton--Raphson: tolerance, max.\ iterations & \multicolumn{3}{l}{$10^{-8}$, $50$, with line search} \\
            & full order: linear solver & \multicolumn{3}{l}{BiCGSTAB \cite{vandervorst1992}, block Jacobi preconditioner ($3\times 3$)} \\
            & full order: Krylov tolerance, max.\ iterations & $10^{-8}$, $2\,000$ & $10^{-8}$, $2\,000$ & $10^{-9}$, $5\,000$ \\
            & reduced order: linear solver & \multicolumn{3}{l}{dense direct} \\
        \midrule
        Reduction
            & bootstrap epochs $N_{\mathrm{boot}}$, canary interval $N_{\mathrm{canary}}$ & $2$, $50$ & $2$, $50$ & $2$, $50$ \\
            & maximum number of snapshots $N_{\mathrm{snap}}^{\max}$ & $30$ & $30$ & $30$ \\
            & snapshot update $\varepsilon_{\mathrm{proj}}$, \cref{eq:proj_error} & $5\cdot 10^{-3}$ & $5\cdot 10^{-3}$ & $5\cdot 10^{-3}$ \\
            & POD tolerance $\varepsilon_{\mathrm{POD}}$, \cref{eq:Nr} & $10^{-4}$ & $10^{-6}$ & $10^{-4}$ \\
            & ECSW tolerance $\varepsilon_{\mathrm{ECSW}}$, \cref{eq:ecsw_error} & $10^{-4}$ & $10^{-4}$ & $10^{-4}$ \\
            & model selection $\varepsilon_{\mathcal{L}}$, $\varepsilon_{\mathcal{L},\mathrm{rel}}$, \cref{eq:gate_condition} & \multicolumn{3}{l}{$1.5\cdot 10^{-2}$, $5\cdot 10^{-2}$} \\
            & capacity of the padded set $\tilde{\mathcal{E}}$ & $384$ & $1\,024$ & $96$, then $192$ \\
        \bottomrule
    \end{tabularx}
\end{table}


\subsection{Case 1: Heterogeneous material system}
\label{sec:case1}

Our first example is a heterogeneous specimen from the Mechanical MNIST Challenge \cite{Lejeune2020,rausch2026mnistchallenge}.
In every specimen of this dataset, an inclusion shaped like a handwritten digit is embedded in a compliant surrounding phase, the matrix, and is expected to respond more stiffly than the latter.
The plates are loaded in uniaxial tension, and the reaction force of the testing machine as well as the in-plane displacements on one face, obtained by digital image correlation (DIC), are provided for a sequence of load steps.
We select sample 22, which has also served as training specimen for the full-order identification in \cite{knipper2026}.
Since both phases are present in a single specimen, the total reaction force depends on two constitutive laws at once.
This raises two questions: Can the reaction force of a single test separate the two phases, and does the adaptive reduction remain reliable while the optimizer distributes the stiffness between them?

\begin{figure}[tbp]
    \centering
    \begin{tikzpicture}[>={Stealth[length=2.2mm]}]
        \node (geo) at (0,0) {\includegraphics[width=5cm]{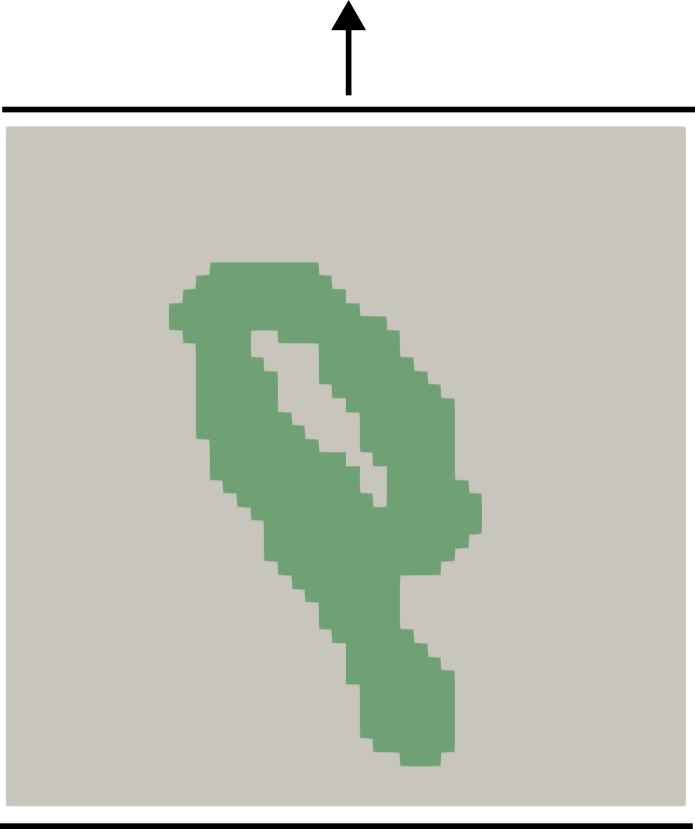}};
        \node[above] at (0,3.0) {$\bm{u}=(u_x(x,t),\,u_y(x,t),\,0)$};
        \node[below] at (0,-3.0) {$\bm{u}=(u_x(x,t),\,u_y(x,t),\,0)$};
        \draw[->, thick] (3.6,-3.0) -- (4.6,-3.0) node[right] {$x$};
        \draw[->, thick] (3.6,-3.0) -- (3.6,-2.0) node[above] {$y$};
    \end{tikzpicture}
    \caption{\textbf{Case~1, sample 22 of the Mechanical MNIST Challenge \cite{Lejeune2020,rausch2026mnistchallenge} in the reference configuration.} The inclusion is shown in green and the matrix in grey. The plate of $40\times 40\times 2$~mm is discretized by $50\times 50\times 1$ hexahedral elements, whose nodes on the observed face coincide with the DIC points. On the top and bottom edges, the in-plane displacements $u_x(x,t)$ and $u_y(x,t)$ measured by DIC are prescribed for $N_{\mathrm{t}}=21$ load steps, and the out-of-plane displacement is fixed. The reaction force on the top edge is the only quantity that enters the loss.}
    \label{fig:mnist_geometry}
\end{figure}

\cref{fig:mnist_geometry} shows the specimen in the reference configuration.
The plate is discretized by $50\times 50\times 1$ trilinear hexahedral elements, i.e., $n_{\mathrm{el}} = 2\,500$ elements with one element across the thickness.
Following \cite{knipper2026}, the mesh is aligned with the DIC grid, such that the nodes on the observed face coincide with the DIC points and the measured displacements can be compared without interpolation.
Each element is assigned to exactly one of the two phases, and each phase is described by its own network of the form \cref{eq:energy}, i.e., $\bm{\theta} = [\bm{\theta}_{\mathrm{M}}, \bm{\theta}_{\mathrm{I}}]$.
On the top and the bottom edge, we prescribe the in-plane displacements measured by DIC at the corresponding row of nodes, while the out-of-plane displacement is fixed.
Since DIC observes only one face, both node layers of an edge receive the same value, which corresponds to the assumption of a uniform motion across the thickness of $2$~mm.
The loading path comprises $N_{\mathrm{t}} = 21$ load steps up to a prescribed displacement of approximately $7.5$~mm.

In the following, we first summarize the loss and the hyperparameters, before we discuss the training history together with the adaptivity of the reduction.
We then assess the identified material, analyze the trial space and the ECSW rule, and conclude with a parameter study of the reduction.

\paragraph{Loss and hyperparameters.}
We identify both phases exclusively from the reaction force, i.e., we set $\alpha = 0$ in \cref{eq:loss}.
The measured displacements thus never enter the training and serve as an independent check of the identified material.
To obtain a loss of order one irrespective of the units, we divide the measured force by its largest absolute value $\bar{f} := \max_n |f^{\mathrm{exp}}_n| = 15.94$~N, such that the loss reads
\begin{equation}
    \mathcal{L} = \frac{1}{N_{\mathrm{t}}} \sum_{n=1}^{N_{\mathrm{t}}} \Big[ \mathcal{H}(\mathbf{d}_n,\bm{\theta}) - f^{\mathrm{exp}}_n/\bar{f} \Big]^2 .
    \label{eq:mnist_loss}
\end{equation}
Note that we scale only the target and not the model response.
Since the plate is loaded exclusively by prescribed displacements, the external potential vanishes, and a positive factor in front of $\psi$ leaves the minimizer of \cref{eq:bilevel} unchanged, while it scales all reaction forces by the same factor.
Consequently, the networks learn the scaled energy $\psi/\bar{f}$ without any effect on the displacement field, and the physical stresses and forces follow from a multiplication by $\bar{f}$.
Both phases share the same architecture, two hidden layers with eight neurons each, and differ only in their parameters.
We initialize them with different random keys, since two identical networks would start at a symmetric point of the loss landscape that only the data could resolve.
The non-negativity of the weights is enforced by a projection after each update.
We train with AdamW \cite{loshchilov2019} at a learning rate of $2\cdot 10^{-3}$ and the default weight decay of $10^{-4}$ and clip the gradient by its global norm \cite{pascanu2013} for $1\,000$ epochs, each of which comprises all load steps.
The full-order reference run uses identical settings and the same initial parameters, and differs only in that every epoch is solved in the full space.
All remaining settings are listed in \cref{tab:hyperparameters}.

\begin{figure}[tbp]
    \centering
    \includegraphics[width=0.8\textwidth]{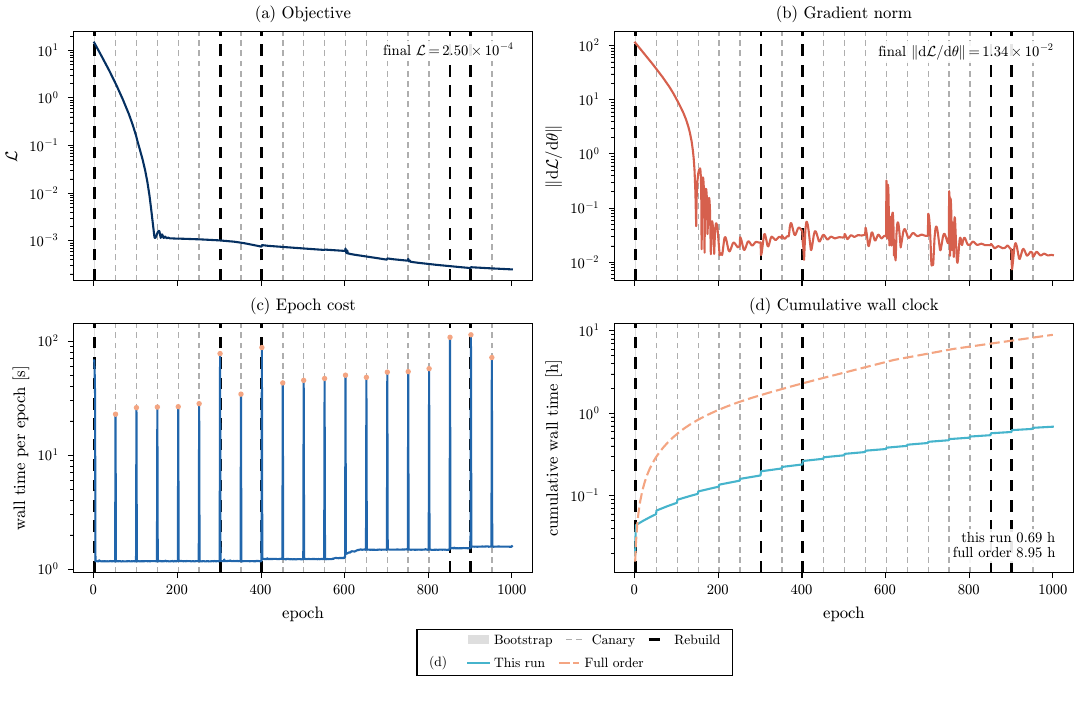}
    \caption{\textbf{Case~1, training history of the hyperreduced identification.} (a) The loss decreases by about four orders of magnitude within the first $150$ epochs and reaches $\mathcal{L}=2.50\cdot10^{-4}$, compared with $4.09\cdot10^{-4}$ of the full-order reference. (b) The gradient norm drops from about $10^{2}$ to $1.34\cdot10^{-2}$; rebuilds cause no visible disturbance. (c) Wall time per epoch: reduced epochs take $1.2$ to $1.6$~s, canary epochs (orange markers) between $23$ and $72$~s, and canary epochs with a rebuild up to $114$~s. (d) The cumulative wall-clock time of $0.69$~h is $13$ times smaller than the $8.95$~h of the full-order reference (dashed). Grey dashed lines mark canary epochs and black dashed lines rebuilds of the reduced model.}
    \label{fig:mnist_training}
\end{figure}

\paragraph{Training history and adaptivity of the reduction.}
\cref{fig:mnist_training}(a) shows that the loss decreases by roughly four orders of magnitude within the first $150$ epochs.
Afterwards, it continues to decrease slowly and reaches $\mathcal{L} = 2.50\cdot 10^{-4}$ after $1\,000$ epochs, while the gradient norm in \cref{fig:mnist_training}(b) drops from about $10^{2}$ to $1.34\cdot 10^{-2}$.
The full-order reference follows a nearly identical path during the first $300$ epochs and ends at $\mathcal{L} = 4.09\cdot 10^{-4}$.
That both runs separate afterwards is expected: the gradient in \cref{eq:reduced_adjoint} is the exact gradient of the reduced model, and small differences to the full-order gradient accumulate over hundreds of Adam updates.
In terms of the force, both final losses correspond to a root-mean-square deviation of about $2\,\%$ of $\bar{f}$ or less.

\begin{figure}[tbp]
    \centering
    \includegraphics[width=0.8\textwidth]{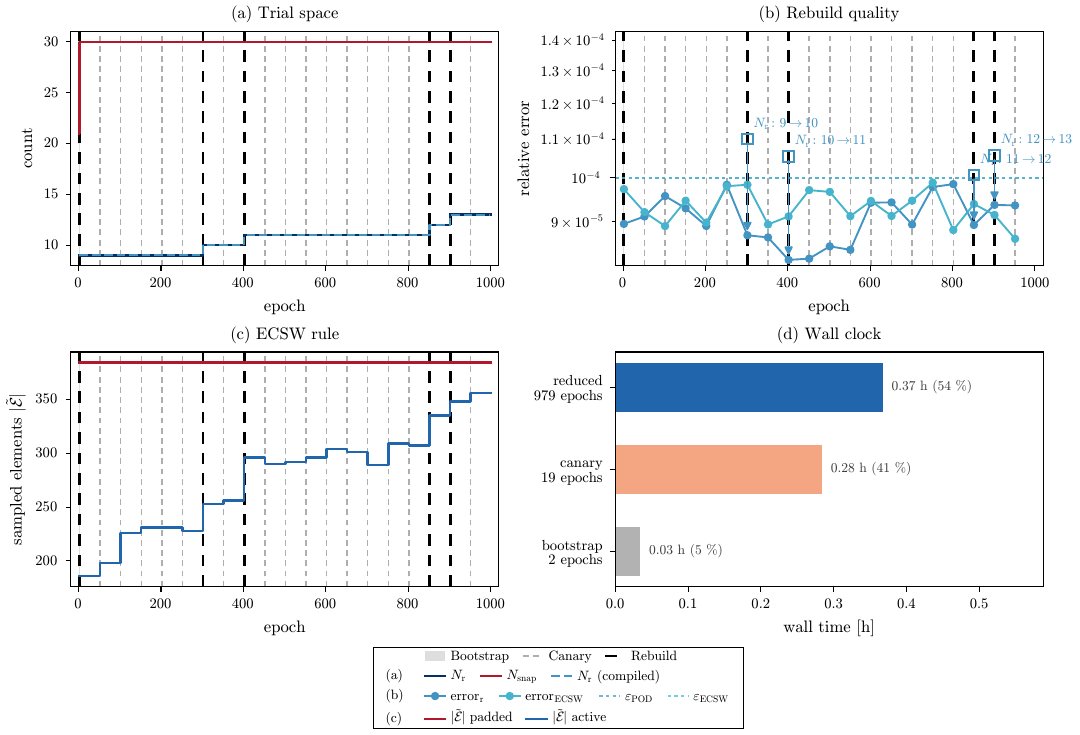}
    \caption{\textbf{Case~1, adaptivity of the reduction.} (a) The snapshot matrix reaches its capacity of $N_{\mathrm{snap}}^{\max}=30$ directly after the two bootstrap epochs, while the dimension of the trial space grows stepwise from $N_{\mathrm{r}}=9$ to $N_{\mathrm{r}}=13$. (b) Training errors of POD and ECSW at each canary epoch; squares mark the POD error of the updated snapshots with the previous dimension, which exceeds $\varepsilon_{\mathrm{POD}}=10^{-4}$ at four canary epochs and triggers the growth of $N_{\mathrm{r}}$. (c) The number of active ECSW elements grows from $186$ to $356$, whereas the padded capacity remains at $384$, such that no change of the ECSW rule requires a recompilation. (d) Split of the wall-clock time: the $19$ canary epochs consume $41\,\%$ of the total time, almost as much as the $979$ reduced epochs.}
    \label{fig:mnist_reduction}
\end{figure}

The cost per epoch in \cref{fig:mnist_training}(c) reveals three distinct levels.
A reduced epoch takes about $1.2$~s at the beginning and $1.6$~s at the end of the training.
A canary epoch, which comprises a full-order solve, the snapshot update and the recomputation of the reduction, takes between $23$ and $72$~s, and an epoch with a rebuild of the reduced model between $78$ and $114$~s.
For comparison, a full-order epoch of the reference run requires $33.4$~s in the median, rising from about $20$~s at the beginning to about $50$~s towards the end of the training; the canary epochs follow the same trend, as they are dominated by the full-order solve.
The median epoch, $1.23$~s against $33.4$~s, is thus accelerated by a factor of $27$.
Over the entire training, the wall-clock time decreases from $8.95$~h to $0.69$~h, see \cref{fig:mnist_training}(d), i.e., by a factor of $13$.
The gap between both factors is explained by \cref{fig:mnist_reduction}(d): although the $19$ canary epochs make up less than $2\,\%$ of all epochs, they account for $41\,\%$ of the wall-clock time, while the $979$ reduced epochs account for $54\,\%$.
Hence, once the reduced solve is cheap, the full-order checks rather than the reduced model limit the overall speed-up.
Moreover, the reduced model was accepted at every canary epoch, such that no epoch after the bootstrap fell back to the full-order model.

\cref{fig:mnist_reduction}(a) shows how the reduction adapts to the evolving parameters.
The two bootstrap epochs provide $42$ equilibrium states, such that the snapshot matrix reaches its capacity of $N_{\mathrm{snap}}^{\max} = 30$ right after the bootstrap.
From then on, every state accepted at a canary epoch replaces the oldest snapshot, and the snapshot matrix follows the most recent part of the optimization path.
At each canary epoch, those states of the full-order trajectory whose projection error \cref{eq:proj_error} exceeds $\varepsilon_{\mathrm{proj}}$ are added, but at least the state with the largest error; in the present run, between $1$ and $21$ of the $21$ states per canary epoch.
The dimension of the trial space grows stepwise from $N_{\mathrm{r}} = 9$ to $N_{\mathrm{r}} = 13$ at the canary epochs $302$, $402$, $852$ and $902$.
At exactly these canary epochs, the projection error of the updated snapshots, evaluated with the previous dimension, exceeds $\varepsilon_{\mathrm{POD}}$, see the squares in \cref{fig:mnist_reduction}(b).
In other words, the new snapshots carry deformation content that the previous trial space cannot represent within the tolerance, and \cref{eq:Nr} adds one mode.
At all other canary epochs, both $\mathrm{error}_{\mathrm{r}}$ and $\mathrm{error}_{\mathrm{ECSW}}$ remain below $10^{-4}$ with the current dimension.
Interestingly, nine modes suffice during the steep descent of the loss.
We attribute this to the scaling argument above: as long as the optimizer mainly adjusts the overall stiffness level, the displacement field barely changes, and the existing snapshots remain representative.
Additional modes become necessary only later, when the ratio between the phases and the nonlinearity of the response change, which alters the pattern of deformation itself.

The reduced problem is compiled just-in-time \cite{bradbury2018}, and every change in the dimension of an array triggers a new compilation.
We refer to such an update as a rebuild, marked by the black dashed lines in \cref{fig:mnist_training,fig:mnist_reduction}.
In contrast, an update that keeps all dimensions merely replaces the entries of the compiled arrays.
Since the number of sampled elements changes at almost every canary epoch, from $186$ elements after the bootstrap to $356$ elements at the end, see \cref{fig:mnist_reduction}(c), we store $\tilde{\mathcal{E}}$ in an array of fixed capacity and fill the unused entries with an element of zero weight.
The capacity is chosen at the bootstrap as the smallest size of a fixed sequence $4, 8, 16, 24, 32, 48, 64, 96, 128, 192, 256, 384, 512, 768, 1\,024, \dots$ that is not smaller than $\max\{1.5\,n, n+4\}$, where $n$ denotes the number of active elements; here, $1.5\cdot 186 = 279$ yields $384$ elements.
Only if a later rule outgrows this capacity, it is raised by the same rule, and it is never decreased.
These padded elements are evaluated, but by $w^e = 0$ they contribute neither to $\tilde{\mathbf{r}}_{\mathrm{r}}$ nor to $\tilde{\mathbf{K}}_{\mathrm{r}}$ in \cref{eq:ecsw}.
Hence, as long as the active set fits into this capacity, a new ECSW rule requires no recompilation.
The reduced basis, in contrast, is not padded, so that each increase of $N_{\mathrm{r}}$ results in a rebuild; a decrease is not admitted, for the same reason.
In the present run, the active set remains below the capacity throughout, and all four rebuilds after the initial construction stem from the growth of the trial space.
The padding comes at the price of evaluating up to $198$ elements without contribution at the beginning and $28$ at the end.
In return, it avoids a recompilation at $15$ of the $19$ canary epochs, each of which costs about $45$ to $55$~s here, as the comparison of the canary epochs with and without a rebuild in \cref{fig:mnist_training}(c) shows.

\begin{figure}[tbp]
    \centering
    \includegraphics[width=\textwidth]{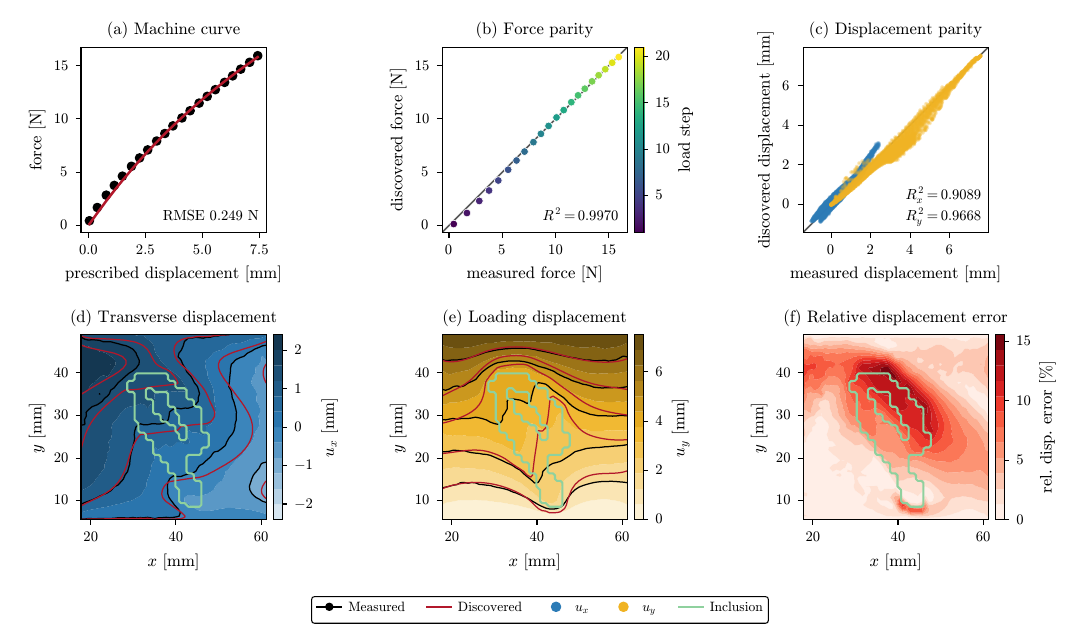}\\[8pt]
    \begin{minipage}[b]{0.2\textwidth}
        \centering
        \includegraphics[width=\linewidth]{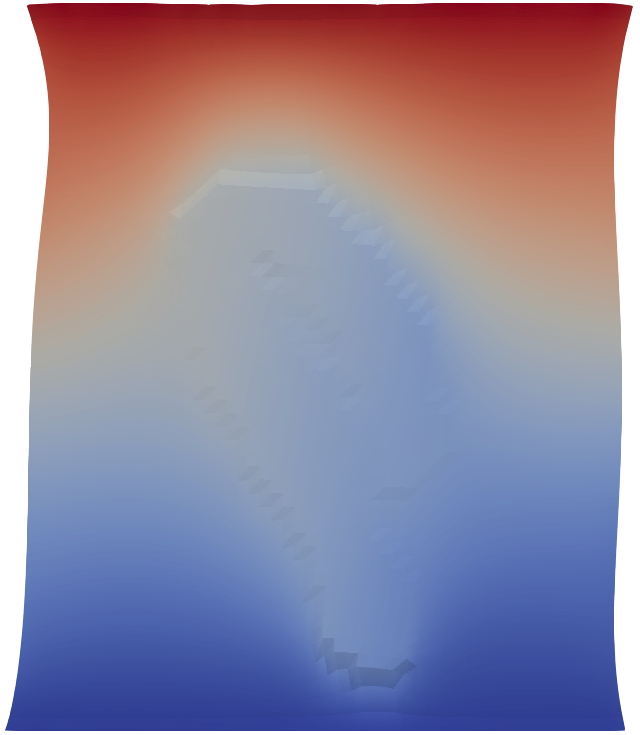}\\[2pt]
        \hspace*{-0.9cm}
        \begin{tikzpicture}
            \begin{axis}[
                width=1.3\linewidth, height=1.6cm, hide axis,
                colormap name=coolwarmC, colorbar horizontal,
                point meta min=-0.014, point meta max=7.5,
                colorbar style={
                    xlabel={$u_y$ [mm]},
                    xtick={-0.014,2,4,6,7.5},
                    xticklabels={$-1.4\cdot10^{-2}$,2,4,6,$7.5$},
                    tick label style={font=\scriptsize},
                    label style={font=\footnotesize}
                }
            ]
                \addplot [draw=none, forget plot] coordinates {(0,0)};
            \end{axis}
        \end{tikzpicture}\par\smallskip
        \makebox[\linewidth][c]{\small (g) Displacement $u_y$}
    \end{minipage}
    \hspace*{3cm}
    \begin{minipage}[b]{0.2\textwidth}
        \centering
        \includegraphics[width=\linewidth]{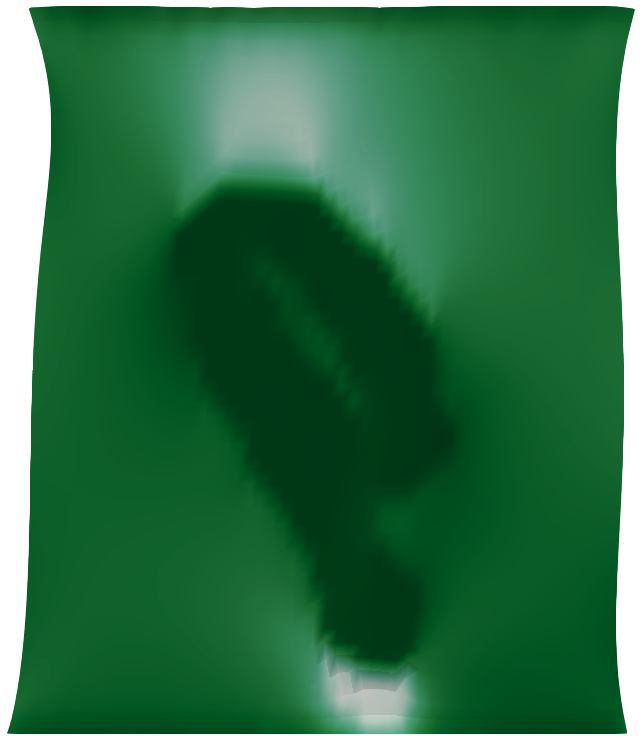}\\[2pt]
        \hspace*{-0.4cm}
        \begin{tikzpicture}
            \begin{axis}[
                width=1.3\linewidth, height=1.6cm, hide axis,
                colormap name=greensRevC, colorbar horizontal,
                point meta min=-0.00027, point meta max=0.867,
                colorbar style={
                    xlabel={$F_{yy}$ [-]},
                    xtick={-0.00027,0.2,0.4,0.6,0.867},
                    xticklabels={$1.0$,$1.2$,$1.4$,$1.6$,$1.87$},
                    tick label style={font=\scriptsize},
                    label style={font=\footnotesize}
                }
            ]
                \addplot [draw=none, forget plot] coordinates {(0,0)};
            \end{axis}
        \end{tikzpicture}\par\smallskip
        \makebox[\linewidth][c]{\small (h) Deformation gradient $F_{yy}$}
    \end{minipage}

    \caption{\textbf{Case~1, identified material evaluated by a full-order solve, compared with the measurement.} (a) The machine curve is reproduced with a root-mean-square error of $0.249$~N, and (b) the force parity yields $R^2=0.9970$. (c) The displacements, which are not part of the loss, agree with the DIC measurement with $R^2_x=0.909$ and $R^2_y=0.967$. (d)--(e) Contour lines of the measured (black) and discovered (red) transverse and loading displacement nearly coincide away from the inclusion. (f) The relative displacement error remains below $5\,\%$ in most of the plate and concentrates along the upper flank of the inclusion (green outline) with up to $15\,\%$. (g)--(h) Deformed configuration at the last load step: the inclusion moves almost rigidly and remains almost undeformed with $F_{yy}\approx 1.0$, whereas the matrix above and below the inclusion is stretched up to $F_{yy}=1.87$. Since only the reaction force enters the loss, this contrast is learned from the structural response alone.}
    \label{fig:mnist_evaluation}
\end{figure}

\paragraph{Identified material.}
\cref{fig:mnist_evaluation} compares the response of the identified material with the measurement.
The machine curve in \cref{fig:mnist_evaluation}(a) is reproduced over all load steps with a root-mean-square error of $0.249$~N at a maximum force of $15.9$~N, i.e., a relative error of $2.5\,\%$, and the force parity in \cref{fig:mnist_evaluation}(b) yields $R^2 = 0.9970$.
More informative, however, are the displacements, which the loss never contained.
Their relative error amounts to $10.4\,\%$, and the displacement parity in \cref{fig:mnist_evaluation}(c) yields $R^2 = 0.909$ for the transverse and $R^2 = 0.967$ for the loading direction, and the contour lines of the discovered and the measured displacement in \cref{fig:mnist_evaluation}(d)--(e) largely coincide.
The relative error in \cref{fig:mnist_evaluation}(f) remains below $5\,\%$ in large parts of the plate and reaches up to $15\,\%$ in a band along the upper flank of the inclusion and at its lower tip close to the clamped edge.
For the full-order reference, the force error is slightly larger with $0.326$~N or $3.2\,\%$, whereas the displacement agreement is better with a relative error of $4.7\,\%$, $R^2 = 0.976$ and $R^2 = 0.993$, and a pointwise maximum of about $8\,\%$.
Both parameter sets thus reproduce the force comparably well but distribute the stiffness slightly differently between the phases, which becomes visible only in the displacement field.
Since both parameter sets are evaluated by the same full-order solve, this difference is not an error of the reduced model; rather, two optimization paths end in different minima of comparable loss.
In \cite{knipper2026}, the displacements entered the loss explicitly, which constrains the choice between such minima at the expense of requiring full-field data during training; for the same specimen, they report a force error of about $3.5\,\%$ and a displacement error of about $2.8\,\%$.
The force error of the hyperreduced identification is thus of the same order, whereas its displacement error is larger, as expected for a loss that contains no displacements.

\cref{fig:mnist_evaluation}(g)--(h) shows the displacement in loading direction and the corresponding component $F_{yy}$ of the deformation gradient on the deformed configuration at the last load step.
The stretch differs markedly between the phases: the inclusion remains almost undeformed with $F_{yy}\approx 1.0$, while the matrix above and below the inclusion is stretched up to $F_{yy} = 1.87$.
Since both networks start from random parameters drawn from the same distribution, such a contrast can only arise if the optimizer has assigned a substantially stiffer response to the inclusion than to the matrix, which is consistent with the description of the specimens \cite{knipper2026}.
Together with the agreement of the displacement field, this indicates that the force alone suffices to separate both phases, at least qualitatively.
A quantitative comparison of the individual phase laws is not possible here, since the data contain no stress--strain response of the isolated phases.

\begin{figure}[tbp]
    \centering
    \includegraphics[width=0.8\textwidth]{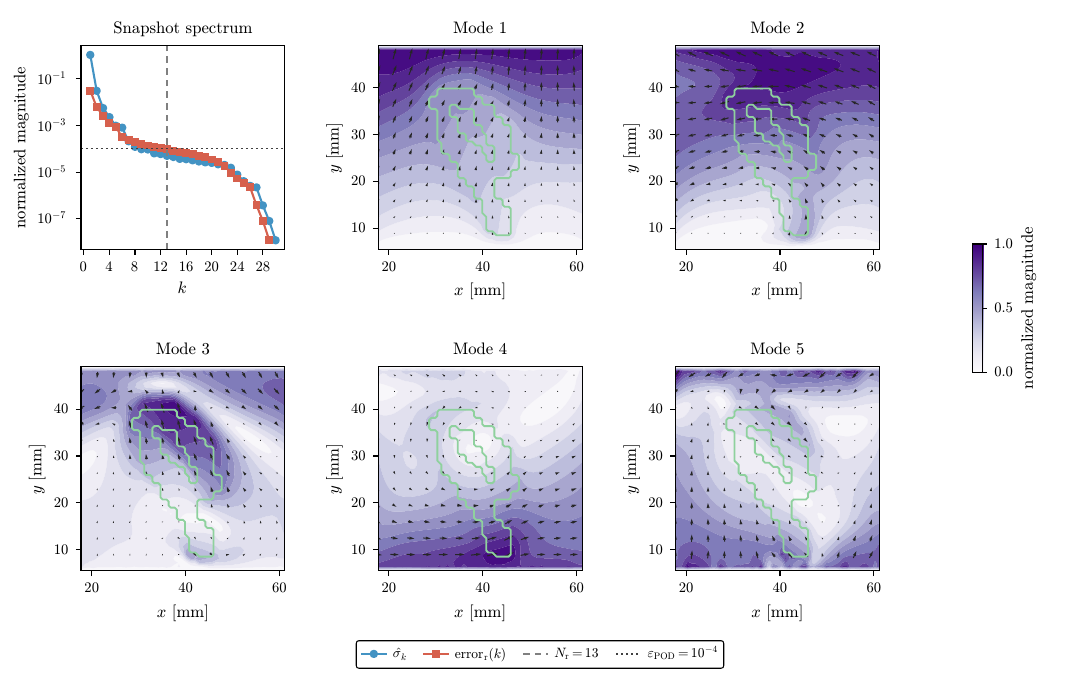}
    \caption{\textbf{Case~1, trial space at the end of the training.} Top left: normalized singular values $\hat{\sigma}_k$ of the snapshot matrix and projection error $\mathrm{error}_{\mathrm{r}}(k)$. After a decay of four orders of magnitude within eight modes, the spectrum forms a plateau, in which the projection error crosses $\varepsilon_{\mathrm{POD}}=10^{-4}$ at $N_{\mathrm{r}}=13$. Remaining panels: magnitude (color) and direction (arrows) of the first five modes. Mode~1 describes the stretching in loading direction, mode~2 a lateral motion of the upper part, mode~3 the deformation around the upper part of the inclusion, mode~4 the region around its lower tip, and mode~5 boundary layers along the clamped edges.}
    \label{fig:mnist_modes}
\end{figure}

\paragraph{Trial space and hyperreduction.}
\cref{fig:mnist_modes} shows the singular values of the final snapshot matrix together with the projection error \cref{eq:pod_error}.
The normalized singular values decay by about four orders of magnitude within the first eight modes, followed by a plateau between $10^{-4}$ and $10^{-5}$ that extends up to $k\approx 22$, before they drop sharply.
The projection error falls below $\varepsilon_{\mathrm{POD}} = 10^{-4}$ within this plateau at $k = 13$, which is the final dimension $N_{\mathrm{r}} = 13$ in \cref{fig:mnist_reduction}(a).
Since the threshold intersects the flat part of the spectrum, small changes of the snapshots shift the selected dimension by one mode, which explains the stepwise growth of $N_{\mathrm{r}}$ during training.
For the same reason, the choice of $\varepsilon_{\mathrm{POD}}$ strongly affects the size of the trial space, as the study below confirms.
The first mode represents the stretching in loading direction, whose distribution is distorted by the stiffer inclusion.
The second mode describes a lateral motion of the upper part of the plate, while the third mode concentrates on the upper part of the inclusion and the adjacent matrix.
Notably, this is the region in which the displacement error in \cref{fig:mnist_evaluation}(f) is largest, i.e., the region whose deformation is most sensitive to the ratio between both phases.
The fourth mode captures the lower tip of the inclusion close to the clamped edge, and the fifth mode boundary layers along both clamped edges, which presumably stem from the non-uniform measured edge displacements.

\begin{figure}[tbp]
    \centering
    \begin{tikzpicture}
        \node[anchor=west, inner sep=0] (ecswimg) at (0,0) {\includegraphics[height=6cm]{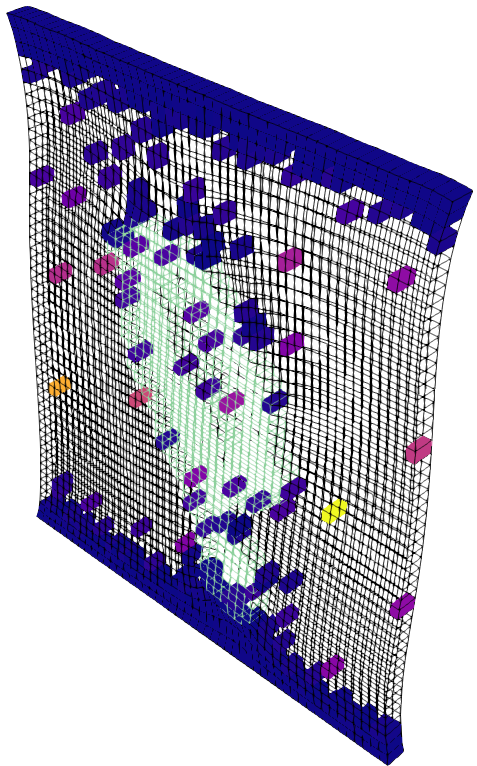}};
        \begin{axis}[
            at={(ecswimg.east)}, anchor=west, xshift=0.8cm,
            width=1.8cm, height=6cm, hide axis,
            colormap name=plasmaC, colorbar,
            point meta min=0, point meta max=220,
            colorbar style={
                width=0.35cm,
                ylabel={ECSW weight $w^e$},
                ytick={0,55,110,165,220},
                yticklabels={0,55,110,165,220},
                tick label style={font=\footnotesize},
                ylabel style={font=\small}
            }
        ]
            \addplot [draw=none, forget plot] coordinates {(0,0)};
        \end{axis}
    \end{tikzpicture}
    \caption{\textbf{Case~1, final ECSW rule on the deformed mesh.} The remaining mesh is shown as wireframe and the inclusion in green. The $356$ active elements (about $14\,\%$ of the mesh) are colored by their weight; the $28$ padded entries of the set of $384$ carry zero weight and add no element. The element rows along both clamped edges are selected completely with weights close to one, and further elements concentrate in and around the inclusion, whereas the matrix away from the inclusion is represented by few elements with weights of up to about $220$.}
    \label{fig:mnist_ecsw}
\end{figure}

\cref{fig:mnist_ecsw} shows the final ECSW rule.
Of the $384$ entries of the padded set, $356$ refer to active elements, i.e., about $14\,\%$ of the mesh, of which $21$ belong to the inclusion.
The remaining $28$ entries are padded with zero weight and repeatedly refer to one element, such that they add no element to the figure.
The sampled elements cluster along both clamped edges as well as in and around the inclusion, in particular along the interface between the phases.
In the matrix away from the inclusion, in contrast, only a few elements are selected, some of them with weights of up to $220$.
This reflects the principle of ECSW: in regions of nearly uniform deformation, a single element with a large weight represents many similar elements, whereas regions with strong gradients and a change of material require many elements with moderate weights.
Notably, the element rows along both clamped edges are selected completely, with weights between $0.994$ and $1.006$, i.e., they are integrated practically exactly.
This is a consequence of the loss: since the trial space spans only the free degrees of freedom, whereas the reaction force lives on the prescribed ones, the ECSW training additionally contains one row per snapshot that reproduces the reaction force entering the loss.
The elements adjacent to the loaded edge contribute directly to this reaction, and the boundary layers of the fifth mode involve both edges.
Note that the sampled set necessarily contains elements of both phases.
Otherwise, the reduced residual and thus the gradient in \cref{eq:reduced_adjoint} would not depend on the parameters of the missing phase, and this phase could not be trained.

\begin{figure}[tbp]
    \centering
    \includegraphics[width=0.8\textwidth]{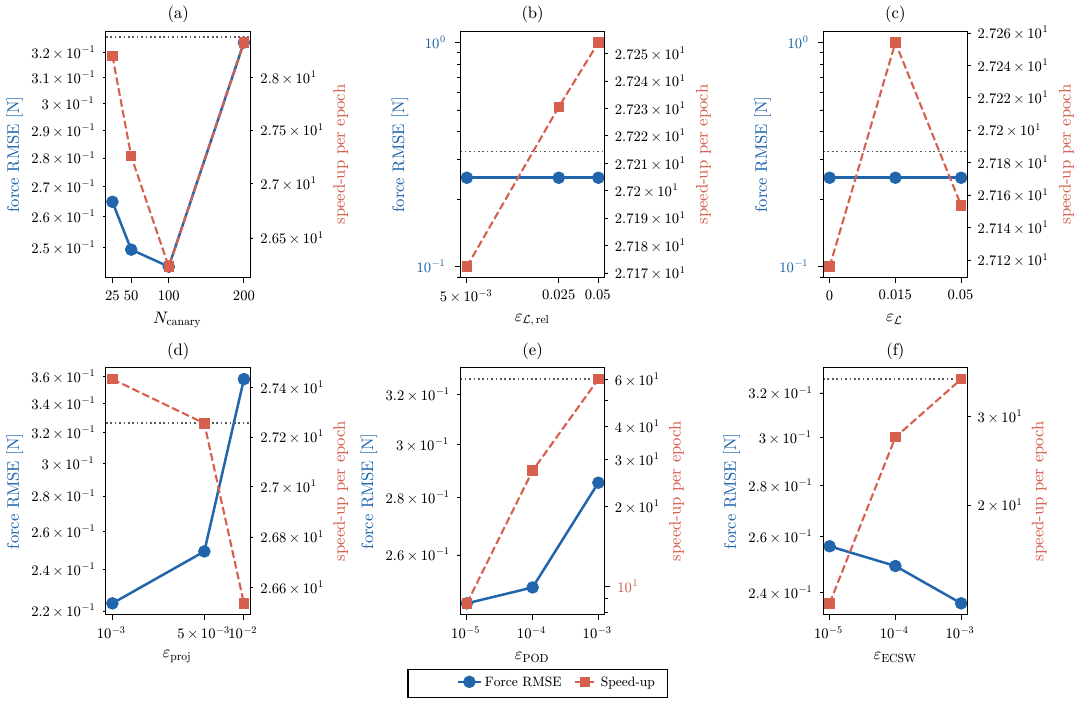}
    \caption{\textbf{Case~1, one-at-a-time variation of the reduction parameters around the base configuration of \cref{tab:hyperparameters}.} Shown are the force error of the identified material (blue, left axis) and the speed-up per epoch against the full-order reference (red, right axis); the dotted line marks the force error of the full-order reference. (a) Canary interval $N_{\mathrm{canary}}$: only the largest interval deteriorates the accuracy. (b)--(c) Tolerances of the model selection: no influence in the tested range. (d) Snapshot update $\varepsilon_{\mathrm{proj}}$: stricter tolerances improve the accuracy at constant cost. (e) POD tolerance $\varepsilon_{\mathrm{POD}}$ and (f) ECSW tolerance $\varepsilon_{\mathrm{ECSW}}$: the speed-up varies between $8.6$ and $60$, while the force error remains close to the base configuration.}
    \label{fig:mnist_sweeps}
\end{figure}

\begin{figure}[tbp]
    \centering
    \includegraphics[width=0.8\textwidth]{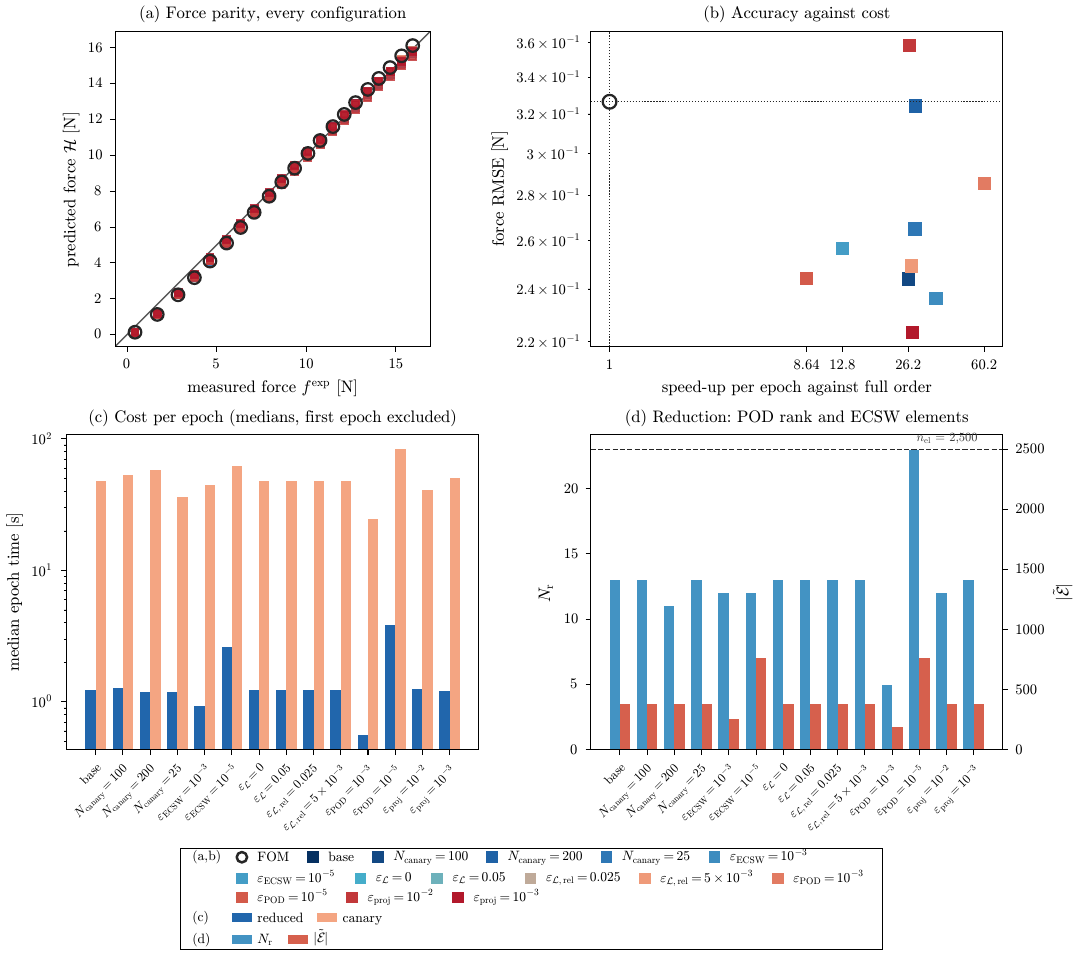}
    \caption{\textbf{Case~1, comparison of all configurations of the parameter study.} (a) Force parity: all reduced runs (squares) and the full-order reference (circles) reproduce the measured force nearly identically. (b) Force error against the speed-up per epoch: all reduced runs except $\varepsilon_{\mathrm{proj}}=10^{-2}$ reach an error at or below that of the full-order reference (dotted lines) at speed-ups between $8.6$ and $60$. (c) Median epoch times: reduced epochs require $0.5$ to $4$~s, canary epochs between $25$ and $85$~s. (d) Final dimension of the trial space $N_{\mathrm{r}}$ and size of the padded ECSW set $|\tilde{\mathcal{E}}|$, compared with the $n_{\mathrm{el}}=2\,500$ elements of the mesh; only the tolerances of POD and ECSW change these sizes considerably.}
    \label{fig:mnist_study}
\end{figure}

\paragraph{Influence of the reduction parameters.}
Finally, we investigate how the parameters of the reduction affect accuracy and cost.
To this end, we vary one parameter at a time around the base configuration in \cref{tab:hyperparameters}, which results in $13$ additional runs.
\cref{fig:mnist_sweeps} shows the force error and the speed-up per epoch for each parameter, and \cref{fig:mnist_study} compares all runs with each other and with the full-order reference.
Here, the speed-up per epoch denotes the ratio of the median epoch times of the full-order and the reduced run, excluding the first epoch.
Since the few canary epochs do not affect the median, it isolates the efficiency of the reduced solve, whereas the cost of the canary epochs, discussed above, comes on top.

All configurations reproduce the measured force, see \cref{fig:mnist_study}(a), with root-mean-square errors between $0.22$ and $0.36$~N, compared with $0.33$~N for the full-order reference.
The tolerances of the model selection, $\varepsilon_{\mathcal{L}}$ and $\varepsilon_{\mathcal{L},\mathrm{rel}}$, have no influence in the tested range: all runs yield the same force error, and the speed-up varies by less than $1\,\%$, see \cref{fig:mnist_sweeps}(b)--(c).
Hence, the model selection takes the same decisions as in the base configuration, in which the reduced model was accepted at every canary epoch, and the criterion \cref{eq:gate_condition} acts as a safeguard that is not triggered in this example.
The canary interval affects the accuracy only for the largest value, see \cref{fig:mnist_sweeps}(a): for $N_{\mathrm{canary}} \leq 100$, the error lies between $0.244$ and $0.265$~N, whereas for $N_{\mathrm{canary}} = 200$ it increases to the level of the full-order reference, and the trial space grows only to $N_{\mathrm{r}} = 11$.
The speed-up per epoch remains between $26$ and $28$, since the median is insensitive to the canary epochs, while the overall cost grows with their number.
A similar trend appears for the snapshot update in \cref{fig:mnist_sweeps}(d): the strictest tolerance $\varepsilon_{\mathrm{proj}} = 10^{-3}$ yields the smallest error of all runs with $0.223$~N, and the loosest one, $\varepsilon_{\mathrm{proj}} = 10^{-2}$, the only error above the full-order reference with $0.358$~N.
Both observations indicate that the accuracy depends primarily on how up to date the snapshots are.

The cost, in contrast, is governed by the tolerances of POD and ECSW, see \cref{fig:mnist_sweeps}(e)--(f) and \cref{fig:mnist_study}(c)--(d).
For $\varepsilon_{\mathrm{POD}} = 10^{-3}$, five modes suffice, the padded set shrinks to $192$ elements, and the speed-up per epoch reaches $60$ at a force error of $0.285$~N.
For $\varepsilon_{\mathrm{POD}} = 10^{-5}$, the trial space grows to $N_{\mathrm{r}} = 23$ and the padded set to $768$ elements, which reduces the speed-up to $8.6$ without a noticeable gain in accuracy.
This sensitivity follows directly from the plateau of the singular values in \cref{fig:mnist_modes}.
Likewise, $\varepsilon_{\mathrm{ECSW}} = 10^{-3}$ reduces the padded set to $256$ elements and increases the speed-up to $35.5$ at an error of $0.236$~N, whereas $\varepsilon_{\mathrm{ECSW}} = 10^{-5}$ doubles the padded set to $768$ elements and decreases the speed-up to $12.8$, again without improving the force.
Overall, the differences in accuracy between the reduced runs are of the same order as the difference between the base configuration and the full-order reference, and they show no systematic trade-off against the cost.
We therefore conclude that, in this example, the remaining error stems from the non-uniqueness of the force-based identification rather than from the reduction, and that the tolerances of POD and ECSW can be relaxed considerably as long as the snapshots are updated regularly.

In summary, a single force--displacement curve suffices to identify a material that separates both phases and reproduces the unseen displacement field, while the adaptive reduction lowers the cost of the training by a factor of $13$ without degrading the force accuracy.
In the next case, we examine whether these observations persist for a perforated specimen.


\subsection{Case 2: Homogeneous material, perforated specimen}
\label{sec:case2}

\begin{figure}[tbp]
    \centering
    \begin{tikzpicture}[>={Stealth[length=2.2mm]}]
        \node (geo) at (0,0) {\includegraphics[width=0.4\textwidth]{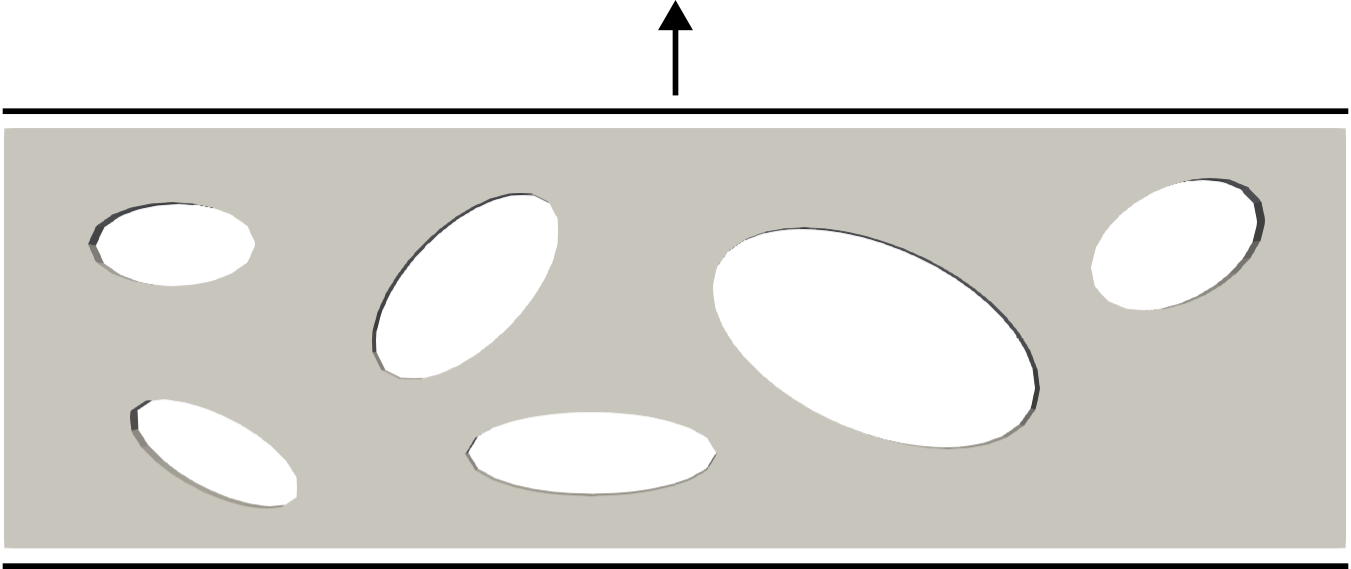}};
        \node[above] at (geo.north) {$\bm{u}=(u_x(x,t),\,u_y(x,t),\,0)$};
        \node[below] at (geo.south) {$\bm{u}=(u_x(x,t),\,u_y(x,t),\,0)$};
        \draw[->, thick] ([xshift=0.3cm,yshift=-0.4cm]geo.south east) -- ++(1,0) node[right] {$x$};
        \draw[->, thick] ([xshift=0.3cm,yshift=-0.4cm]geo.south east) -- ++(0,1) node[above] {$y$};
    \end{tikzpicture}
    \caption{\textbf{Case~2, region of interest of the perforated specimen TT1 of \citet{abbasi2026} in the reference configuration.} The membrane of $160\times 50\times 2.5$~mm with six elliptical holes is discretized by $1\,518$ hexahedral elements with one element across the thickness. On the top and bottom edges, B-spline fits of the measured in-plane displacements $u_x(x,t)$ and $u_y(x,t)$ are prescribed for $N_{\mathrm{t}}=51$ load steps, and the out-of-plane displacement is fixed. The reaction force on the bottom edge enters both losses, the DIC displacements at $11$ load steps only the second one.}
    \label{fig:abbasi_geometry}
\end{figure}

Next, we turn to the opposite situation: a single material, but a strongly heterogeneous deformation.
We use the tensile test TT1 of the open dataset by \citet{abbasi2026}, in which a perforated polymer membrane with six elliptical holes is stretched in tension, see \cref{fig:abbasi_geometry}.
Along the narrow ligaments between the holes, the material is locally stretched far beyond the nominal strain, while the regions above and below the holes are barely loaded.
For each recorded frame, the dataset provides the machine force and the DIC displacements, and both have also served for the full-order identification in \cite{knipper2026}.
Unlike in Case~1, the displacement field serves here as more than a check.
We compare two identifications that differ only in their loss: one from the force alone and one from the force and the displacement field.
This allows us to ask what the force already implies about the field, what the field adds to the identification, and how the reduction copes with a loss that contains the displacements themselves.

The region of interest of $160\times 50\times 2.5$~mm is discretized with the mesh provided with the dataset, which consists of $1\,518$ hexahedral elements with one element across the thickness, $3\,402$ nodes and $10\,206$ degrees of freedom, of which $N_{\mathrm{f}} = 9\,234$ are free.
The material of TT1 is usually treated as incompressible \cite{abbasi2026,knipper2026}.
We do not impose this assumption, since the energy of \cref{sec:neural_elasticity} is compressible, but we use an element-wise constant volume ratio (Q1/P0) \cite{nagtegaal1974,simo1985} to avoid volumetric locking should the identified material approach the incompressible limit, see \cref{app:q1p0}.
Since the element potential is differentiated with the volume ratio as a function of the nodal displacements, the variational structure of \cref{eq:residual} is preserved.
On the top and bottom edges, we prescribe B-spline fits of the measured in-plane displacements, as in \cite{knipper2026}, together with a vanishing out-of-plane displacement.
The loading path consists of $N_{\mathrm{t}} = 51$ load steps, essentially every second DIC frame up to frame $100$, of which $11$ carry a reference displacement field.
The force is evaluated as the reaction on the bottom edge, and the material is represented by a single network with the architecture of Case~1.

In the following, we first introduce the two losses, then discuss training and reduction, and assess the identified material on the training and on an unseen specimen.
We then compare the initial moduli of both identified materials with the common assumption of incompressibility, and conclude with the trial space and the ECSW rule.

\paragraph{Loss and hyperparameters.}
The first loss contains only the force, as in Case~1.
Here, however, we divide both the measured and the predicted force by $\bar{f} = \max_n|f^{\mathrm{exp}}_n| = 236.9$~N, such that the identified energy carries physical units,
\begin{equation}
    \mathcal{L}_{\mathrm{F}} = \frac{1}{N_{\mathrm{t}}} \sum_{n=1}^{N_{\mathrm{t}}} \left[ \frac{\mathcal{H}(\mathbf{d}_n,\bm{\theta}) - f^{\mathrm{exp}}_n}{\bar{f}} \right]^2 .
    \label{eq:abbasi_loss_force}
\end{equation}
Both scalings share the same minimizer in terms of the physical energy, but they differ in the magnitude of the loss and thus in the meaning of the absolute tolerance $\varepsilon_{\mathcal{L}}$, which we calibrated for the present scaling.
The second loss adds the measured displacements at the set $\mathcal{D}$ of the $11$ observed load steps,
\begin{equation}
    \mathcal{L}_{\mathrm{F}+\mathrm{u}} = \mathcal{L}_{\mathrm{F}} + \frac{\alpha}{2\,N_{\mathrm{obs}}\,|\mathcal{D}|} \sum_{n\in\mathcal{D}} \sum_{k=1}^{N_{\mathrm{obs}}} \big\lVert \bm{u}^h_n(\bm{X}_k) - \bm{u}^{\mathrm{exp}}_n(\bm{X}_k) \big\rVert^2 ,
    \label{eq:abbasi_loss_fd}
\end{equation}
where $\bm{X}_k$ denote the $N_{\mathrm{obs}}$ DIC points on the observed face, and the factor two accounts for the two in-plane components, such that the second term is the mean squared displacement error in mm$^2$.
In contrast to \cref{eq:loss}, the force and the displacements are thus evaluated on different sets of load steps.
We choose $\alpha$ such that both contributions are equal at the initial parameters, which yields $\alpha = 0.434$~mm$^{-2}$, since $\alpha$ converts units as much as it weights the contributions.
Recall from Case~1 that the displacement field is invariant under a scaling of the energy for a purely displacement-driven problem.
Hence, the displacement term alone cannot determine the stiffness level, while the force term determines it.
Both terms are thus complementary: the force fixes the level of the response, and the displacement constrains its spatial distribution.
Consequently, both terms can be reduced at the same time only if model and measurement are consistent with each other.
Both runs start from the same initial parameters, run over $700$ epochs at a learning rate of $2\cdot 10^{-2}$, and use $\varepsilon_{\mathrm{POD}} = 10^{-6}$; all further settings are listed in \cref{tab:hyperparameters}.
No full-order training run is available for this case.

\begin{figure}[tbp]
    \centering
    \includegraphics[width=0.8\textwidth]{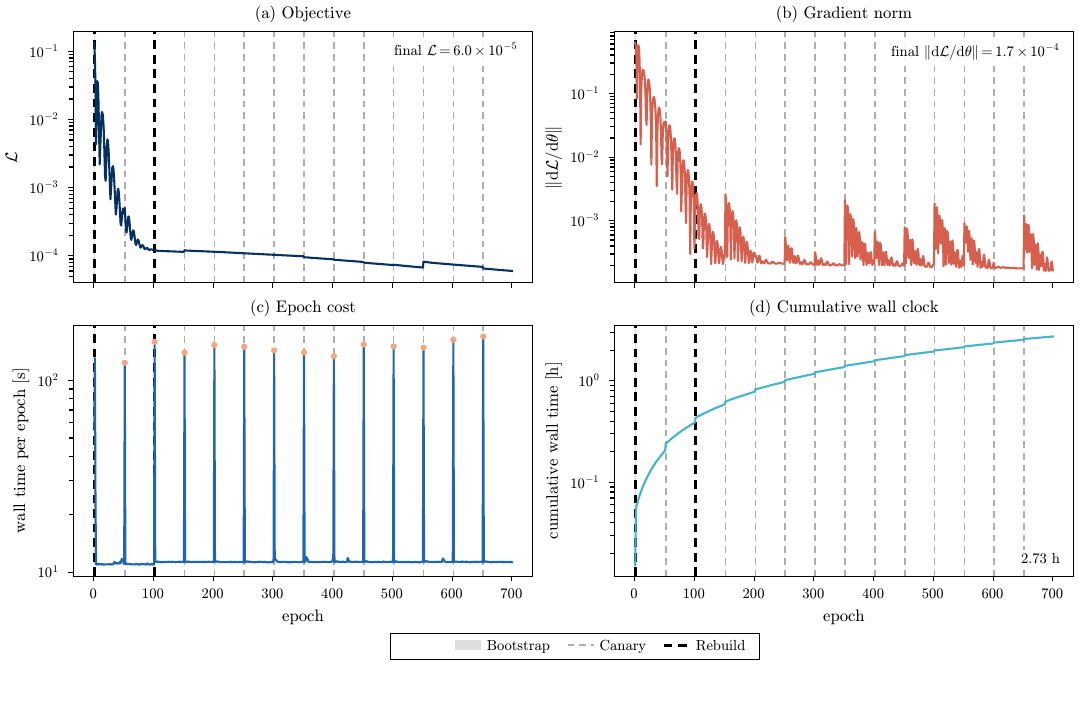}
    \caption{\textbf{Case~2, training history of the force-based identification on TT1.} (a) The loss decreases by three orders of magnitude within $100$ epochs and reaches $\mathcal{L}=6.0\cdot10^{-5}$. (b) The gradient norm shows short peaks at canary epochs, when the ECSW rule is updated, which decay within a few epochs. (c) Reduced epochs take about $11$~s, canary epochs between $123$ and $169$~s. (d) The training requires $2.73$~h in total. Grey dashed lines mark canary epochs and black dashed lines rebuilds.}
    \label{fig:abbasi_training_force}
\end{figure}

\begin{figure}[tbp]
    \centering
    \includegraphics[width=0.8\textwidth]{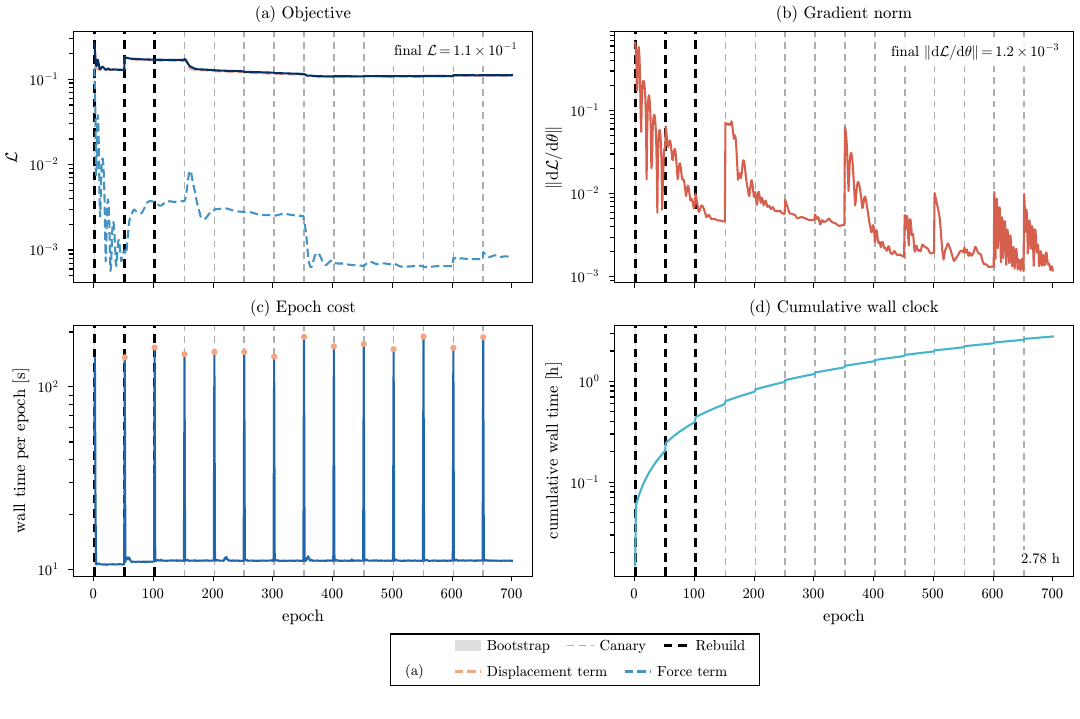}\\[4pt]
    \includegraphics[width=0.8\textwidth]{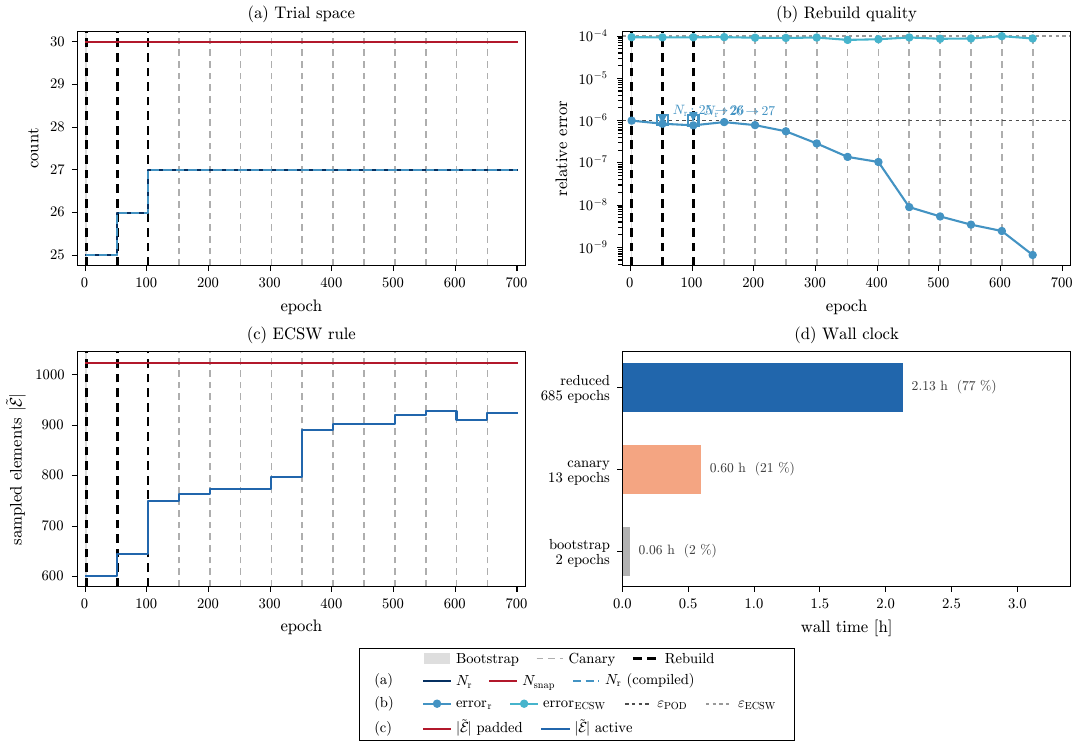}
    \caption{\textbf{Case~2, identification from force and displacements.} Top: training history; the total loss (solid) is dominated by the displacement term, which levels off at about $1.1\cdot10^{-1}$, while the force term (dashed blue) falls below $10^{-3}$. The jump of the displacement term at the first canary epoch shows that the outdated trial space had underestimated the displacement misfit. Bottom: adaptivity of the reduction; the trial space grows from $N_{\mathrm{r}}=25$ to $27$ at epochs $52$ and $102$, and ECSW requires up to $928$ active elements. The total wall-clock time of $2.78$~h is nearly identical to the force-based run.}
    \label{fig:abbasi_reduction_fd}
    \label{fig:abbasi_training_fd}
\end{figure}

\paragraph{Training history and adaptivity of the reduction.}
For the force-based run, \cref{fig:abbasi_training_force}(a) shows a decrease of the loss from about $10^{-1}$ to $10^{-4}$ within the first $100$ epochs, accompanied by pronounced oscillations, and a slow decrease to $\mathcal{L} = 6.0\cdot 10^{-5}$ thereafter.
The gradient norm in \cref{fig:abbasi_training_force}(b) exhibits short peaks at several canary epochs, i.e., whenever an updated ECSW rule slightly changes the reduced landscape, but these peaks decay within a few epochs and leave no trace in the loss.
For the run with both channels, \cref{fig:abbasi_training_fd}(a) shows that the total loss is dominated by the displacement term, which levels off at about $1.1\cdot 10^{-1}$, while the force term falls below $10^{-3}$.
At the first canary epoch, epoch~$52$, the displacement term jumps from $0.128$ to $0.179$.
At this epoch, the full-order trajectory deviates from the current trial space by a projection error of $1.3\cdot 10^{-2}$, i.e., well above $\varepsilon_{\mathrm{proj}} = 5\cdot 10^{-3}$, such that all $51$ states are added to the snapshot matrix and the trial space is enlarged from $25$ to $26$ modes.
Since the displacements enter this loss through the reconstruction $\mathbf{d}_{\mathrm{f},0}+\mathbf{V}\mathbf{q}$, an outdated trial space directly affects the loss value, and the reduced model had apparently underestimated the displacement misfit during the first $50$ epochs.

\begin{figure}[tbp]
    \centering
    \includegraphics[width=0.8\textwidth]{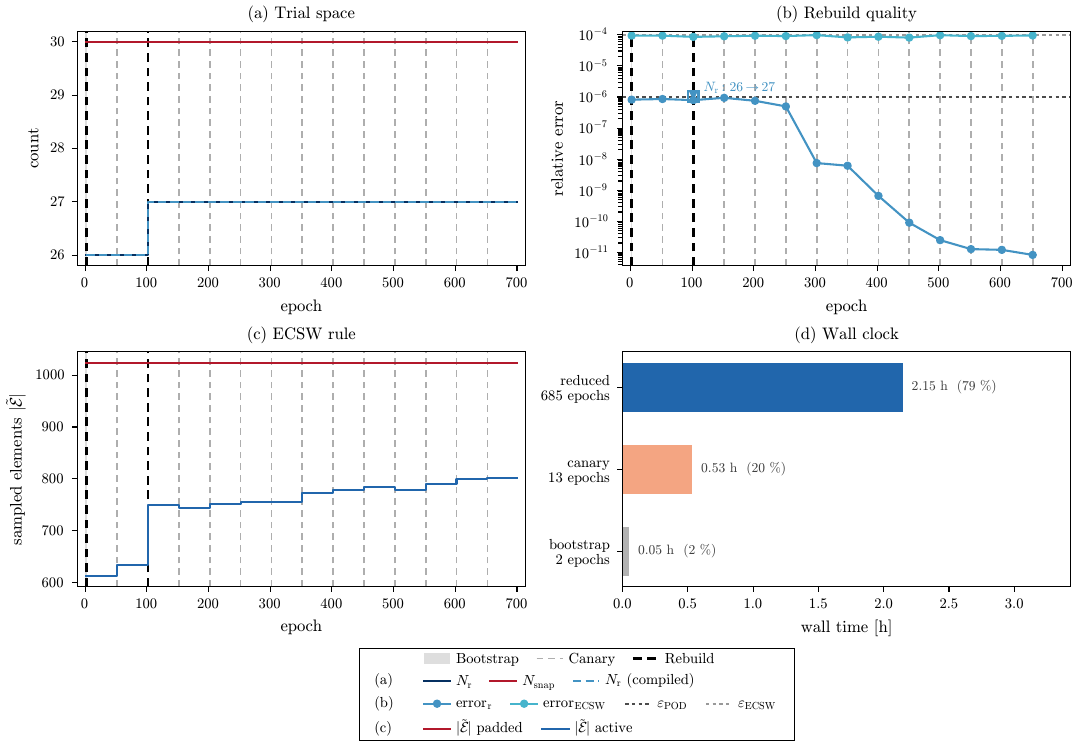}
    \caption{\textbf{Case~2, adaptivity of the reduction in the force-based identification.} (a) The snapshot matrix is full from the first bootstrap epoch on, since one epoch contains $51$ states, and the trial space grows once from $N_{\mathrm{r}}=26$ to $27$. (b) After epoch~$300$, the projection error of the snapshots drops to about $10^{-11}$, as the snapshots of the converging material become nearly linearly dependent. (c) ECSW requires $613$ to $801$ of the $1\,518$ elements, stored in a padded set of $1\,024$ entries. (d) The reduced epochs dominate the wall-clock time with $79\,\%$.}
    \label{fig:abbasi_reduction}
\end{figure}

The reduced epochs take about $11$~s in both runs and thus about ten times longer than in Case~1.
Three effects add up.
First, the loading path contains $51$ instead of $21$ load steps.
Second, the tighter tolerance $\varepsilon_{\mathrm{POD}} = 10^{-6}$ leads to $N_{\mathrm{r}} = 27$ modes, see \cref{fig:abbasi_reduction}(a), and thus to $N_{\mathrm{r}} + N_{\mathrm{r}}^2 = 756$ quantities per snapshot that ECSW has to reproduce.
Third, the deformation concentrates in many small regions around the holes, and ECSW requires $613$ to $801$ active elements, i.e., up to $53\,\%$ of the mesh, stored in a padded set of $1\,024$ elements, see \cref{fig:abbasi_reduction}(c).
The hyperreduction is therefore far less effective than for the plate of Case~1.
The canary epochs take between $123$ and $169$~s in the force-based and between $145$ and $189$~s in the two-channel run.
Their cost is dominated by the update of the reduction rather than by the full-order solve, since the first epoch, which is solved in full order, took only $55$~s including the compilation.
Without a full-order reference run, we can therefore only state that a reduced epoch is several times cheaper than a full-order epoch at this mesh size; the $685$ reduced epochs account for $79\,\%$ of the total wall-clock time of $2.73$~h, and the $13$ canary epochs for $20\,\%$, see \cref{fig:abbasi_reduction}(d).

The adaptivity differs from Case~1 in one respect.
Already a single epoch provides $51$ states, i.e., more than the snapshot capacity, such that the snapshot matrix is full from the first bootstrap epoch on.
The trial space grows only once in the force-based run, from $26$ to $27$ modes at epoch~$102$, and twice in the two-channel run, at epochs $52$ and $102$, see \cref{fig:abbasi_reduction_fd}.
Afterwards, the projection error of the snapshots decreases by several orders of magnitude to about $10^{-11}$, see \cref{fig:abbasi_reduction}(b).
As the parameters converge, the replaced snapshots stem from nearly the same material, and the snapshot matrix loses rank.
The dimension of the trial space, however, is never decreased, since a smaller trial space would require another compilation, and it remains at $N_{\mathrm{r}} = 27$.

\begin{figure}[tbp]
    \centering
    \includegraphics[width=\textwidth]{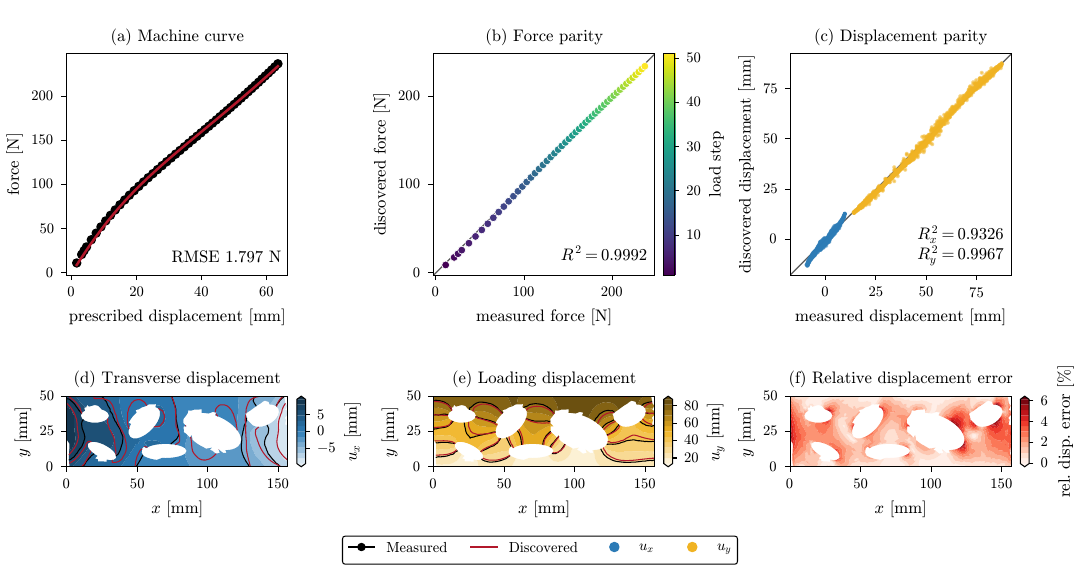}\\[4pt]
    \includegraphics[width=\textwidth]{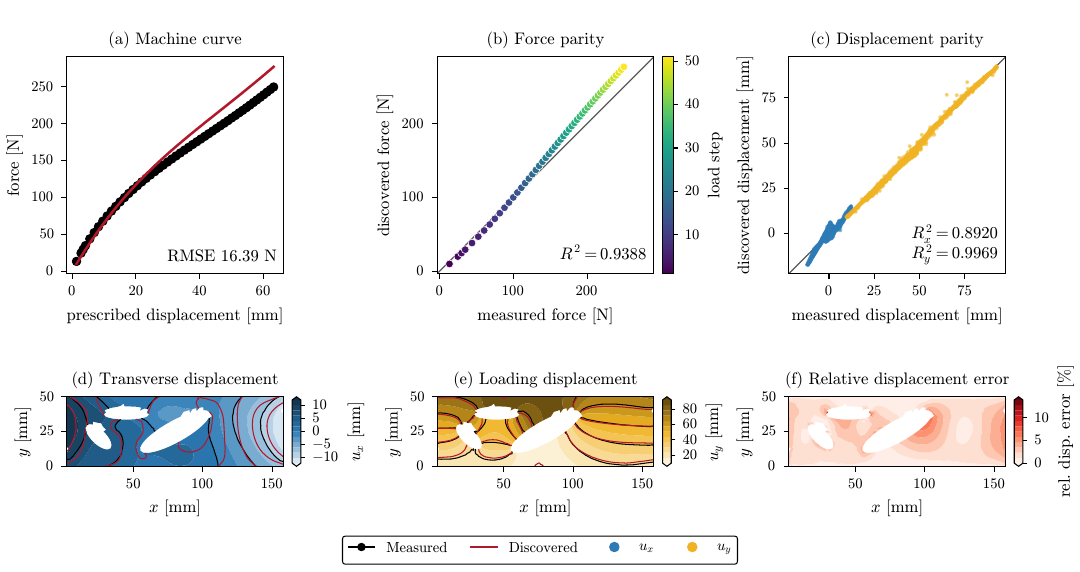}
    \caption{\textbf{Case~2, material identified from the force alone, evaluated by a full-order solve on the training specimen TT1 (top) and on the unseen specimen TT2 (bottom).} Top: (a)--(b) The machine curve of TT1 is reproduced with a root-mean-square error of $1.80$~N and $R^2=0.9992$. (c) The displacements, which are not part of the loss, agree with $R^2_x=0.933$ and $R^2_y=0.997$. (d)--(e) Measured (black) and discovered (red) contour lines nearly coincide in loading direction and deviate slightly in transverse direction. (f) The relative displacement error remains below $6\,\%$ and concentrates around the largest hole and at the right edge. Bottom: For TT2, the material overestimates the force at large displacements by about $10\,\%$ (root-mean-square error of $16.4$~N), in line with the full-order results of \cite{knipper2026}, and predicts the displacement field with $R^2_x=0.892$ and $R^2_y=0.997$. The corresponding results for the material identified from force and displacements are shown in \cref{fig:abbasi_eval_fd}.}
    \label{fig:abbasi_eval_force}
\end{figure}

\paragraph{Identified material.}
The upper half of \cref{fig:abbasi_eval_force} evaluates the material identified from the force alone on the training specimen TT1.
The machine curve is reproduced with a root-mean-square error of $1.80$~N at a maximum force of $236.9$~N, i.e., a relative error of $1.2\,\%$, and the force parity yields $R^2 = 0.9992$.
For comparison, \cite{knipper2026} report a force error of about $0.5\,\%$ for their full-order identification from the same specimen with displacements in the loss.
The displacement field, which is a prediction in this run, agrees with the measurement with a relative error of $3.2\,\%$ and $R^2 = 0.933$ in transverse and $R^2 = 0.997$ in loading direction.
The pointwise error stays below $6\,\%$ and concentrates at the right part of the specimen and around the largest hole.
Adding the displacements to the loss changes this picture in a characteristic way, see \cref{fig:abbasi_eval_fd} in \cref{app:case2_fd}.
The displacement error decreases to $2.1\,\%$, with $R^2 = 0.991$ and $R^2 = 0.998$, whereas the force error increases to $4.2\,\%$, i.e., $6.47$~N, with $R^2 = 0.990$.
In particular, the identified material now underestimates the force at small and overestimates it at large displacements.
Hence, the two measurements do not determine exactly the same material within the present model.
We attribute this mainly to modeling assumptions, such as the nominal geometry of the holes, the thickness, and the prescribed edge displacements, which affect force and field differently; a similar conclusion was drawn in \cite{knipper2026}.

The lower halves of \cref{fig:abbasi_eval_force,fig:abbasi_eval_fd} apply both materials to the second specimen TT2 of the dataset, which has a different arrangement of holes and was not used for training.
Both materials overestimate the force at large displacements by about $10\,\%$, with root-mean-square errors of $16.4$~N and $16.5$~N and $R^2 \approx 0.94$.
This overestimation is independent of whether the displacements entered the loss, and it agrees with the overestimation of about $9\,\%$ reported for TT1-trained models in \cite{knipper2026}, which were identified without any reduction.
We therefore attribute it to differences between both specimens rather than to the reduction.
For the displacement field, in contrast, the additional information pays off: the material identified from both channels predicts the field of TT2 with $R^2 = 0.998$ in both directions, compared with $R^2 = 0.892$ and $R^2 = 0.997$ for the force-based material.

At the last load step of the force-based run, the component $F_{yy}$ of the deformation gradient ranges from $0.64$ at the crowns of the holes, where the material is compressed in loading direction, to $3.94$ in the vertical ligaments between the holes.
A single test thus samples compression as well as large tension in loading direction.

\paragraph{Compressibility of the identified material.}
The energy of \cref{sec:neural_elasticity} does not enforce incompressibility, such that we can ask which volumetric stiffness the identifications have found.
To this end, we compare the initial bulk and shear moduli, which follow from the identified energy along a purely volumetric and a purely isochoric path \citep{ogden1997},
\begin{equation}
    K = \frac{\mathrm{d}^2}{\mathrm{d}\varepsilon^2}\,\psi\big(\exp(\varepsilon/3)\,\bm{I}\big)\Big|_{\varepsilon=0},
    \qquad
    \mu = \frac{\mathrm{d}^2}{\mathrm{d}\gamma^2}\,\psi\big(\bm{I}+\gamma\,\bm{e}_1\otimes\bm{e}_2\big)\Big|_{\gamma=0},
    \label{eq:linear_moduli}
\end{equation}
where $\varepsilon = \ln J$ denotes the volumetric strain and $\gamma$ the shear.
Both identifications yield nearly the same shear modulus, $\mu = 0.58$~MPa from the force and $\mu = 0.61$~MPa from force and displacements, but different bulk moduli of $K = 4.35$~MPa and $K = 1.16$~MPa.
This corresponds to Poisson ratios of $0.44$ and $0.28$, i.e., $K/\mu = 7.5$ and $1.9$, whereas a nearly incompressible material would require $K/\mu$ of the order of $10^{3}$.
Hence, neither identification supports the common assumption of incompressibility, and the displacement field mainly affects the volumetric response, presumably through the observed lateral contraction.
We note, however, that the out-of-plane kinematics are only coarsely represented by one element across the thickness and the fixed out-of-plane displacement on the clamped edges.

\begin{figure}[tbp]
    \centering
    \includegraphics[width=0.8\textwidth]{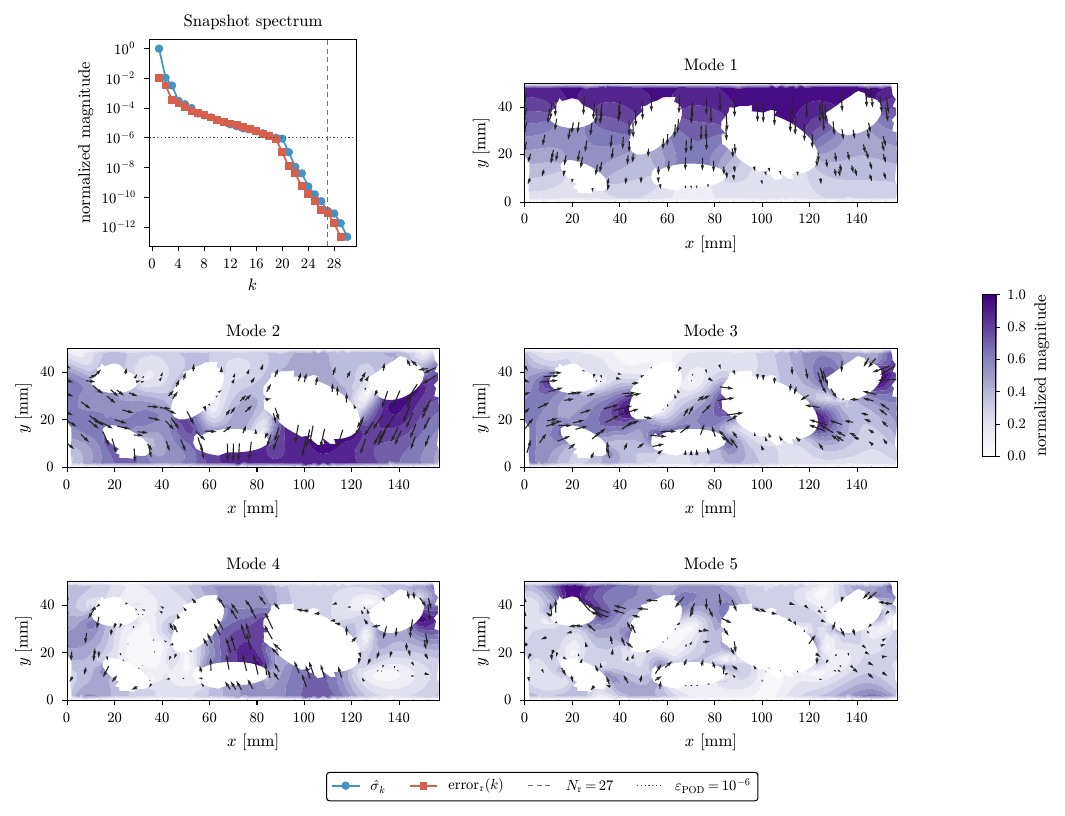}
    \caption{\textbf{Case~2, trial space at the end of the force-based identification.} Top left: normalized singular values $\hat{\sigma}_k$ and projection error $\mathrm{error}_{\mathrm{r}}(k)$; the spectrum decays slowly over almost twenty modes, and the dimension $N_{\mathrm{r}}=27$, set by $\varepsilon_{\mathrm{POD}}=10^{-6}$ earlier in the training, is never decreased. Remaining panels: magnitude (color) and direction (arrows) of the first five modes; mode~1 describes the stretching in loading direction, while the higher modes resolve the motion of individual ligaments between the holes.}
    \label{fig:abbasi_modes}
\end{figure}

\begin{figure}[tbp]
    \centering
    \begin{tikzpicture}
        \node[anchor=west, inner sep=0] (ecswimg) at (0,0) {\includegraphics[height=6cm]{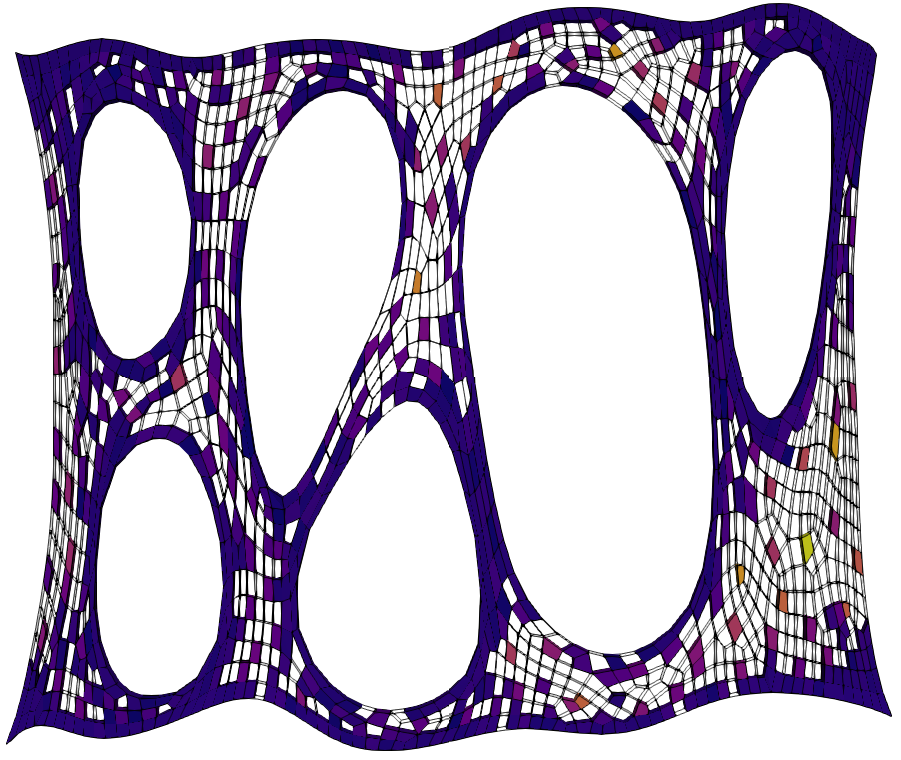}};
        \begin{axis}[
            at={(ecswimg.east)}, anchor=west, xshift=0.8cm,
            width=1.8cm, height=6cm, hide axis,
            colormap name=plasmaC, colorbar,
            point meta min=0, point meta max=17,
            colorbar style={
                width=0.35cm,
                ylabel={ECSW weight $w^e$},
                ytick={0,5,10,17},
                yticklabels={0,5,10,17},
                tick label style={font=\footnotesize},
                ylabel style={font=\small}
            }
        ]
            \addplot [draw=none, forget plot] coordinates {(0,0)};
        \end{axis}
    \end{tikzpicture}
    \caption{\textbf{Case~2, final ECSW rule of the force-based identification on the deformed mesh of TT1.} The $801$ active elements ($53\,\%$ of the mesh) are colored by their weight, the remaining mesh is shown as wireframe. They line both clamped edges and the boundaries of all holes, where the deformation localizes. With weights of at most $17$, the hyperreduction is considerably less sparse than for the plate of Case~1.}
    \label{fig:abbasi_ecsw}
\end{figure}

\paragraph{Trial space and hyperreduction.}
The spectrum of the final snapshot matrix in \cref{fig:abbasi_modes} drops by two orders of magnitude within the first two modes and then decays slowly over a range of almost twenty modes, before it falls sharply to $10^{-12}$.
As in Case~1, the tolerance intersects this slowly decaying part, which explains why the strict choice $\varepsilon_{\mathrm{POD}} = 10^{-6}$ results in a large trial space.
The first mode represents the stretching in loading direction, and the higher modes resolve the motion of individual ligaments between the holes, i.e., the regions in which the deformation localizes.
The final ECSW rule in \cref{fig:abbasi_ecsw} reflects the same localization.
Of the $1\,024$ entries of the padded set, $801$ refer to active elements, and the remaining $223$ entries carry zero weight and add no element to the figure.
The sampled elements line both clamped edges and the boundaries of all holes, and only the interior of the wider ligaments is represented by few elements.
Compared with Case~1, the weights remain small, with a maximum of about $17$, since hardly any region deforms uniformly enough to be represented by a single heavily weighted element.
In summary, the reduction remains accurate for a perforated specimen and both kinds of loss, yet its efficiency is limited by the localization of the deformation. In the last case, we turn to a problem in which the full-order solve dominates the cost.

\subsection{Case 3: Heterogeneous material system with spherical inclusions}
\label{sec:case3}

Our final example is a three-dimensional block with two stiff spherical inclusions, for which we generate the data synthetically.
This serves two purposes.
First, the reference material is known, and the reference data are free of noise and of uncertainties in geometry and boundary conditions, such that any deviation stems from the identification and the reduction alone.
Second, the deformation is genuinely three-dimensional, and the problem is about three times larger than the plate of Case~1, such that the full-order solve becomes the dominant cost of the training.

\begin{figure}[tbp]
\centering
\begin{tikzpicture}[>={Stealth[length=2.2mm]}]
    \node (geo3d) at (0,0) {\includegraphics[width=0.3\textwidth]{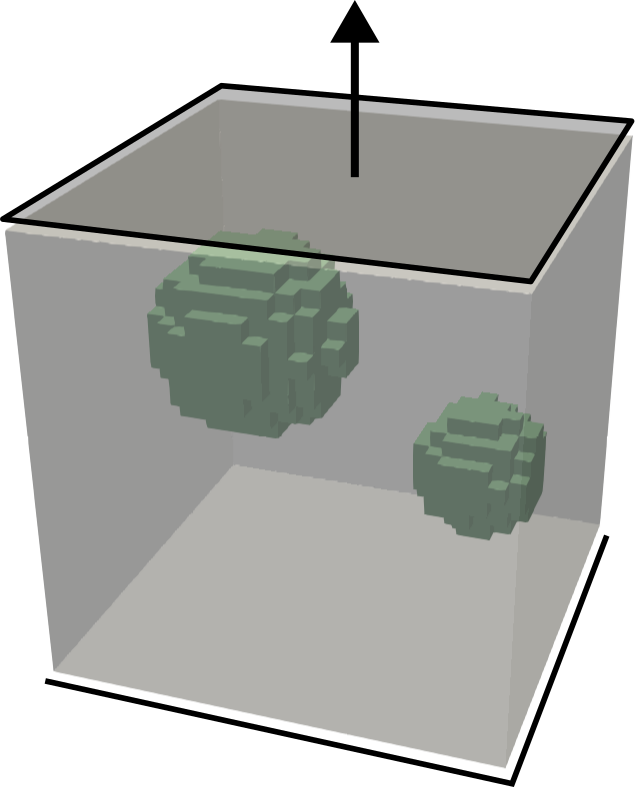}};
    \node[above] at (geo3d.north) {$\bm{u}=(0,\,\bar{u}_y(t),\,0)$};
    \node[below] at (geo3d.south) {$\bm{u}=(0,\,0,\,0)$};
    \draw[->, thick] ([xshift=0.4cm,yshift=-0.2cm]geo3d.south east) -- ++(1,0) node[right] {$x$};
    \draw[->, thick] ([xshift=0.4cm,yshift=-0.2cm]geo3d.south east) -- ++(0,1) node[above] {$y$};
    \draw[->, thick] ([xshift=0.4cm,yshift=-0.2cm]geo3d.south east) -- ++(-0.6,-0.45) node[below left] {$z$};
\end{tikzpicture}\\[6pt]
\begin{tikzpicture}[scale=0.33, line cap=round]

\begin{scope}[shift={(4,0)}]
  \fill[matrixfill] (0,0) rectangle (10,10);
  \fill[phasefill] (2.5,2.5) circle (1.5);
  \fill[phasefill] (6.5,6.5) circle (2.0);
  \draw[very thick,phaseline] (2.5,2.5) circle (1.5);
  \draw[very thick,phaseline] (6.5,6.5) circle (2.0);
  \draw[very thick] (0,0) rectangle (10,10);
  \cmark{2.5}{2.5}
  \cmark{6.5}{6.5}
  \draw[thin] (2.5,2.5) -- (4.0,2.5);
  \node[above,font=\small] at (3.25,2.5) {$R_1$};
  \draw[thin] (6.5,6.5) -- (8.5,6.5);
  \node[above,font=\small] at (7.5,6.5) {$R_2$};
  \hdim{0}{2.5}{-1.5}{2{,}5}
  \hdim{2.5}{6.5}{-1.5}{4{,}0}
  \hdim{6.5}{10}{-1.5}{3{,}5}
  \hdim{0}{10}{-3.2}{10}
  \vdim{0}{2.5}{-1.5}{2{,}5}
  \vdim{2.5}{6.5}{-1.5}{4{,}0}
  \vdim{6.5}{10}{-1.5}{3{,}5}
  \vdim{0}{10}{-3.2}{10}
  \node[font=\bfseries] at (5,12.2) {plane (x--z)};
\end{scope}

\begin{scope}[shift={(20,0)}]
  \fill[matrixfill] (0,0) rectangle (10,10);
  \fill[phasefill] (2.5,2.5) circle (1.5);
  \fill[phasefill] (6.5,7.0) circle (2.0);
  \draw[very thick,phaseline] (2.5,2.5) circle (1.5);
  \draw[very thick,phaseline] (6.5,7.0) circle (2.0);
  \draw[very thick] (0,0) rectangle (10,10);
  \cmark{2.5}{2.5}
  \cmark{6.5}{7.0}
  \draw[thin] (2.5,2.5) -- (4.0,2.5);
  \node[above,font=\small] at (3.25,2.5) {$R_1$};
  \draw[thin] (6.5,7.0) -- (8.5,7.0);
  \node[above,font=\small] at (7.5,7.0) {$R_2$};
  \hdim{0}{2.5}{-1.5}{2{,}5}
  \hdim{2.5}{6.5}{-1.5}{4{,}0}
  \hdim{6.5}{10}{-1.5}{3{,}5}
  \hdim{0}{10}{-3.2}{10}
  \vdim{0}{2.5}{-1.5}{2{,}5}
  \vdim{2.5}{7.0}{-1.5}{4{,}5}
  \vdim{7.0}{10}{-1.5}{3{,}0}
  \vdim{0}{10}{-3.2}{10}
  \node[font=\bfseries] at (5,12.2) {plane (x--y)};
\end{scope}

\begin{scope}[shift={(36,0)}]
  \fill[matrixfill] (0,0) rectangle (10,10);
  \fill[phasefill] (2.5,2.5) circle (1.5);
  \fill[phasefill] (7.0,6.5) circle (2.0);
  \draw[very thick,phaseline] (2.5,2.5) circle (1.5);
  \draw[very thick,phaseline] (7.0,6.5) circle (2.0);
  \draw[very thick] (0,0) rectangle (10,10);
  \cmark{2.5}{2.5}
  \cmark{7.0}{6.5}
  \draw[thin] (2.5,2.5) -- (4.0,2.5);
  \node[above,font=\small] at (3.25,2.5) {$R_1$};
  \draw[thin] (7.0,6.5) -- (9.0,6.5);
  \node[above,font=\small] at (8.0,6.5) {$R_2$};
  \hdim{0}{2.5}{-1.5}{2{,}5}
  \hdim{2.5}{7.0}{-1.5}{4{,}5}
  \hdim{7.0}{10}{-1.5}{3{,}0}
  \hdim{0}{10}{-3.2}{10}
  \vdim{0}{2.5}{-1.5}{2{,}5}
  \vdim{2.5}{6.5}{-1.5}{4{,}0}
  \vdim{6.5}{10}{-1.5}{3{,}5}
  \vdim{0}{10}{-3.2}{10}
  \node[font=\bfseries] at (5,12.2) {plane (y--z)};
\end{scope}

\end{tikzpicture}
\caption{\textbf{Case~3, cube with two spherical inclusions in the reference configuration.} The cube of edge length $10$~mm contains two spheres of radii $R_1=1.5$~mm and $R_2=2.0$~mm (green) and is discretized by $25^3$ hexahedral elements. Top: three-dimensional view; the bottom face is clamped, and the top face is displaced by $\bar{u}_y(t)$ up to $3$~mm in loading direction $y$ in $N_{\mathrm{t}}=10$ load steps, with the transverse displacements fixed. Bottom: front view (x--z), top view (x--y) and side view (y--z) with dimensions in mm. The inclusions are ten times stiffer than the matrix.}
\label{fig:case3_geometry}
\end{figure}

\cref{fig:case3_geometry} shows the geometry.
A cube with an edge length of $10$~mm contains two spheres with radii $R_1 = 1.5$~mm and $R_2 = 2.0$~mm.
We discretize the cube by $25\times 25\times 25$ hexahedral elements, i.e., $n_{\mathrm{el}} = 15\,625$ elements with $52\,728$ degrees of freedom, of which $N_{\mathrm{f}} = 48\,672$ are free, and assign every element whose barycenter lies inside a sphere to the inclusion phase, which yields $734$ inclusion elements.
The bottom face is clamped, while the top face is clamped in the transverse directions and displaced by $3$~mm in loading direction in $N_{\mathrm{t}} = 10$ load steps, which corresponds to a nominal stretch of $1.3$.
The reference data stem from a full-order solve with the compressible Neo-Hookean energy
\begin{equation}
    \psi_{\mathrm{NH}}(\bm{F}) = \frac{\mu}{2}\big[\operatorname{tr}\bm{C} - 3 - 2\ln J\big] + \frac{\Lambda}{4}\big[J^2 - 1 - 2\ln J\big],
    \label{eq:neo_hooke}
\end{equation}
where $\mu$ denotes the shear modulus and $\Lambda$ the first Lam\'e parameter.
We choose $\mu = 1$~MPa and $\Lambda = 4$~MPa for the matrix and $\mu = 10$~MPa and $\Lambda = 40$~MPa for the inclusions, i.e., a stiffness ratio of ten and a Poisson ratio of $0.4$ in both phases.
As in Case~1, each phase is represented by its own network, and only the total reaction force on the top face enters the loss, whose target we scale by $\bar{f} = \max_n|f^{\mathrm{exp}}_n| = 85.61$~N,
\begin{equation}
    \mathcal{L} = \frac{1}{N_{\mathrm{t}}} \sum_{n=1}^{N_{\mathrm{t}}} \Big[ \mathcal{H}(\mathbf{d}_n,\bm{\theta}) - f^{\mathrm{exp}}_n/\bar{f} \Big]^2 ,
    \label{eq:threeD_loss}
\end{equation}
where $f^{\mathrm{exp}}_n$ now denotes the reaction force of the reference solve at load step $t_n$.
As in Case~1, the networks thus learn the scaled energy $\psi/\bar{f}$.
The complete displacement field of the reference solve serves exclusively for validation.
Note that the reference data are computed on the same mesh that is used for the identification.
All further settings follow Case~1, see \cref{tab:hyperparameters}, except that we train for $400$ epochs.

\begin{figure}[tbp]
    \centering
    \includegraphics[width=0.8\textwidth]{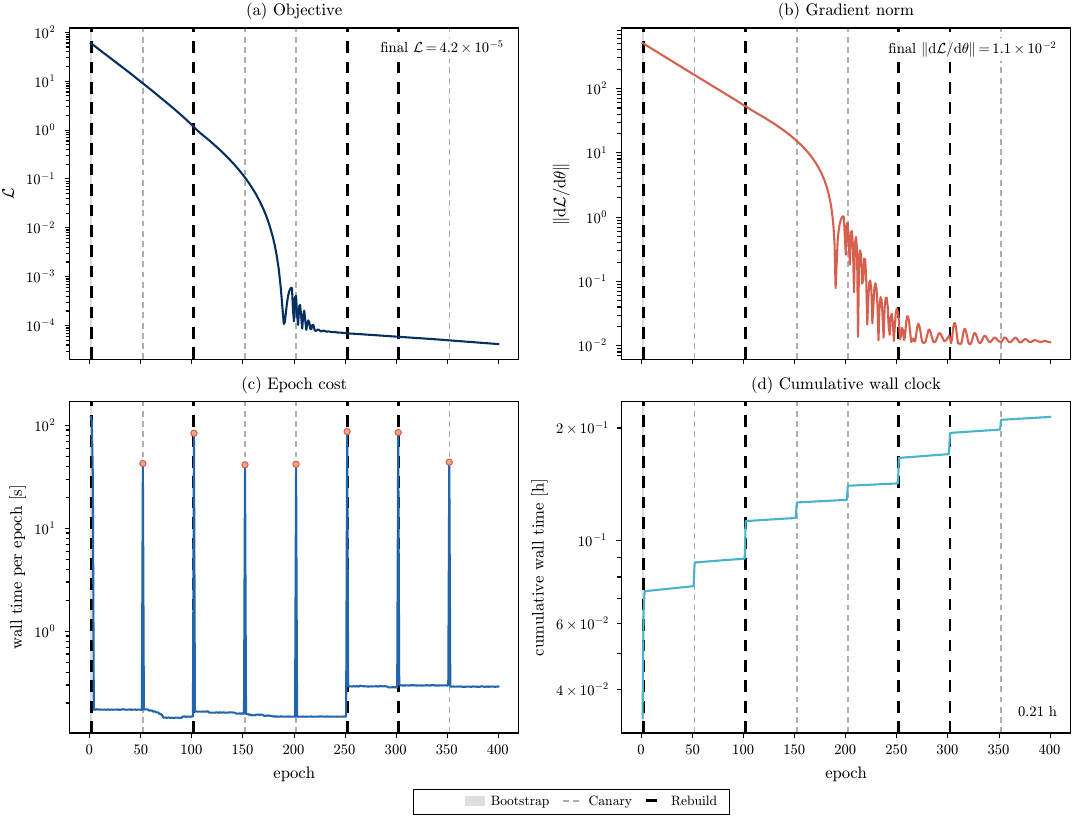}
    \caption{\textbf{Case~3, training history of the hyperreduced identification.} (a) The loss decreases from about $60$ to $\mathcal{L}=4.2\cdot10^{-5}$, with a steep drop around epoch~$190$. (b) The gradient norm decreases from about $5\cdot10^{2}$ to $1.1\cdot10^{-2}$. (c) Reduced epochs take $0.17$ to $0.3$~s, canary epochs about $40$~s and canary epochs with a rebuild about $85$~s, i.e., a reduced epoch is more than two orders of magnitude cheaper than a canary epoch. (d) The training requires $0.21$~h in total, most of which is spent in the full-order epochs.}
    \label{fig:threeD_training}
\end{figure}

\begin{figure}[tbp]
    \centering
    \includegraphics[width=0.8\textwidth]{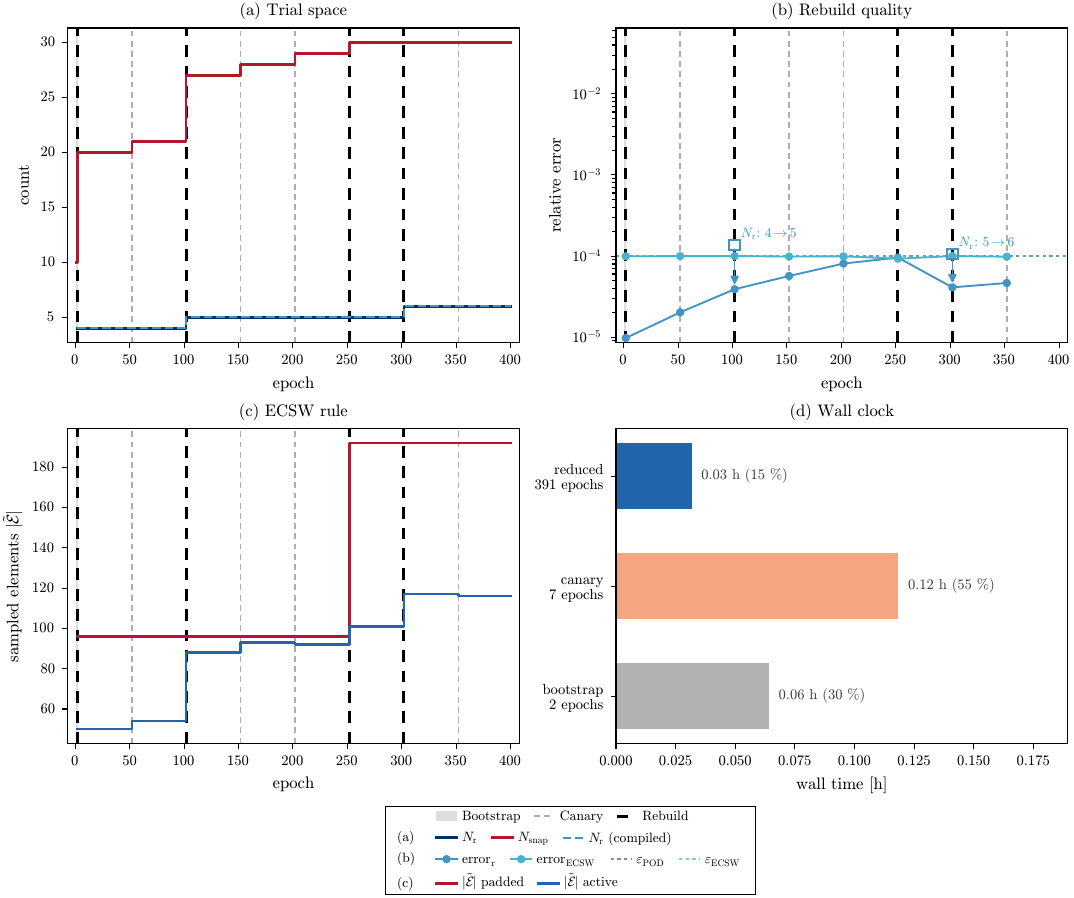}
    \caption{\textbf{Case~3, adaptivity of the reduction.} (a) With ten load steps per epoch, the snapshot matrix reaches its capacity of $30$ only at epoch~$252$; the trial space grows from $N_{\mathrm{r}}=4$ to $6$. (b) At epochs $102$ and $302$, the POD error with the previous dimension exceeds $\varepsilon_{\mathrm{POD}}=10^{-4}$ (squares). (c) At epoch~$252$, the active ECSW set exceeds the padded capacity of $96$ elements, which is then raised to $192$; this rebuild changes no dimension of the trial space. (d) Bootstrap and canary epochs account for $85\,\%$ of the wall-clock time, the $391$ reduced epochs for $15\,\%$.}
    \label{fig:threeD_reduction}
\end{figure}

\paragraph{Training history and adaptivity of the reduction.}
\cref{fig:threeD_training}(a) shows a nearly exponential decrease of the loss from about $60$ to $10^{-1}$ within the first $150$ epochs, followed by a steep drop and decaying oscillations around epoch~$190$, and a slow decrease to $\mathcal{L} = 4.2\cdot 10^{-5}$ at epoch~$400$.
In contrast to the experimental cases, a reduced epoch is now more than two orders of magnitude cheaper than a canary epoch, see \cref{fig:threeD_training}(c).
A reduced epoch takes between $0.17$ and $0.3$~s, a canary epoch about $40$~s, and an epoch with a rebuild about $85$~s.
Consequently, the $391$ reduced epochs amount to only $15\,\%$ of the total wall-clock time of $0.21$~h, whereas the two bootstrap epochs account for $30\,\%$ and the seven canary epochs for $55\,\%$, see \cref{fig:threeD_reduction}(d).
Since a canary epoch comprises a full-order solve and the update of the reduction, a full-order epoch costs at most about $40$~s, and a full-order training over $400$ epochs would take up to about four and a half hours, i.e., up to twenty times longer.
Here, the speed-up is limited almost entirely by the full-order epochs that the adaptive strategy requires.

The adaptivity in \cref{fig:threeD_reduction} shows a second reason for a rebuild.
Since each epoch comprises only ten load steps, the snapshot matrix fills slowly and reaches its capacity of $30$ only at epoch~$252$.
The trial space grows from four to five modes at epoch~$102$ and to six modes at epoch~$302$, in both cases because the projection error of the updated snapshots exceeds $\varepsilon_{\mathrm{POD}}$ with the previous dimension, as in Case~1.
At epoch~$252$, however, the dimension remains unchanged, and the rebuild is caused by the ECSW rule: the number of active elements grows from $92$ to $101$ and thus exceeds the padded capacity of $96$ elements.
The capacity is raised to the next size that provides the headroom introduced in Case~1, i.e., to $192$ elements, which requires a recompilation.
This rebuild also shows the price of the padding, since the cost of a reduced epoch increases from about $0.17$~s to $0.3$~s, although the number of active elements grows by only about ten percent.
At the end of the training, $116$ of the $15\,625$ elements are active, i.e., less than $1\,\%$ of the mesh.

\begin{figure}[tbp]
    \centering
    \includegraphics[width=\textwidth]{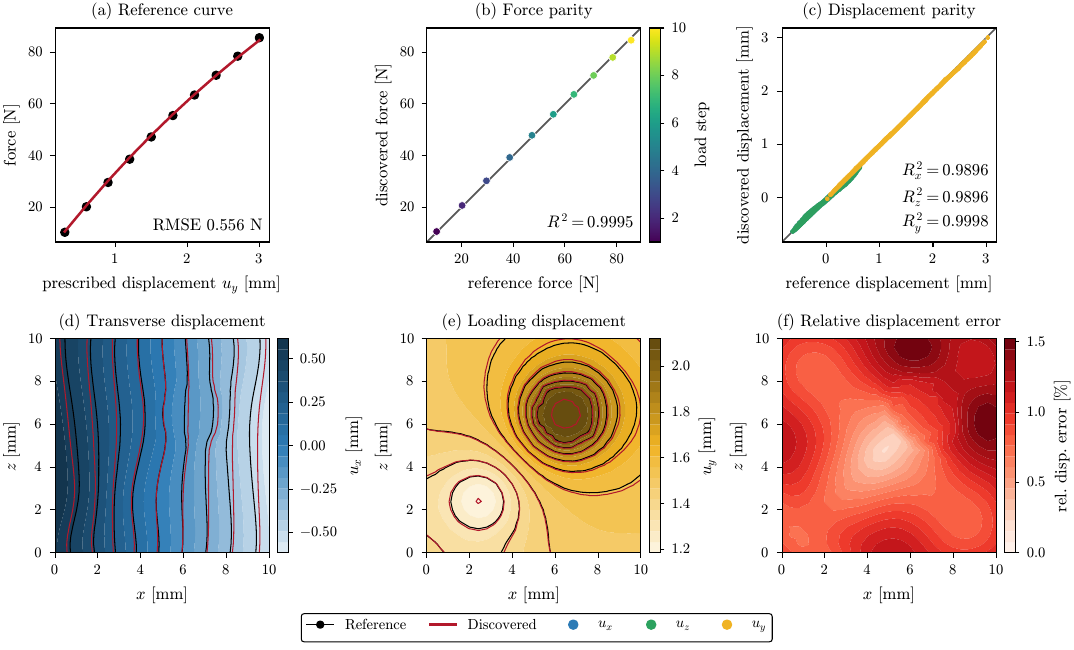}\\[8pt]
    \begin{minipage}[b]{0.2\textwidth}
        \centering
        \includegraphics[width=\linewidth]{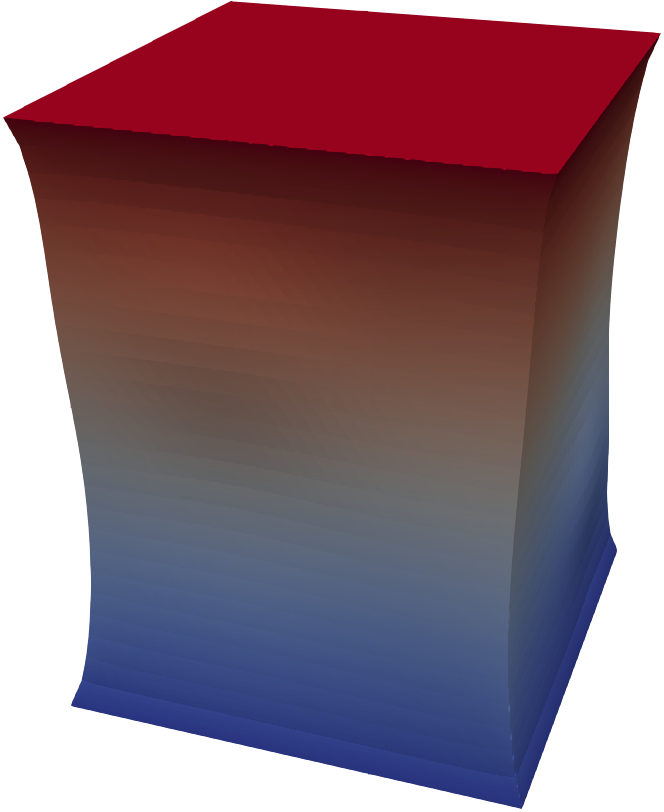}\\[2pt]
        \begin{tikzpicture}
            \begin{axis}[
                width=\linewidth, height=1.6cm, hide axis,
                colormap name=coolwarmC, colorbar horizontal,
                point meta min=0, point meta max=3,
                colorbar style={
                    xlabel={$u_y$ [mm]},
                    xtick={0,1,2,3},
                    tick label style={font=\scriptsize},
                    label style={font=\footnotesize}
                }
            ]
                \addplot [draw=none, forget plot] coordinates {(0,0)};
            \end{axis}
        \end{tikzpicture}\par\smallskip
        \makebox[\linewidth][c]{\small (g) Displacement $u_y$}
    \end{minipage}
    \hspace*{2cm}
    \begin{minipage}[b]{0.2\textwidth}
        \centering
        \includegraphics[width=\linewidth]{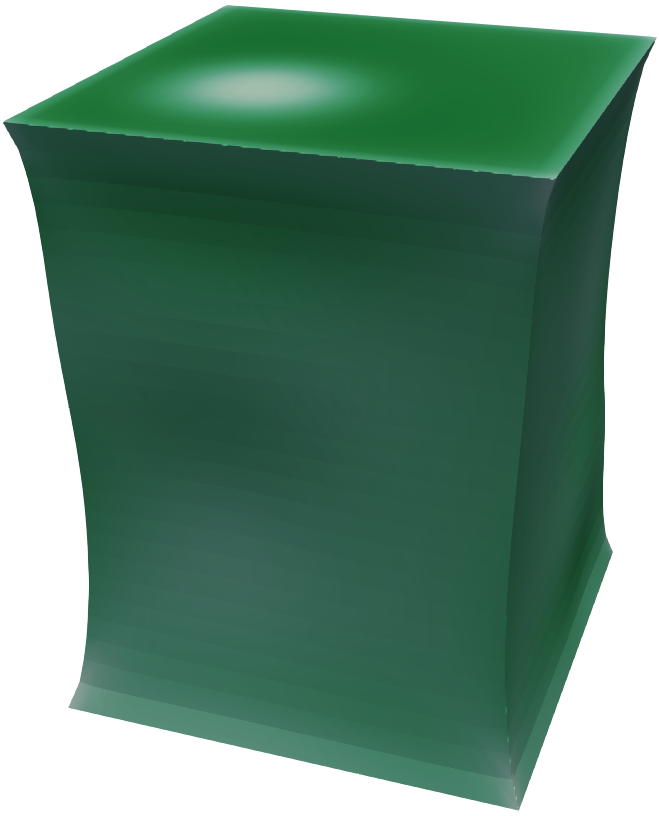}\\[2pt]
        \begin{tikzpicture}
            \begin{axis}[
                width=\linewidth, height=1.6cm, hide axis,
                colormap name=greensRevC, colorbar horizontal,
                point meta min=0.044, point meta max=0.610,
                colorbar style={
                    xlabel={$F_{yy}$ [-]},
                    xtick={0.044,0.2,0.4,0.610},
                    xticklabels={$1.044$,$1.2$,$1.4$,$1.61$},
                    tick label style={font=\scriptsize},
                    label style={font=\footnotesize}
                }
            ]
                \addplot [draw=none, forget plot] coordinates {(0,0)};
            \end{axis}
        \end{tikzpicture}\par\smallskip
        \makebox[\linewidth][c]{\small (h) Deformation gradient $F_{yy}$}
    \end{minipage}

    \caption{\textbf{Case~3, identified material compared with the synthetic reference.} (a)--(b) The force is reproduced with a root-mean-square error of $0.556$~N and $R^2=0.9995$. (c) All three displacement components, which are not part of the loss, agree with $R^2\geq 0.990$. (d)--(e) Reference (black) and discovered (red) contour lines of the transverse and loading displacement in the mid-plane $y=5$~mm nearly coincide, also around both inclusions. (f) The relative displacement error remains below $1.5\,\%$. (g)--(h) Deformed configuration at the last load step: the displacement in loading direction increases from the clamped bottom face to the top face, which is displaced by $3$~mm, and $F_{yy}$ varies between $1.044$ and $1.61$ on the surface, with the largest stretches in the vicinity of the clamped faces.}
    \label{fig:threeD_evaluation}
\end{figure}

\paragraph{Identified material.}
\cref{fig:threeD_evaluation} compares the identified material with the reference.
The force--displacement curve is reproduced with a root-mean-square error of $0.556$~N at a maximum force of about $86$~N, and the force parity yields $R^2 = 0.9995$.
The relative force error amounts to $1.0\,\%$.
The displacement field, which did not enter the loss, agrees with the reference with a relative error of $3.2\,\%$ over all nodes and load steps and with $R^2 = 0.990$ in both transverse directions and $R^2 = 0.9998$ in loading direction.
In the mid-plane $y = 5$~mm shown in \cref{fig:threeD_evaluation}(d)--(f), the pointwise error remains below $1.5\,\%$.
In particular, the contour lines of the loading displacement in \cref{fig:threeD_evaluation}(e) follow the reference closely around both inclusions, which move almost rigidly within the softer matrix.
\cref{fig:threeD_evaluation}(g)--(h) shows the displacement in loading direction and the component $F_{yy}$ on the deformed configuration.
The stretch varies between $F_{yy} = 1.044$ and $1.61$, with the largest values in the vicinity of the clamped faces.
Unlike that of the plate in Case~1, the force--displacement curve of the block is almost linear and samples only a moderate range of stretches.
Nevertheless, the identified material reproduces the interior field of both phases, which indicates that the force suffices to separate both phases also in three dimensions.

\begin{figure}[tbp]
    \centering
    \includegraphics[width=0.8\textwidth]{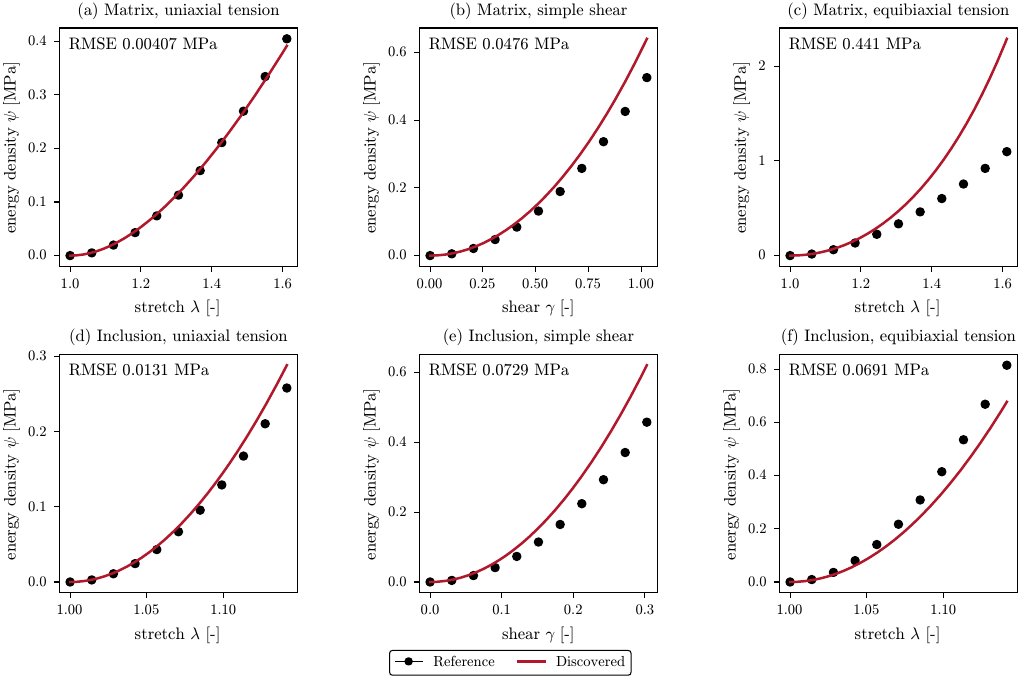}
    \caption{\textbf{Case~3, energy density of the identified (red) and the reference Neo-Hookean material \cref{eq:neo_hooke} (black) at a single material point.} Top: matrix; bottom: inclusion; columns: uniaxial tension, simple shear and equibiaxial tension, each over the range the phase experiences in the reference solve, with the deformation gradients \cref{eq:materialpoint_F}; in tension, the unprescribed diagonal components follow from vanishing normal stresses. The matrix law is reproduced in uniaxial tension (root-mean-square error of $0.004$~MPa), which dominates the response of the block, whereas it is too stiff in simple shear ($0.048$~MPa) and in equibiaxial tension ($0.44$~MPa), which the experiment hardly activates. The stiffer and less deformed inclusion is identified less accurately in all three load cases ($0.013$, $0.073$ and $0.069$~MPa).}
    \label{fig:threeD_materialpoint}
\end{figure}

\paragraph{Comparison with the reference material.}
Since the reference energies are known, we finally compare the identified and the reference energy of both phases at a single material point, see \cref{fig:threeD_materialpoint}.
We consider uniaxial tension, simple shear, and equibiaxial tension, given by
\begin{equation}
    \bm{F}_{\mathrm{ut}} =
    \begin{bmatrix}
        \xi & 0 & 0 \\
        0 & \lambda & 0 \\
        0 & 0 & \xi
    \end{bmatrix},
    \qquad
    \bm{F}_{\mathrm{ss}} =
    \begin{bmatrix}
        1 & 0 & 0 \\
        \gamma & 1 & 0 \\
        0 & 0 & 1
    \end{bmatrix},
    \qquad
    \bm{F}_{\mathrm{eb}} =
    \begin{bmatrix}
        \lambda & 0 & 0 \\
        0 & \xi & 0 \\
        0 & 0 & \lambda
    \end{bmatrix},
    \label{eq:materialpoint_F}
\end{equation}
where only $\lambda$ or $\gamma$ is prescribed and $\xi$ follows from vanishing normal stresses.
Each load case covers the range that the respective phase experiences in the reference solve.
For the matrix, the identified energy coincides with the reference in uniaxial tension, whereas it is too stiff in simple shear and, by a factor of about two, in equibiaxial tension.
The inclusion deviates in all three load cases.
This reflects what the experiment probes: the block is stretched essentially uniaxially, and the force constrains the energy only along this path, but not its extrapolation to other states, such as simple shear.
Moreover, the force measures only the axial response and not the lateral contraction.
Indeed, at the end of the uniaxial range, the identified matrix attains the reference energy with $J = 1.03$ instead of $1.08$, i.e., it is less compressible than the reference.
The volumetric response thus remains largely undetermined, which becomes apparent in equibiaxial tension, where the identified matrix expands to $J = 1.66$ instead of $1.18$.
The stiff inclusions, in turn, deform little and contribute only weakly to the force, such that their energy is less well determined, in line with the observation that heterogeneous experiments do not automatically improve identifiability \cite{knipper2026}.
Note that this limitation concerns the information content of the data and not the reduction.

\begin{figure}[tbp]
    \centering
    \includegraphics[width=0.8\textwidth]{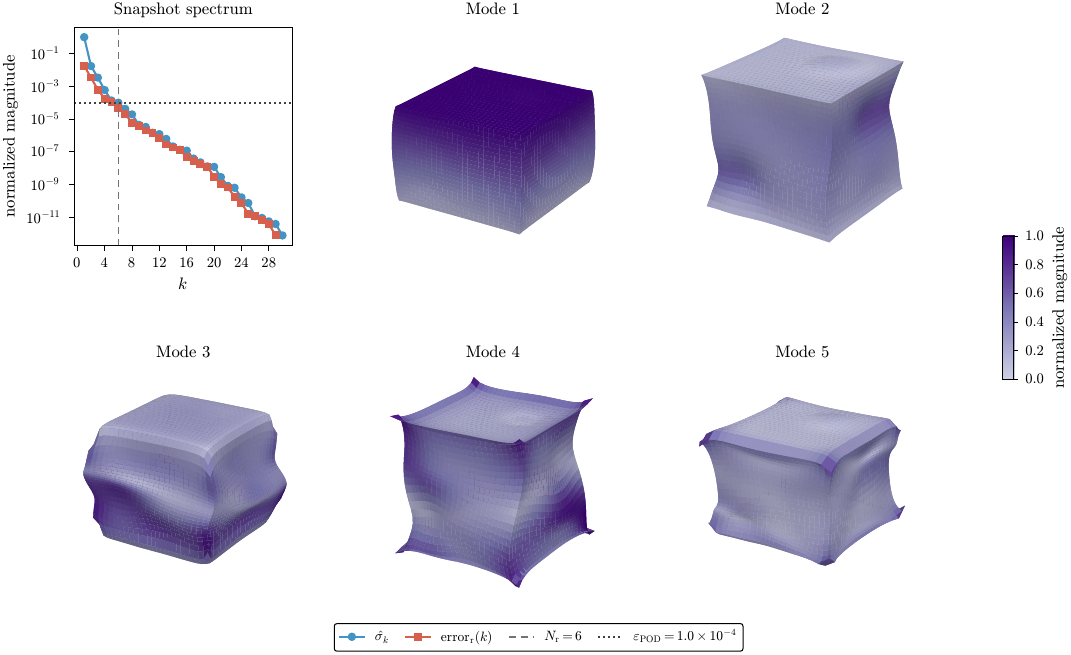}
    \caption{\textbf{Case~3, trial space at the end of the training.} Top left: normalized singular values $\hat{\sigma}_k$ and projection error $\mathrm{error}_{\mathrm{r}}(k)$; in contrast to the experimental cases, the spectrum decays steadily, and $\varepsilon_{\mathrm{POD}}=10^{-4}$ is reached with $N_{\mathrm{r}}=6$ modes. Remaining panels: magnitude of the first five modes on the deformed block; mode~1 describes the uniaxial stretching, the higher modes the bulging of the lateral faces and the deformation at the edges of the clamped faces.}
    \label{fig:threeD_modes}
\end{figure}

\begin{figure}[tbp]
    \centering
    \begin{tikzpicture}
        \node[anchor=west, inner sep=0] (ecswimg) at (0,0) {\includegraphics[height=6cm]{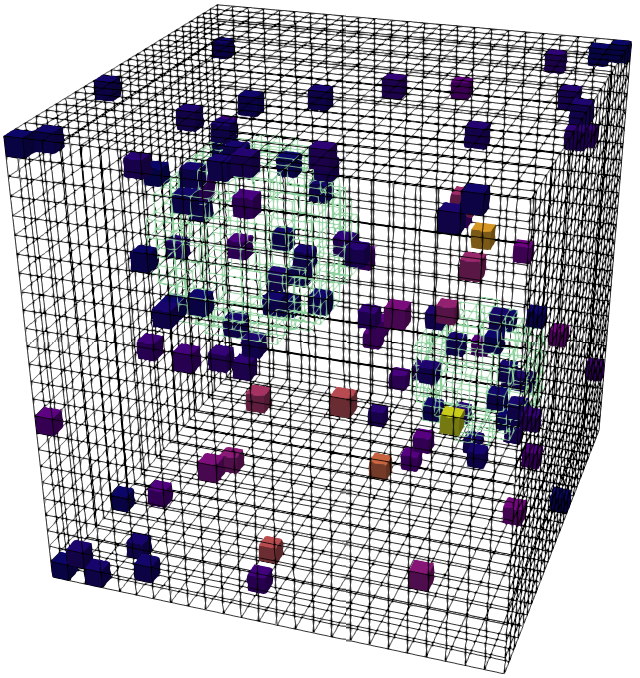}};
        \begin{axis}[
            at={(ecswimg.east)}, anchor=west, xshift=0.8cm,
            width=1.8cm, height=6cm, hide axis,
            colormap name=plasmaC, colorbar,
            point meta min=0, point meta max=1200,
            colorbar style={
                width=0.35cm,
                ylabel={ECSW weight $w^e$},
                ytick={0,500,1000,1200},
                yticklabels={0,500,1000,1200},
                tick label style={font=\footnotesize},
                ylabel style={font=\small}
            }
        ]
            \addplot [draw=none, forget plot] coordinates {(0,0)};
        \end{axis}
    \end{tikzpicture}
    \caption{\textbf{Case~3, final ECSW rule.} The mesh is shown as wireframe and the inclusions in green. Only $116$ of the $15\,625$ elements carry a positive weight. They cluster inside and around both inclusions and at the corners of the clamped faces, whereas the matrix is represented by a few elements with weights of up to about $1\,200$. The $76$ padded entries carry zero weight and add no element to the figure.}
    \label{fig:threeD_ecsw}
\end{figure}

\paragraph{Trial space and hyperreduction.}
In contrast to the experimental cases, the singular values in \cref{fig:threeD_modes} decay steadily over twelve orders of magnitude without a plateau, and the tolerance $\varepsilon_{\mathrm{POD}} = 10^{-4}$ is reached with only six modes.
We presume that the plateaus of Cases~1 and~2 stem from the measured boundary data, whose small irregularities enter every snapshot, whereas the boundary conditions of the synthetic block are smooth.
The first mode represents the uniaxial stretching of the block with lateral contraction, and the higher modes describe the bulging of the lateral faces, the deformation at the edges of the clamped faces and the perturbation by the inclusions.
The final ECSW rule in \cref{fig:threeD_ecsw} selects $116$ elements, of which $21$ belong to the inclusions, stored in a padded set of $192$ entries.
The active elements cluster inside and around both inclusions and at the corners of the clamped faces.
The matrix away from the inclusions, in contrast, is represented by a few elements with weights of up to about $1\,200$, each of which stands for a large region of nearly uniform deformation.
With less than $1\,\%$ of the elements, this is the sparsest rule of all three cases. Together with the small trial space and the ten load steps, it explains why the reduced epochs of the largest problem are also the cheapest.
In summary, the reduction pays off most where the full-order model is most expensive, while the accuracy of the identified material is limited by the information content of the data rather than by the reduction.

\section{Discussion}
\label{sec:discussion}

\paragraph{Hyperreduction accelerates the discovery of neural materials without degrading the identified material.}
The central result of this work is that the adaptive combination of POD and ECSW lowers the cost of the adjoint-based discovery by an order of magnitude, while the identified material, evaluated by a full-order solve, reproduces the structural response as accurately as a material identified with the full-order model.
For the plate of Case~1, the reduction accelerates the median epoch by a factor of $27$ and the entire training by a factor of $13$ (\cref{fig:mnist_training}(c)--(d)), while the force error of $2.5\,\%$ remains below that of the full-order reference (\cref{fig:mnist_evaluation}).
Across all variations of the reduction parameters, the speed-up per epoch ranges from $8.6$ to $60$ without a systematic loss of accuracy (\cref{fig:mnist_study}(b)).
For the three-dimensional block of Case~3, a reduced epoch is more than two orders of magnitude cheaper than a canary epoch (\cref{fig:threeD_training}(c)), from which we estimate that a full-order training would take up to twenty times longer.
Since six modes and less than $1\,\%$ of the elements suffice (\cref{fig:threeD_modes,fig:threeD_ecsw}), the largest problem has the cheapest reduced epochs, i.e., the gain grows with the cost of the full-order model.
This answers the question raised in the introduction: the reduced model remains accurate while the weights evolve, provided that it is adapted along the optimization path.
In Case~1, it was accepted at every canary epoch, and the reduced and the full-order training follow nearly the same path and end at losses of comparable magnitude (\cref{fig:mnist_training}(a)).
The progressive construction of reduced models along the optimization path is established in PDE-constrained design optimization with few parameters \citep{zahr2015,amsallem2015}; our results show that it carries over to neural materials, whose many weights and biases change at every update.
Moreover, the reduction is complementary to differentiable and GPU-accelerated full-order solvers \citep{xue2023jaxfem,knipper2026}, which lower the cost of a single solve, but not the number of degrees of freedom that enter it.
How much is gained, however, depends on two factors discussed next: the reducibility of the deformation and the cost of the full-order checks.

\paragraph{Reducibility is governed by the localization of the deformation rather than by the size of the problem.}
The perforated membrane of Case~2 is the smallest of the three meshes, yet it requires the largest reduced model, with $27$ modes and up to $53\,\%$ of the elements (\cref{fig:abbasi_reduction}), whereas the block of Case~3, with three times as many degrees of freedom as the plate of Case~1, requires the smallest.
The decisive factor is the localization of the deformation.
In Case~2, the stretch concentrates in the ligaments between the holes, where $F_{yy}$ reaches $3.94$, the higher modes resolve individual ligaments, and the singular values decay slowly (\cref{fig:abbasi_modes}).
Since ECSW reproduces $N_{\mathrm{r}} + N_{\mathrm{r}}^2$ projected quantities per snapshot, see \cref{eq:ecsw_system}, a large trial space directly enlarges the sampled set, and hardly any region deforms uniformly enough to be represented by a single heavily weighted element (\cref{fig:abbasi_ecsw}).
In addition, the measured boundary data presumably cause the plateaus in the spectra of Cases~1 and~2, which are absent for the synthetic block (\cref{fig:mnist_modes,fig:threeD_modes}), and Case~2 employs the stricter tolerance $\varepsilon_{\mathrm{POD}} = 10^{-6}$, which affects the size of the reduced model more than any other parameter in the study of Case~1 (\cref{fig:mnist_sweeps}(e)).
Consequently, the update of the reduction dominates the canary epochs of Case~2, and without a full-order reference, the speed-up for this specimen remains open.
Case~2 also shows that a loss which observes the displacement field reacts more sensitively to an outdated trial space than a loss which observes only the reaction force (\cref{fig:abbasi_training_fd}).
These observations reflect the known limitation of linear subspaces for localized deformation \citep{benner2015}.
Nonlinear manifolds \citep{lee2020,scheunemann2025,faust2025manifold}, hyperreduction in strain space \citep{wulfinghoff2025e3c,faust2026strain}, and component-wise reduced models \citep{ritzert2026} offer possible routes, whose combination with the adjoint-based discovery remains to be investigated.

\paragraph{Up-to-date snapshots govern the accuracy, full-order checks govern the cost.}
The parameter study of Case~1 separates the parameters of the reduction into two groups: the canary interval and the snapshot tolerance, which control how up to date the snapshots are, determine the accuracy, whereas the tolerances of POD and ECSW determine the cost (\cref{fig:mnist_sweeps}).
The model selection was never triggered and acts as a safeguard.
Once the reduced solve is cheap, the full-order checks dominate the cost: the canary epochs of Case~1 make up less than $2\,\%$ of all epochs, but $41\,\%$ of the wall-clock time (\cref{fig:mnist_reduction}(d)), and the bootstrap and canary epochs of Case~3 even $85\,\%$ (\cref{fig:threeD_reduction}(d)).
Even if the reduced epochs were free, the training of Case~1 could thus be accelerated by at most a factor of about $28$ for $N_{\mathrm{canary}} = 50$.
Recompilations add to this cost whenever a dimension of the reduced model changes, which the padded ECSW set avoids at most, but not all canary epochs (\cref{fig:mnist_reduction}(c), \cref{fig:threeD_reduction}(c)).
Two directions follow.
First, the canary interval could be chosen adaptively, since the weights change less as the training converges, cf.\ \cref{sec:workflow}.
Second, the full-order check could be complemented or replaced by an error indicator of the reduced solution, in the spirit of adaptive hyperreduction \citep{ryckelynck2005} or of estimators based on the constitutive relation error \citep{ladeveze1983,bhattacharyya2026}.
Both would spend the expensive full-order solves where the reduced model is actually at risk.

\paragraph{The reaction force fixes the stiffness level, but not the material uniquely.}
Since all identified materials are evaluated by the same full-order solve, their remaining deviations reflect the information content of the data rather than the reduction.
Three observations support this view.
First, in Case~1, the reduced and the full-order training reproduce the force comparably well, but distribute the stiffness differently between the phases, which only the displacement field reveals (\cref{fig:mnist_evaluation}).
Second, in Case~2, adding the displacements to the loss trades accuracy of the force for accuracy of the field (\cref{fig:abbasi_eval_force,fig:abbasi_eval_fd}, top), improves the predicted field of the unseen specimen (\cref{fig:abbasi_eval_force,fig:abbasi_eval_fd}, bottom), and changes mainly the bulk modulus, which the force alone barely constrains.
Third, in Case~3, the identified matrix coincides with the reference along the uniaxial path that the block probes, whereas its response in simple shear and equibiaxial tension, as well as the stiff inclusions, remain less well determined (\cref{fig:threeD_materialpoint}).
All three follow from the scaling argument of \cref{sec:case1}: the force fixes the level of the response, the displacement field its spatial distribution, and both constrain the energy only along the activated deformation states.
This agrees with the observations that the coverage of deformation states governs the generalization of the identified model \citep{friedrichs2026} and that heterogeneous specimens do not automatically improve identifiability \citep{knipper2026}, see also \citep{roemer2025}.
In particular, neither identification of Case~2 supports the incompressibility commonly assumed for this material \citep{abbasi2026,knipper2026}, although the coarse out-of-plane kinematics call for caution.
A natural remedy is the combination of several specimens and load cases within a single loss \citep{friedrichs2026,tan2026}.
Since this multiplies the number of nonlinear solves per epoch, it is precisely the setting in which the reduction pays off.

\paragraph{Extension to anisotropy and inelasticity.}
In the present work, we restricted ourselves to isotropic hyperelasticity.
For anisotropic hyperelastic materials, the formulation of \cref{sec:methods} carries over directly: the energy additionally depends on structural tensors \citep{klein2022,holthusen2026complement}, the inner problem remains a minimization, ECSW still approximates a weighted sum of element energies, and the reduced tangent remains symmetric.
Recent works on the discovery of anisotropic materials \citep{flaschel2026adaptive,klein2027group,riemer2026construction} and on reduced models for anisotropic problems \citep{kehls2025ae} provide a starting point.
Two aspects require attention.
First, spatially varying preferred directions act like additional phases: as in Case~1, the sampled set must contain elements of every phase, since otherwise the gradient \cref{eq:reduced_adjoint} does not depend on the corresponding parameters (\cref{fig:mnist_ecsw}).
Second, additional invariants increase the number of parameters that the data have to determine, which aggravates the non-uniqueness discussed above.
Inelasticity poses a more fundamental challenge.
The load steps no longer decouple, see \cref{sec:inverse_discovery}, but are connected by the evolution of internal variables, such that the adjoint problem has to be solved backward along the loading path.
Moreover, ECSW requires the history only at the sampled elements; once the rule changes at a canary epoch, the history of newly sampled elements is unknown and has to be reconstructed, e.g., from the preceding full-order trajectory.
Incremental variational formulations, in which each load step minimizes an incremental potential composed of the energy and a dissipation potential \citep{holthusen2026generalized,flaschel2025convex}, preserve the minimization structure on which the present reduction relies.
Hyperreduction has been applied to forward simulations of viscoelastic, elastoplastic, and damaging solids \citep{trainotti2024,agouzal2024,kehls2026,ashkpour2026}, and neural materials are available for viscoelasticity \citep{holthusen2024,rosenkranz2024,kalina2026}, plasticity \citep{boes2026,vandervelden2026}, anisotropic inelasticity \citep{holthusen2026complement}, and thermomechanics \citep{holthusen2026thermo}.
Their discovery from structural data has so far relied on full-order solvers or on local equilibrium evaluations \citep{mahnken1996,benady2024,riemer2026,gavris2025,ferreira2026,spencer2026}.
Whether the adaptive reduction remains reliable when the trial space has to represent path-dependent states, and how auxiliary networks for the internal variables \citep{holthusen2026complement} interact with the hyperreduction, remains open for future work.

\section{Conclusion}
\label{sec:conclusion}

This study shows that hyperreduction makes finite element model updating of neural materials computationally affordable, while preserving the advantages that distinguish it from residual-based approaches: it accommodates arbitrary geometries and boundary conditions, tolerates sparse and noisy data, and does not require full-field measurements.
The approach embeds a polyconvex neural network into a finite element model, projects the equilibrium onto a POD subspace, and approximates the internal energy by a weighted sum over a small subset of elements.
Adjoint sensitivities of the reduced model provide the exact gradient with respect to the weights and biases, while bootstrap and canary epochs adapt the reduction to the evolving network along the optimization path.

The numerical investigations on two experimental and one synthetic example demonstrate that the reduction lowers the cost of the training by an order of magnitude without degrading the identified material.
In particular, the reduced model remains accurate while the weights evolve, and its gain grows with the size of the full-order problem.
Moreover, a single reaction force suffices to separate the phases of heterogeneous specimens and to predict displacement fields that never entered the loss.

At the same time, the study highlights two limitations.
The efficiency of the reduction decreases when the deformation localizes, and the full-order checks bound the overall speed-up.
In addition, the accuracy of the identified material is limited by the information content of the data rather than by the reduction, since the reaction force constrains the energy only along the deformation states activated by the experiment.

Future work may focus on adaptive full-order checks, nonlinear trial manifolds for localized deformation, and the extension to anisotropic and inelastic materials.

\section*{Acknowledgements}
Hagen Holthusen gratefully acknowledges financial support by the Deutsche Forschungsgemeinschaft (DFG, German Research Foundation) through the projects 495926269 and 417002380.
Kevin Linka gratefully acknowledges financial support by the Deutsche Forschungsgemeinschaft (DFG, German Research Foundation) through the project 517243167.

\section*{Declaration of generative AI and AI-assisted technologies in the writing process}
During the preparation of this work, the authors used Claude (Anthropic) to draft and revise text passages, to improve language and readability, and to assist in the implementation and documentation of the source code. All generated text and code were reviewed, verified, and edited by the authors, including all statements, numerical values, and references, such that the manuscript reflects the authors' own analysis and conclusions. The authors take full responsibility for the content of the published article.

\section*{Data availability}
The data generated in this study will be made publicly available upon publication.
The experimental data of Cases~1 and~2 are openly available from \cite{Lejeune2020,rausch2026mnistchallenge} and \cite{abbasi2026}, respectively.

\FloatBarrier

\appendix


\section{Polyconvexity and coercivity of the energy}
\label{app:polyconvexity_coercivity}

The energy consists of the neural network $\mathcal N$, a stress normalization $\psi_{\mathrm{nrm}}$ and a growth term $\psi_{\mathrm{gr}}$,
\begin{equation}
  \psi(\bm F) = \mathcal N(\bm x) + \psi_{\mathrm{nrm}}(J) + \psi_{\mathrm{gr}}(\bm F),
  \qquad
  \bm x = \big[\bar I_1,\, \bar I_2,\, \bar J,\, -\bar J\big],
  \label{eq:app_energy}
\end{equation}
with the normalized invariants
\begin{equation}
  I_1 = \tfrac13\operatorname{tr}\bm C = \tfrac13\|\bm F\|^2, \qquad
  I_2 = \tfrac13\operatorname{tr}(\operatorname{cof}\bm C) = \tfrac13\|\operatorname{cof}\bm F\|^2, \qquad
  J = \det\bm F > 0,
  \label{eq:app_invariants}
\end{equation}
and $\bar I_1 = I_1 - 1$, $\bar I_2 = I_2 - 1$, $\bar J = J - 1$, such that $\bm x = \bm 0$ at $\bm F = \bm I$. The network is designed as an Input Convex Neural Network (ICNN) \cite{amos_input_2017}, which is additionally zero-anchored $\mathcal N(\bm 0) = 0$ \cite{holthusen2026complement}.
As an ICNN, it has non-negative weights and convex and non-decreasing activations. 
Thus, it is therefore convex and non-decreasing in each input. Its derivatives at the reference configuration are denoted by
\begin{equation}
  a_1 = \frac{\partial\mathcal N}{\partial \bar I_1}\bigg|_{\bm F = \bm I} \ge 0,
  \quad
  a_2 = \frac{\partial\mathcal N}{\partial \bar I_2}\bigg|_{\bm F = \bm I} \ge 0,
  \quad
  a_J = \frac{\mathrm d}{\mathrm d\bar J}\,\mathcal N\big(0,\, 0,\, \bar J,\, -\bar J\big)\bigg|_{\bar J = 0} \in \mathbb R .
  \label{eq:app_derivatives}
\end{equation}
Since the network is non-decreasing in each input, $a_1, a_2 \ge 0$. By the chain rule, $a_J$ is the derivative of $\mathcal N$ with respect to its third input minus that with respect to its fourth input. As a difference of two non-negative numbers, it may take either sign, which allows for both tensile and compressive volumetric stresses.

\paragraph{Normalizations and growth term.}
We compare two stress normalizations,
\begin{equation}
  \psi_{\mathrm{nrm}}^{\mathrm{lin}} = -\big[\tfrac23 a_1 + \tfrac43 a_2 + a_J\big]\,\bar J ,
  \qquad
  \psi_{\mathrm{nrm}}^{\ln} = -\big[\tfrac23 a_1 + \tfrac43 a_2\big]\ln J - a_J\,\bar J ,
  \label{eq:app_normalizations}
\end{equation}
where the former is in line with \cite{linden_neural_2023}.
Further, we employ the growth term with $\epsilon > 0$,
\begin{equation}
  \psi_{\mathrm{gr}} = \epsilon\Big[\big[I_1 - 1 - \tfrac23\ln J\big] + \big[I_2 - 1 - \tfrac43\ln J\big] + \big[J - 1 - \ln J\big]\Big] = \epsilon\big[ I_1 + I_2 + J - 3 - 3\ln J \big].
  \label{eq:app_growth}
\end{equation}
All three terms within this growth energy vanish at $\bm F = \bm I$. With $\partial_{\bm F} I_1 = \tfrac23\bm I$, $\partial_{\bm F} I_2 = \tfrac43\bm I$ and $\partial_{\bm F} J = \partial_{\bm F}\ln J = \bm I$ at $\bm F = \bm I$, the Piola stress of each contribution at the reference configuration reads
\begin{equation}
  \begin{aligned}
    \partial_{\bm F}\mathcal N &= \big(\tfrac23 a_1 + \tfrac43 a_2 + a_J\big)\bm I,
    \qquad
    \partial_{\bm F}\psi_{\mathrm{gr}} = \bm 0, \\
    \partial_{\bm F}\psi_{\mathrm{nrm}}^{\mathrm{lin}} &= \partial_{\bm F}\psi_{\mathrm{nrm}}^{\ln} = -\big(\tfrac23 a_1 + \tfrac43 a_2 + a_J\big)\bm I .
  \end{aligned}
  \label{eq:app_reference_stress}
\end{equation}
Hence $\psi(\bm I) = 0$ and $\bm P(\bm I) = \bm 0$ for both normalizations. In the following, we first show polyconvexity and then coercivity. In both parts, network, normalization and growth term are treated one after another.

\subsection{Polyconvexity}

An energy is polyconvex if it is a convex function of $\bm F$, $\operatorname{cof}\bm F$ and $\det\bm F$ \cite{ball1977}. As a sum of convex functions is convex, each contribution is checked separately. By Eq.~\eqref{eq:app_invariants}, $I_1$ is convex in $\bm F$, $I_2$ is convex in $\operatorname{cof}\bm F$, and $J = \det\bm F$ is itself an argument.

\paragraph{Neural network.}
The argument rests on a simple composition rule: a function that is convex and non-decreasing in each argument, applied to convex functions, is again convex \cite{amos_input_2017,Rockafellar1970}. Every layer computes $\bm{h}_\ell = \varphi_\ell(\bm{W}_\ell\bm{h}_{\ell-1} + \bm{S}_\ell\bm{x} + \bm{b}_\ell) - \varphi_\ell(\bm{b}_\ell)$ with non-negative $\bm{W}_\ell$, $\bm{S}_\ell$ and convex, non-decreasing $\varphi_\ell$. The argument of $\varphi_1$ is affine in $\bm{x}$ with non-negative weights, and hence convex and non-decreasing in $\bm{x}$, which carries over to $\bm{h}_1$. In every further layer, the argument is a non-negative combination of the entries of $\bm{h}_{\ell-1}$ and $\bm{x}$, so the same reasoning applies. Thus, $\mathcal{N}$ is convex and non-decreasing in each entry of $\bm{x}$, where the anchoring term $\varphi_\ell(\bm{b}_\ell)$ is constant in $\bm{x}$. Finally, $I_1 - 1$ and $I_2 - 1$ are convex in $\bm{F}$ and $\operatorname{cof}\bm{F}$, respectively, so the composition rule applies once more, while $\pm[J-1]$ are affine in $J$, for which convexity of $\mathcal{N}$ alone suffices. Hence, $\mathcal{N}(\bm{x}(\bm{F});\bm{\theta})$ is convex in $(\bm{F},\operatorname{cof}\bm{F},J)$, i.e., polyconvex.

\paragraph{Normalization.}
As functions of $J = \det\bm F$, the two normalizations satisfy
\begin{equation}
  \frac{\mathrm d^2\psi_{\mathrm{nrm}}^{\mathrm{lin}}}{\mathrm dJ^2} = 0,
  \qquad
  \frac{\mathrm d^2\psi_{\mathrm{nrm}}^{\ln}}{\mathrm dJ^2} = \frac{\tfrac23 a_1 + \tfrac43 a_2}{J^2} \ge 0 .
  \label{eq:app_nrm_convex}
\end{equation}
The linear normalization is affine and thus convex for any sign of its coefficient. The logarithmic normalization is convex because $a_1, a_2 \ge 0$ by Eq.~\eqref{eq:app_derivatives}; here, the monotonicity of the network is essential.

\paragraph{Growth term.}
With $\epsilon > 0$, $\psi_{\mathrm{gr}}$ is a sum of the convex terms $I_1$, $I_2$, $J$ and $-3\ln J$.

Consequently, the overall energy~\eqref{eq:app_energy} is polyconvex for both normalizations.

\subsection{Coercivity}

Before we investigate the coercivity by a growth condition, let us introduce two essential inequalities.
Let $\lambda_1, \lambda_2, \lambda_3 > 0$ be the singular values of $\bm F$, so that $J = \lambda_1\lambda_2\lambda_3$. 
By the arithmetic-geometric mean, we obtain
\begin{equation}
  \begin{aligned}
    I_1 &= \tfrac13\big(\lambda_1^2 + \lambda_2^2 + \lambda_3^2\big) \ge \big(\lambda_1^2\lambda_2^2\lambda_3^2\big)^{1/3} = J^{2/3}, \\
    I_2 &= \tfrac13\big(\lambda_1^2\lambda_2^2 + \lambda_2^2\lambda_3^2 + \lambda_3^2\lambda_1^2\big) \ge \big(\lambda_1^4\lambda_2^4\lambda_3^4\big)^{1/3} = J^{4/3} .
  \end{aligned}
  \label{eq:app_amgm}
\end{equation}
Moreover, the function $y(x)=x - 1 - \ln x$ attains its minimum zero at $x = 1$, since its derivative $1 - 1/x$ changes sign there. Thus
\begin{equation}
  y(x)=x - 1 - \ln x \ge 0 \qquad \forall\, x > 0 .
  \label{eq:app_log}
\end{equation}

\paragraph{Coercivity and existence of minimizers.}
Existence of minimizers follows from polyconvexity of $\psi$ together with a
coercivity estimate of the form, cf. \cite{ball1977,Schroeder2010CISM},
\begin{equation}
  \psi(\bm F) \ge c_1\big(\|\bm F\|^p + \|\operatorname{cof}\bm F\|^q + J^r\big) + c_0,
  \qquad p \ge 2,\quad q \ge \tfrac{p}{p-1},\quad r > 1,
  \label{eq:app_ball}
\end{equation}
with $c_1 > 0$, and the condition $\psi \to \infty$ as $J \to 0^+$.
We will show the simpler bound 
\begin{equation}
  \psi \ge c\,(I_1 + I_2) + c_0  
\end{equation}
with $c > 0$.
This is sufficient, since with \cref{eq:app_invariants} and
$J^{4/3} \le I_2$ from \cref{eq:app_amgm},
\begin{equation}
  c\,(I_1 + I_2)
  = \tfrac{c}{3}\|\bm F\|^2 + \tfrac{c}{6}\|\operatorname{cof}\bm F\|^2 + \tfrac{c}{2} I_2
  \ge \tfrac{c}{6}\big(\|\bm F\|^2 + \|\operatorname{cof}\bm F\|^2 + J^{4/3}\big) ,
  \label{eq:app_reduction}
\end{equation}
which is \cref{eq:app_ball} with $p = q = 2$ and $r = 4/3$.

\paragraph{Growth term.}
We first show that each bracket in \cref{eq:app_growth} is non-negative and, for the first two, bounded below by half the respective invariant. For the first bracket we use $I_1 \ge J^{2/3}$ from \cref{eq:app_amgm} and \cref{eq:app_log} with $x = J^{2/3}$, i.e., $x - 1 - \ln x \ge 0$,
\begin{equation}
  I_1 - 1 - \tfrac23\ln J = \big(J^{2/3} - 1 - \ln J^{2/3}\big) + \big(I_1 - J^{2/3}\big) \ge 0 .
  \label{eq:app_bracket_nonneg}
\end{equation}
Using $I_1 \ge J^{2/3}$ for only one half of $I_1$ and $\tfrac12 x - \ln x \ge 1 - \ln 2 > 0$ for all $x > 0$ (minimum at $x = 2$), we further obtain
\begin{equation}
  I_1 - 1 - \tfrac23\ln J \ge \tfrac12 I_1 + \big(\tfrac12 J^{2/3} - \ln J^{2/3}\big) - 1 \ge \tfrac12 I_1 - 1 .
  \label{eq:app_bracket_lower}
\end{equation}
The second bracket follows in the same way, i.e. $I_2-1-\tfrac43\ln J\geq \tfrac12 I_2-1$, and the third, $J - 1 - \ln J$, is non-negative by \cref{eq:app_log} and tends to infinity for $J \to 0^+$.
Summing the three brackets, with the first two bounded by Eq.~\eqref{eq:app_bracket_lower} and the third kept as it is, yields
\begin{equation}
  \psi_{\mathrm{gr}} \ge \tfrac{\epsilon}{2}\,(I_1 + I_2) - 2\epsilon + \epsilon\,(J - 1 - \ln J) .
  \label{eq:app_growth_bound}
\end{equation}
The first term provides the required growth, and the last one is non-negative and tends to infinity for $J \to 0^+$, so the growth term alone satisfies the desired bound.
It remains to show that network and normalization together do not destroy it, i.e., that their sum is bounded from below.

\paragraph{Neural Network.}
From the convexity of $\mathcal N$, cf.~\cite{Rockafellar1970}, we obtain
\begin{equation}
  \mathcal N(\bm x) \;\ge\; \mathcal N(\bm 0) + \nabla\mathcal N(\bm 0)\cdot\bm x \;=\; a_1\bar I_1 + a_2\bar I_2 + a_J\bar J ,
  \label{eq:app_tangent}
\end{equation}
since $\mathcal N(\bm 0) = 0$.
This bound is sharp, as equality holds for a network that consists of the output skip connection only, i.e. linear in $\bm x$. A trained network can therefore be (almost) linear, so we cannot rely on its curvature to provide additional growth. The coercivity of $\psi$ must instead come from the explicit growth term $\psi_{\mathrm{gr}}$.

\paragraph{Linear normalization.}
Adding $\psi_{\mathrm{nrm}}^{\mathrm{lin}}$ to \cref{eq:app_tangent}, the terms with $a_J$ cancel and the remaining terms can be grouped by the weights $a_1$ and $a_2$,
\begin{equation}
  \mathcal N + \psi_{\mathrm{nrm}}^{\mathrm{lin}}
  \;\ge\; a_1\big[\bar I_1 - \tfrac23\bar J\big] + a_2\big[\bar I_2 - \tfrac43\bar J\big] .
  \label{eq:app_lin_split}
\end{equation}
Both brackets contain a term that is linear in $J$, so the question is whether $I_1$ and $I_2$ grow fast enough to compensate it. For the second bracket, $I_2 \ge J^{4/3}$ from \cref{eq:app_amgm} gives
\begin{equation}
  \bar I_2 - \tfrac43\bar J \;\ge\; J^{4/3} - 1 - \tfrac43(J-1) \;\ge\; 0 .
  \label{eq:app_lin_second}
\end{equation}
For the first bracket, Eq.~\eqref{eq:app_amgm} only provides $I_1 \ge J^{2/3}$, i.e., $J \le I_1^{3/2}$, so $J$ may grow faster than $I_1$. A pure dilatation $\bm F = \lambda\bm I$ with $I_1 = \lambda^2$ and $J = \lambda^3$ shows that this happens,
\begin{equation}
  \bar I_1 - \tfrac23\bar J \;=\; I_1 - \tfrac13 - \tfrac23 J \;=\; \lambda^2 - \tfrac13 - \tfrac23\lambda^3 \;\xrightarrow{\lambda\to\infty}\; -\infty .
  \label{eq:app_lin_first}
\end{equation}
Network and linear normalization together are therefore not bounded from below, and the growth term has to provide the missing coercivity. We add $\psi_{\mathrm{gr}}$, use $a_1, a_2 \ge 0$, drop the non-negative second bracket and the non-negative term $a_1 I_1$ of the first one, and bound $\psi_{\mathrm{gr}}$ by \cref{eq:app_growth_bound} without the non-negative term $\epsilon\,(J - 1 - \ln J)$,
\begin{equation}
  \begin{aligned}
    \psi
    \;\ge\; \tfrac{\epsilon}{2}\,(I_1 + I_2) - 2\epsilon + a_1\big[I_1 - \tfrac13 - \tfrac23 J\big] \;\ge\; \tfrac{\epsilon}{2}\,(I_1 + I_2) - 2\epsilon - \tfrac{a_1}{3} - \tfrac23 a_1 J .
  \end{aligned}
  \label{eq:app_lin_step1}
\end{equation}
The only negative contribution is $-\tfrac23 a_1 J$. Against it, we have the growth term $\tfrac{\epsilon}{2} I_2 \ge \tfrac{\epsilon}{2} J^{4/3}$, which grows faster than $J$. We reserve one half of it, $\tfrac{\epsilon}{4} I_2$, for this purpose and minimize the resulting expression over $J$,
\begin{equation}
  g(J) = \tfrac{\epsilon}{4}\,J^{4/3} - \tfrac23 a_1 J ,
  \qquad
  g'(J) = 0 \;\Rightarrow\; J^{1/3} = \tfrac{2a_1}{\epsilon} ,
  \qquad
  g \;\ge\; -\frac{4a_1^4}{3\epsilon^3} .
  \label{eq:app_lin_min}
\end{equation}
The remaining terms in Eq.~\eqref{eq:app_lin_step1} are combined via $\tfrac{\epsilon}{2} I_1 + \tfrac{\epsilon}{4} I_2 \ge \tfrac{\epsilon}{4}(I_1 + I_2)$. Together with Eq.~\eqref{eq:app_lin_min}, this yields
\begin{equation}
  \psi \;\ge\; \tfrac{\epsilon}{4}\,(I_1 + I_2) - C_{\mathrm{lin}},
  \qquad
  C_{\mathrm{lin}} = \frac{4a_1^4}{3\epsilon^3} + \frac{a_1}{3} + 2\epsilon .
  \label{eq:app_lin_bound}
\end{equation}
The energy is thus coercive, but the constant grows with $\epsilon^{-3}$ and becomes very large for a small growth term.

\paragraph{Logarithmic normalization.}
Adding $\psi_{\mathrm{nrm}}^{\ln}$ to Eq.~\eqref{eq:app_tangent}, the terms with $a_J$ again cancel,
\begin{equation}
  \begin{aligned}
    \mathcal N + \psi_{\mathrm{nrm}}^{\ln}
    &\;\ge\; a_1\bar I_1 + a_2\bar I_2 + a_J\bar J - \big(\tfrac23 a_1 + \tfrac43 a_2\big)\ln J - a_J\bar J \\
    &\;=\; a_1\big[I_1 - 1 - \tfrac23\ln J\big] + a_2\big[I_2 - 1 - \tfrac43\ln J\big] .
  \end{aligned}
  \label{eq:app_log_split}
\end{equation}
These are exactly the first two brackets of the growth term, which are non-negative by Eq.~\eqref{eq:app_bracket_nonneg}. Since also $a_1, a_2 \ge 0$, network and logarithmic normalization together are non-negative. Adding the growth term and using Eq.~\eqref{eq:app_growth_bound},
\begin{equation}
  \psi \;=\; \underbrace{\mathcal N + \psi_{\mathrm{nrm}}^{\ln}}_{\ge\,0} + \psi_{\mathrm{gr}}
  \;\ge\; \psi_{\mathrm{gr}} \;\ge\; \tfrac{\epsilon}{2}\,(I_1 + I_2) - 2\epsilon ,
  \qquad
  \psi \ge 0 = \psi(\bm I) .
  \label{eq:app_log_final}
\end{equation}
In contrast to \cref{eq:app_lin_bound}, the bound in \cref{eq:app_log_final} does not depend on the trained network weights, since the constant $-2\epsilon$ involves only the growth parameter $\epsilon$, whereas $C_{\mathrm{lin}}$ grows with $a_1^4$.

\section{Mean-dilatation element (Q1/P0)}
\label{app:q1p0}

For nearly incompressible materials, trilinear displacement elements enforce $J \approx 1$ at every quadrature point. These are too many constraints for the available displacement degrees of freedom, and the element locks volumetrically. The mean-dilatation approach \cite{nagtegaal1974,hughes1980,simo1985} reduces the volume change to a single value per element. In the following, we introduce the modified kinematics, derive the resulting displacement-based potential and its variation, and conclude with the consequence for the hyperreduction.

\paragraph{Kinematics.}
For each element $e$, we define the mean dilatation
\begin{equation}
    \Theta^e(\mathbf{d}^e) := \frac{1}{V^e}\int_{\mathcal{B}^e} J\,\mathrm{d}V,
    \label{eq:app_mean_dilatation}
\end{equation}
where $V^e$ denotes the reference volume of element $e$. With this at hand, the modified deformation gradient reads
\begin{equation}
    \hat{\bm{F}} := \Big[\frac{\Theta^e}{J}\Big]^{1/3}\bm{F},
    \qquad
    \det\hat{\bm{F}} = \Theta^e .
    \label{eq:app_Fhat}
\end{equation}
Hence, $\hat{\bm{F}}$ shares the isochoric part $J^{-1/3}\bm{F}$ with $\bm{F}$, but carries the element-wise constant volume change $\Theta^e$ instead of $J$.

\paragraph{Potential.}
We evaluate the energy at $\hat{\bm{F}}$, such that the discrete potential becomes
\begin{equation}
    \Pi^h(\mathbf{d},\bm{\theta}) = \sum_{e=1}^{n_{\mathrm{el}}} \underbrace{\int_{\mathcal{B}^e} \psi\big(\hat{\bm{F}}(\mathbf{d}^e);\bm{\theta}\big)\,\mathrm{d}V}_{=:\,\Pi^{h,e}_{\mathrm{int}}(\mathbf{d}^e,\bm{\theta})} - \;\Pi^h_{\mathrm{ext}}(\mathbf{d}).
    \label{eq:app_q1p0_potential}
\end{equation}
This is a pure displacement formulation: the displacements are interpolated trilinearly (Q1), and no further unknowns occur. Moreover, the construction requires no split of $\psi$ into volumetric and isochoric parts and thus applies directly to the energy of \cref{eq:energy}.

\section{Case 2: Identification from force and displacements}
\label{app:case2_fd}
\setcounter{figure}{0}

\begin{figure}[tbp]
    \centering
    \includegraphics[width=\textwidth]{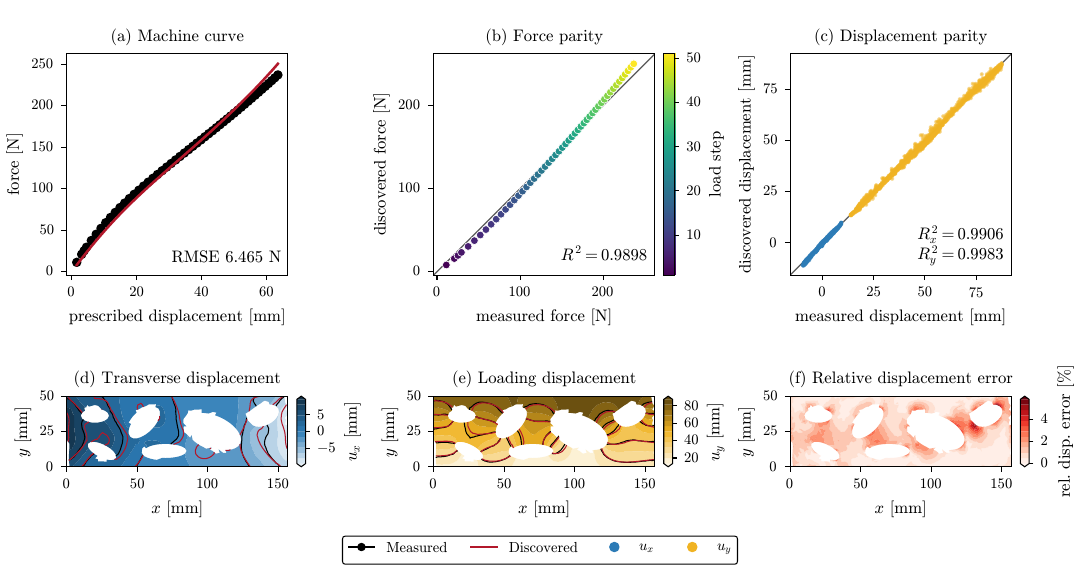}\\[4pt]
    \includegraphics[width=\textwidth]{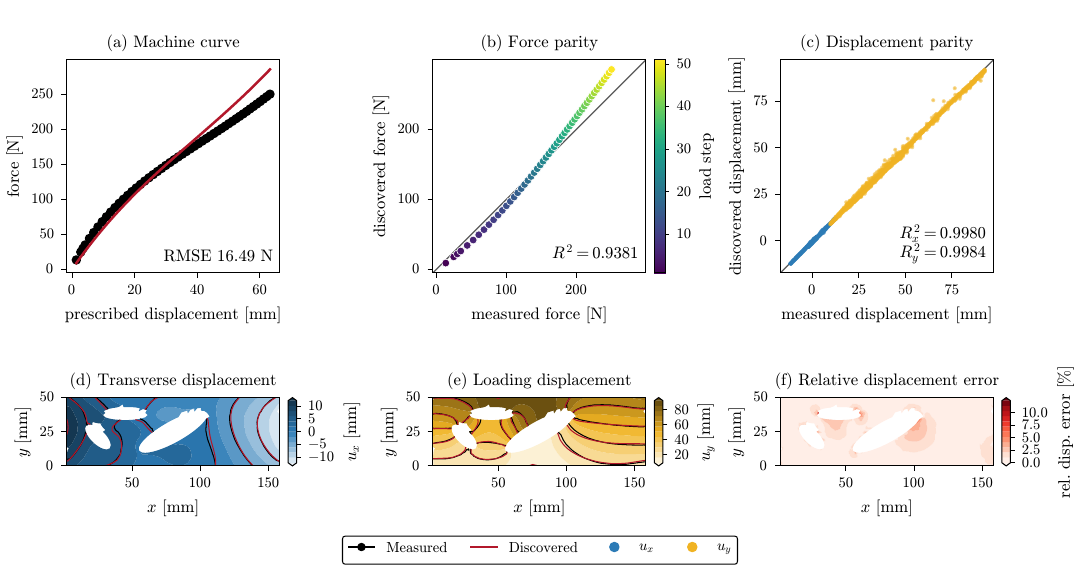}
    \caption{\textbf{Case~2, material identified from force and displacements, evaluated by a full-order solve on the training specimen TT1 (top) and on the unseen specimen TT2 (bottom).} Top: (a)--(b) The force is reproduced less accurately than in \cref{fig:abbasi_eval_force}, with a root-mean-square error of $6.47$~N and $R^2=0.990$; the material is too soft at small and too stiff at large displacements. (c)--(f) In turn, the displacement field agrees with $R^2_x=0.991$ and $R^2_y=0.998$, and the relative error falls below $2\,\%$ in most of the specimen. The trade-off between both channels indicates that force and field are not fully consistent within the present model. Bottom: For TT2, the force at large displacements is overestimated by about $10\,\%$ (root-mean-square error of $16.5$~N), as for the force-based material, whereas the displacement field benefits from the additional data: $R^2_x$ increases from $0.892$ to $0.998$, and the relative error drops below $5\,\%$ almost everywhere.}
    \label{fig:abbasi_eval_fd}
\end{figure}

This appendix complements \cref{sec:case2} by the material identified from the loss \cref{eq:abbasi_loss_fd}, which contains both the reaction force and the displacements.
The training history and the adaptivity of the reduction of this run are shown in \cref{fig:abbasi_training_fd}.
\cref{fig:abbasi_eval_fd} evaluates the identified material by a full-order solve on the training specimen TT1 and on the unseen specimen TT2, in the same way as \cref{fig:abbasi_eval_force} for the force-based material.
Compared with the latter, the additional displacement data improve the agreement of the displacement field on both specimens at the expense of the force on TT1, whereas the force on TT2 is overestimated by about the same amount, see \cref{sec:case2}.

\bibliographystyle{unsrtnat}  
\bibliography{references,references_intro}
\end{document}